\documentclass[12pt]{article}
\usepackage{amssymb,epsf,euscript,graphicx}
\usepackage{subfigure}

 \usepackage[margin=1.0in]{geometry}

 \usepackage{wrapfig}

 \usepackage[hyphens]{url} 
\usepackage{hyperref}  
\hypersetup{breaklinks=true}  

\usepackage{cite}

\usepackage{enumitem}

\usepackage{lipsum} 

\usepackage{mathrsfs}
\usepackage{amsmath} 

\newcommand{\beq}{\begin{equation}}
\newcommand{\eeq}{\end{equation}}
\newcommand{\barr}{\begin{array}}
\newcommand{\earr}{\end{array}}
\newcommand{\beqarr}{\begin{eqnarray}}
\newcommand{\eeqarr}{\end{eqnarray}}
\newcommand{\beqar}{\begin{eqnarray*}}
\newcommand{\eeqar}{\end{eqnarray*}}
\newcommand{\bef}{\begin{figure}}
\newcommand{\eef}{\end{figure}}

\newcommand{\bm}[1]{\mbox{\boldmath$#1$}}

    \usepackage{fancyhdr}
\usepackage{blindtext}
\fancypagestyle{mypagestyle}{
\fancyhf{}
\fancyfoot[C]{\thepage}  

}
\makeatletter
\let\ps@plain\ps@mypagestyle
\makeatother
\fancypagestyle{firstpage}
{
\chead{
\makebox[\linewidth]{\rule{\paperwidth}{2.0pt}}
\textbf{PROJECT DESCRIPTION}
\vspace{-0.2in}
\makebox[\linewidth]{\rule{\paperwidth}{2.0pt}}
}
}

\fancypagestyle{lastpage}
{
\chead{
\makebox[\linewidth]{\rule{\paperwidth}{2.0pt}}
\textbf{BIBLIOGRAPHY}
\vspace{-0.2in}
\makebox[\linewidth]{\rule{\paperwidth}{2.0pt}}
}
}

\begin{document}

\begin{center}
\textbf{\large{A Response to ``Application of Gauss's Principle to the Classical Airfoil Lift Problem"}}\\
By Haithem E. Taha
\end{center}

\begin{abstract}
The classical theory of lift developed by Kutta and Zhukovsky is confined to sharp-edged airfoils. The search for a more general closure condition in potential flow remained elusive for over a century. Recently, a variational theory of lift, inspired by Gauss's principle of least constraint, was proposed as a remedy for this long-standing puzzle. The theory was shown to recover the Kutta condition as a special case for sharp-edged airfoils. However, recent criticism of the variational theory has asserted fundamental issues and discontinuities in its predictions. The present paper demonstrates that these assertions are incorrect and arise from inconsistencies with basic principles of analytical mechanics, the calculus of variations, and ideal-flow aerodynamics, as well as from misapplications of the variational theory itself.

To resolve such misunderstandings, we review foundational concepts principles from analytical mechanics, including least action, Gauss's principle, and Hertz's principle; the definitions of impressed and constraint forces; the \textit{Simplest Problem in the Calculus of Variations}; and the distinction between \textit{actual} work and \textit{virtual} work. We then place these concepts in the context of incompressible fluid mechanics, showing that the Helmholtz decomposition provides a geometric interpretation of the incompressibility constraint. In particular, we demonstrate that, for incompressible flows subject to the no-penetration boundary condition, the pressure force is orthogonal to the entire space of kinematically admissible velocity fields and therefore performs no \textit{virtual} work on the configuration manifold. The pressure force, thus, acts as the \textit{constraint force} required to ensure the continuity and no-penetration constraints. Consequently, extending Gauss's principle of least constraint to this setting yields the principle of minimum pressure gradient.

From an aerodynamic perspective, we show that the classical and variational theories of lift, as well as any theory based on steady, irrotational motion, are necessarily reversible and therefore inapplicable to reversed-flow configurations. Apparent paradoxes associated with reversed airfoils are shown to result from extending ideal-flow theories beyond their domain of applicability, rather than from deficiencies of the variational formulation. Within its proper scope, the variational theory predicts a continuous dependence of circulation and lift on geometric parameters, consistent with classical results and physical intuition.
\end{abstract}


\section{Introduction}
The classical theory of lift by Kutta and Zhukovsky played a central role in the development of modern aviation. It has long served as a cornerstone in forming our body of knowledge for aerodynamic design and analysis. It lies in the heart of the thin airfoil theory that elucidates the effect of camber and flap deflection on lift and moment, which was used to design the skeleton (camber line) of some interesting airfoils such as the NACA 6-series \cite{Schlichting}. Of particular importance to aviation is the Prandtl-Lanchester wing theory \cite{Prandtl_Theory}, which accounts for the effects of aspect ratio, twist and spanwise camber distribution on lift and induced drag. This theory enabled designers to tailor wing geometries to achieve minimum induced drag at a prescribed lift coefficient, simultaneously with good stall characteristics. As Von Karman mentioned, ``\textit{the wing theory became the very basis of the scientific design of all our airplanes, at least as far as the domain of moderate speeds is concerned}" \cite{VonKarman_Book}. Here, it must be emphasized that the wing theory fundamentally incorporates Kutta's two-dimensional theory: it simply assumes that each spanwise section behaves according to Kutta's airfoil theory, but at a smaller angle of attack than the geometric one.

The central role of Kutta's theory extends beyond steady aerodynamics into unsteady lift evolution. From a dynamical-system perspective, linear unsteady aerodynamic models can ultimately be cast into standard forms of linear dynamical systems: step response, frequency response, transfer function, or state space model \cite{Potential_Flow_Lift_Dynamics}. From this perspective, any unsteady aerodynamic theory is primarily concerned with how lift builds up in time toward a steady state, and that steady-state limit is determined by Kutta's theory. In this sense, the two-dimensional Kutta–Zhukovsky theory of lift constitutes a foundational element of aerodynamics and has played an enabling role in the development of aviation.

In the language of Thomas Kuhn, as articulated in \textit{The Structure of Scientific Revolutions} \cite{Kuhn_Structure}, the classical theory of lift defined the \textit{paradigm} within which \textit{normal science} in aerodynamics has been pursued over the past century. Many efforts were devoted to extending its scope and improving its predictive capability. However, despite its remarkable success, the theory has a severely restricted domain of applicability. It applies strictly to (i) steady flows over (ii) streamlined bodies at (iii) small angles of attack, and perhaps the most prohibitive limitation is (iv) the requirement of a mathematically sharp trailing edge.

These limitations give rise to what, in Kuhn's terminology, may be regarded as \textit{anomalies}. No physically realizable airfoil section can possess a mathematically sharp trailing edge, even if the designer and manufacturer aim relentlessly to approximate one. While the aerodynamic consequences of trailing-edge rounding may be modest for some airfoils (often resulting in less than a 10\% reduction in lift when experimental measurements are compared with Kutta's theory), they can be severe for others (e.g., \cite{ross1993computational}). For example, a Zhukovsky airfoil with even slight trailing-edge rounding may lose about 30\% of its lift at angles of attack as small as $2^\circ$, as demonstrated by the computational simulations presented in figures 4(b) and 8(b) in \cite{Kutta_Flat_Plate,PMPG_PoF}, respectively. Similarly, Jones and Ames \cite{ames1942wind} have shown a significant effect of trailing edge geometry on flap hinge moment. These important effects cannot be examined within the classical theory because it is agnostic to trailing edge geometry; it assumes that every airfoil must have a sharp trailing edge. If, nevertheless, we insist on applying it to shapes that do not satisfy this requirement, then we are operating outside its domain of validity; we may sometimes be lucky, but not always.

The situation is exacerbated even more in the unsteady case. It appears that the departure from a sharp trailing edge has lead to serious discrepancies between the classical theory of unsteady aerodynamics and both experimental measurements and computational simulations. Indeed, the literature is replete with concerns about the validity of the Kutta condition in unsteady flow scenarios, see, for example, Sears \cite{Sears_Review,Sears_Two_Models}, Cirghton \cite{Kutta_Crighton}, Xia and Mohseni \cite{Mohseni_Kutta_JFM}, Taha and Rezaei \cite{Viscous_Freq_Resp,Viscous_UVLM_PoF,Viscous_SSM_AIAAJ}, Zhu et al. \cite{Gregory_JFM_Isaacs}, and the references therein.

The fundamental issue lies in the fact that the classical theory of lift by Kutta and Zhukovsky offers little to no guidance for generalization beyond the narrow setting for which it was originally formulated, nor does it suggest systematic relaxations of its underlying assumptions. It was not derived from first principles and is therefore sometimes described as a \textit{mere} mathematical closure applicable only to a specific scenario. Luckily, it was the scenario most relevant to aeronautics during its early development. The \textit{anomalies}, discussed above, thus point to serious deficiencies in the classical framework. Whether they have grown to the level of a \textit{crisis} in the language of Kuhn \cite{Kuhn_Structure} remains undetermined. Nevertheless, Kuhn's historical analysis reminds us that, despite their clarity and seriousness, some practitioners (and possibly a few leaders in the field) may not recognize such anomalies when they appear, owing to deep intellectual commitment to a prevailing paradigm---a pattern that has recurred repeatedly in the history of science.

In response to these deficiencies in the classical theory of lift, Gonzalez and Taha \cite{Variational_Lift_JFM} have proposed a variational theory of lift aimed at resolving some of its limitations, particularly relaxing the overly prohibitive requirement of a sharp trailing edge. The theory was inspired by Gauss's principle of least constraint and its special case of least curvature due to Hertz. Whereas Kutta's theory resolves the chronic non-uniqueness in potential flow by selecting a unique value of circulation that removes the singularity at a sharp trailing edge, the variational theory of lift provides closure by minimizing the total curvature of the flow field. In the limit as the rear portion of an airfoil approaches a mathematically sharp edge, the minimum-curvature circulation converges to Kutta's value. In this sense, Kutta's theory may be viewed as a special case of the variational theory.

As noted above, the variational theory of lift is rooted in Gauss's principle of least constraint and Hertz's principle of least curvature---concepts which have remained entirely absent from the aerodynamic literature; nor have they appeared consistently in modern textbooks of analytical mechanics. Indeed, they are rarely taught in contemporary engineering graduate curricula. In fact, Papastavridis \cite{Papastavridis} remarked, ``\textit{In most of the $20^{rm{th}}$ century English literature, GP [Gauss Principle] has been barely tolerated as a clever but essentially useless academic curiosity, when it was mentioned at all.}" Only a few efforts have employed it in the twenty-first century (e.g., \cite{Udwadia_Kalaba_Book}). In this context, it is perhaps unsurprising that the variational theory of lift (and, in particular, its reliance on Gauss's principle) may be misinterpreted by some aeronautical engineers.

Recently, Peters and Ormiston \cite{Peters_Gauss} have raised concerns about the validity of the variational theory of lift. To any reader familiar with the history of science, such resistance is entirely natural \cite{Resistance}. It is extremely challenging to identify a theory of lasting impact that has entered the scientific canon without encountering sustained skepticism. From the Copernican revolution \cite{Kuhn_Copernican_Revolution}, to Einstein's relativity \cite{wazeck2014einstein}, and Lavoiser's discovery of oxygen \cite{Kuhn_Structure}, no theory was unconditionally welcomed---and nor should it have been. Indeed, such skepticism is a major strength of the scientific enterprise and is essential for safeguarding objectivity in science. A new theory must be subjected to careful scrutiny and serious examination until it either demonstrates clear advantages over the existing paradigm or is ultimately refuted and set aside.

Although the present context does not rise to the stature of the foundational theories cited above, the level of scrutiny should be no less serious. From this perspective, the efforts of Peters and Ormiston \cite{Peters_Gauss} are to be commended. In fact, it may even be exhilarating to witness a rekindle of rich scientific debate within the aeronautical engineering community after nearly a century of relative quiescence, following the prolific exchanges between the Gottingen school (led by Prandtl) and the Cambridge school (led by Rayleigh and Taylor) concerning the classical theory of lift, as virtuosically described by Bloor \cite{Bloor_Enigma}.

However, a careful examination of Ref. \cite{Peters_Gauss} reveals inconsistencies with several basic, well-established concepts from analytical mechanics, geometric mechanics, and the calculus of variations---precisely the disciplines on which the variational theory of lift is grounded. For example, the analysis in Ref. \cite{Peters_Gauss} contradicts the very definitions of constraint and impressed forces \cite{Lanczos_Variational_Mechanics_Book,Papastavridis}. It conflates the concepts of virtual work and actual work. It overlooks the classical Helmholtz orthogonality between divergence-free fields and gradient fields, which is one of the cornerstones of the variational theory. In addition, it fails to recognize a standard solution to what is known as \textit{the simplest problem in the calculus of variations} \cite{Burns_Optimal_Control_Book}. It also includes misinterpretations and inaccurate applications of the variational theory itself. In light of these inconsistencies, the conclusions reached in Ref. \cite{Peters_Gauss} are perhaps not unexpected. The objective of this paper is therefore twofold: first, to clarify the relevant foundational concepts for aeronautical engineering students and practitioners who may not be familiar with them; and second, to identify and explain the specific points at which the analysis in Ref. \cite{Peters_Gauss} departs from these principles.

\section{Technical Background}
\subsection{Classical Airfoil Theory}\label{sec:Classical_Theory}
A brilliant insight of the early pioneers such as Frederick Lanchester was that good aerodynamic designs are those that maintain a smooth flow over their surfaces; i.e., in modern terms, they are streamlined bodies that sustain attached flow and avoid separation. Another remarkable observation, due to Prandtl, was that the flow over such streamlined bodies is largely irrotational, except within a very thin layer adjacent to the surface. Based on these two facts, Lanchester proposed the circulation theory of lift: the flow over a two-dimensional wing may be modelled as an irrotational flow with circulation $\Gamma$; and the resulting lift force is directly proportional to $\Gamma$ through the Kutta-Zhukovksy theorem $L=\rho U \Gamma$.

The flow setup of the circulation theory of lift (i.e, an irrotational flow with circulation) is strictly written in the form
\begin{equation}\label{eq:Family}
\bm{u}(\bm{x})=\bm{u}_0(\bm{x})+\Gamma \bm{u}_1(\bm{x}),
\end{equation}
where $\bm{u}_0$ is the non-circulatory flow and $\bm{u}_1$ is the circulatory flow with unit circulation. They can be clearly understood in the case of a circular cylinder. In this case, $\bm{u}_0$ is the standard potential-flow solution over a circular cylinder whose complex potential is given by  \cite{Karamcheti}
\begin{equation}
F_0(\zeta) =  U_\infty\left[e^{-i\alpha}\zeta + e^{i\alpha}\frac{a^2}{\zeta}\right] ,\
\end{equation}
where $U_\infty$ is the freestream velocity, $\alpha$ is the angle of attack, and $a$ is the cylinder radius. Also, $\bm{u}_1$ is the potential-flow due to a point vortex of unit strength at the center of the cylinder: $\bm{u}_1 = \frac{-1}{2\pi r}\bm{e}_\theta$, where $\bm{e}_\theta$ is a unit vector in the tangential direction. Moreover, the Riemann mapping theorem ensures that any simply connected domain can be (biholomorphically) mapped to the open disc; i.e., the cylinder domain ($\zeta$) can be mapped to a prescribed shape in the $z$-domain through some conformal map $z=f(\zeta)$. Consequently, the flow over any two-dimensional body can be constructed from the flow over a cylinder; the resulting flow will ultimately be in the form (\ref{eq:Family}).

The form (\ref{eq:Family}) of the circulation theory of lift automatically satisfies continuity
\begin{equation}\label{eq:Continuity}
\bm\nabla \cdot \bm{u}(\bm{x})=0
\end{equation}
for all $\bm{x}$ in the domain $\Omega$ exterior of the body, since both $\bm{u}_0$ and $\bm{u}_1$ are potential (i.e., incompressible, irrotational). Also, the form (\ref{eq:Family}) satisfies Euler's equation
\begin{equation}\label{eq:Euler}
\rho \bm{u}\cdot \bm\nabla \bm{u} = -\bm\nabla p
\end{equation}
for any value of $\Gamma$. Moreover, the form (\ref{eq:Family}) satisfies the no-penetration boundary condition $\bm{u}\cdot \bm{n}=0$ on the surface of the body and the far field boundary condition $\lim_{|\bm{x}| \to \infty} \bm{u}(\bm{x})=(U_\infty,0)$ since $\bm{u}_0$ approaches the freestream velocity $U_\infty$ at infinity and $\bm{u}_1$ vanishes there. As such, the family (\ref{eq:Family}) is a legitimate solution of the steady Euler problem for any value of $\Gamma$. In particular, any solution $\bm{u}(\bm{x};\Gamma)$ in the family satisfies the quadruple: (i) momentum conservation (\ref{eq:Euler}), (ii) continuity (\ref{eq:Continuity}), (iii) the no-penetration boundary condition, and (iv) the far field boundary condition.

One may derive the family (\ref{eq:Family}) from a purely mathematical standpoint. Let us ask: what is the set of solutions to Euler's problem that satisfy the four conditions (i)-(iv) listed above? A straightforward mathematical analysis reveals that the set is not finite (i.e., there are infinitely many solutions) and that the family (\ref{eq:Family}) is necessarily contained within the set, since every member of the family satisfies the four conditions. Hence, the \textit{presence} of $\Gamma$ in the family of legitimate solutions of Euler is entirely natural; it need not be externally imposed, as claimed in Ref. \cite{Liu_Steady_Lift}, nor ``\textit{theologically}" assumed, in the words of the late Southwell of Cambridge in his 1930 James Forrest Lecture \cite{Bloor_Enigma}.

This observation directly refutes the common criticism of the classical theory that it ``\textit{assumes the existence of circulation}." Indeed, a rigorous, brute-force mathematical solution of Euler's equation automatically yields infinitely many possible flows---all but one possessing nonzero circulation. An objective researcher must treat all such solutions on equal footing until an additional physical condition tips the balance in favor of a particular flow in the family. The situation is analogous to solving, for example, the cubic equation
\[ x^3 - 6x +11x-6=0.\]
One must obtain all solutions and regard them equally, until an additional physical or mathematical criterion is enforced to select one solution from the family.

The above discussion does not, in any way, suggest that viscosity is unnecessary for the \textit{generation} of circulation, since generation is inherently an unsteady process, whereas the presented analysis is concerned with the steady picture. It does, however, make it clear that viscosity is not required for the \textit{presence} of circulation in the steady-state limit. Indeed, circulatory solutions are perfectly legitimate solutions of the steady Euler equations for ideal flows; they arise naturally from a direct mathematical analysis of Euler's equations, without invoking an external agent such as viscosity. Nevertheless, the non-uniqueness remains unresolved, calling for an additional physical condition that might come from viscous considerations.

This is where the Kutta–Zhukovsky condition plays its indispensable role, by selecting the physical solution from within the family. Without it, the shown analysis would be futile for aeronautical engineering. The steady Euler analysis, presented above, provides the flow field $\bm{u}(\bm{x})$ everywhere, except for only one unknown parameter; and this special parameter solely dictates the lift force. Ironically, the steady Euler equations provide everything except the very quantity that early aeronautical engineers needed---the lift force. It is only by the application of the Kutta-Zhukovsky condition that the circulation, and hence the lift, become uniquely determined, underscoring the \textit{Sine qua non} character of the Kutta condition.

What, then, is the Kutta-Zhukovky condition? For airfoils with sharp trailing edges, the velocity field $\bm{u}$ typically has a singularity at the sharp edge; for every member in the family (\ref{eq:Family}), the velocity $\bm{u}$ goes to infinity at the edge, except for a single solution---this must be the physical solution. It is the only member of the family with bounded velocity field. Indeed, the Kutta condition is incontestably natural for an airfoil with a sharp trailing edge. It is simply a singularity removal condition \cite{Kutta_Crighton}. This observation relegates the commonly asserted viscous nature of the Kutta condition to a secondary role. Viscosity might be necessary for the unsteady evolution toward the steady-state limit. However, if one focuses exclusively on the steady picture (e.g., \cite{Liu_Steady_Lift}), there is no need to invoke viscous considerations for the determination or explanation of lift over airfoils with sharp trailing edges. This point was eloquently articulated by Batchelor (\cite{Batchelor}, p. 437): ``\textit{This fortunate circumstance, that the effect of viscosity acting in the boundary layer initially is to cause the establishment of precisely the value of the circulation that enables effects of viscosity to be ignored ... in the subsequent steady motion, is usually given the name Joukowsky's hypothesis}."

If the airfoil does not have a sharp trailing edge, then we are back to square one: all members of the family  (\ref{eq:Family}) have bounded velocity fields everywhere; there is no singularity to remove. In this case, the Kutta condition becomes tautological and consequently loses its ability to select a unique value of circulation. From this perspective, one cannot but admire the remarkable engineering ingenuity of the early pioneers (Lanchester, Prandlt, Kutta, and Zhukovsky) who devised a theory that is successful in a very specific scenario: steady flow over streamlined bodies with sharp (or nearly sharp) trailing edges at relatively small angles of attack. This is precisely the scenario of primary importance to aviation. Their focus on developing a pragmatic theory that served the immediate needs of their time stands in contrast to the more idealistic aspirations of the Cambridge school, which sought a fully general theory \cite{Bloor_Enigma}. Interestingly, the quest for such a general theory has persisted for nearly a century.

\subsection{Variational Principles in Mechanics}
The history of variational principles can be traced at least to the French philosopher and mathematician Pierre Louis Maupertuis, who was perhaps the first to propose the beautiful \textit{hypothesis} of \textit{Least Action} \cite{Dugas}. He asserted that every motion---whether a planet orbiting the sun, a tree leaf wafting to the ground, or a viola string vibrating---takes place in a way that minimizes a certain fundamental quantity: the action. This philosophical insight effectively transformed mechanics into an optimization problem. If one wishes to analyze the motion of, say, a projectile, one may of course solve Newton's equations of motion. Alternatively, Maupertuis suggested considering \textit{all possible} trajectories and selecting the one that minimizes the total action over the motion. It is a pure optimization problem. It is an alternative way of doing mechanics without invoking Newton's equations. One may even imagine that, in another historical development of mechanics, natural philosophers could have arrived at this principle before $F=ma$, making it the standard way of teaching mechanics at the undergraduate level.

Maupertuis proposed the quantity $S = \int m v ds$ as the action, where $m$ is the mass, $v$ the speed, and $ds$ is the traveled distance. He neither derived this expression from a mathematical argument, nor inferred it from an experimental investigation. This departure from the tradition of rational mechanics, established after Galileo, was the main reason for the scathing criticism directed at Maupertuis and his principle of least action \cite{Dugas}. It was not until Lagrange that a rigorous foundation was provided: under the condition of energy conservation, he proved that every motion indeed renders $S$ \textit{stationary}.

\begin{figure*}
\begin{center}
$\begin{array}{cc}
\subfigure[Schematic of a projectile motion.]{\label{Fig:Projectile_Schematic}\includegraphics[width=7cm]{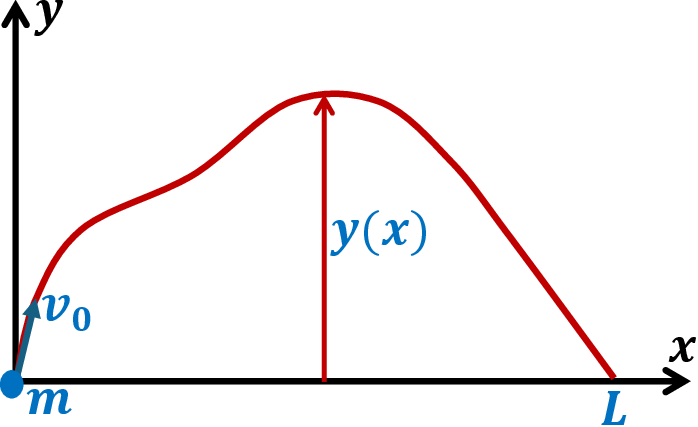}} & \subfigure[Several candidate trajectories.]{\label{Fig:Projectile_Results}\includegraphics[width=8cm]{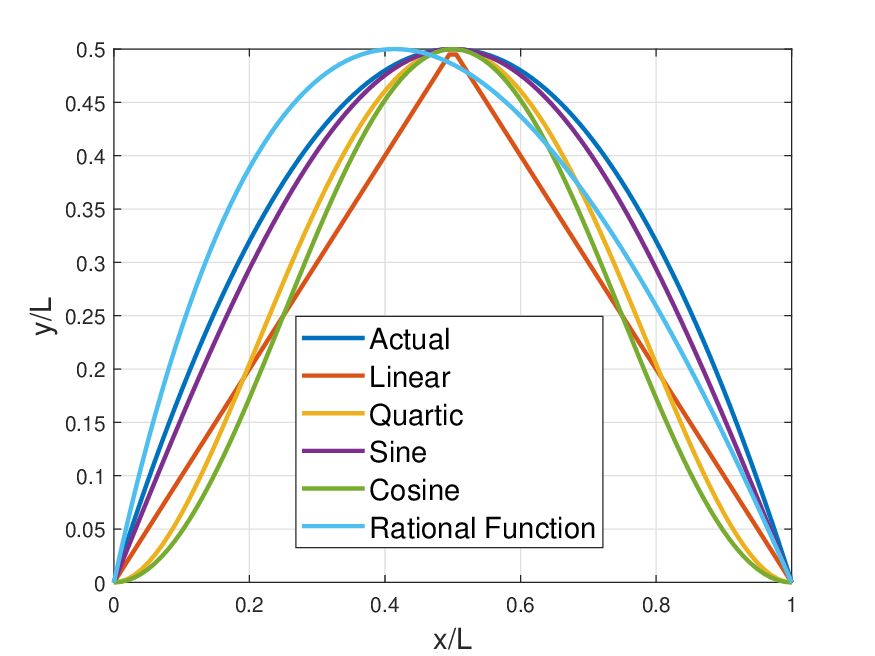}}
\end{array}$
\caption{A Schematic diagram of a projectile motion with several candidate trajectories.}
\label{Fig:Projectile_Example}
\end{center}
\end{figure*}

Although it provides a simple and insightful demonstration, the application of least action to projectile motion is not commonly found in classical textbooks of analytical mechanics. For this reason, we briefly present it here. The example will later acquire a non-trivial significance when connected to the theory of lift. Consider a projectile of mass $m$ launched from, say, zero elevation with initial speed $v_0$, as illustrated in the schematic diagram of Fig. \ref{Fig:Projectile_Schematic}. The projectile trajectory $y(x)$ connecting the two endpoints (0,0), $(L,0)$ is unknown. Energy conservation then implies:
\[ \frac{1}{2}mv_0^2 = \frac{1}{2}mv^2 + mgy, \]
which yields the speed $v$ as a function of height: $v=\sqrt{e_0-gy}$, where $e_0 = \frac{1}{2}v_0^2$. Noting that the length element $ds$ along the trajectory is given by
\[ ds=\sqrt{dx^2+dy^2} = dx\sqrt{1+y'^2}, \]
where $y'\equiv\frac{dy}{dx}$, we write the action integral as
\begin{equation}\label{eq:Action_Projectile}
  S[y(x)] = m\int_0^L \sqrt{ \left[e_0-gy(x)\right]\left[ 1+ y'^2(x) \right] } dx.
\end{equation}
The mechanics problem is thus transformed into the following optimization problem: Find the function $y(x)$ that connects the endpoints (0,0), $(L,0)$ and minimizes the cost functional (\ref{eq:Action_Projectile}). This problem is referred to as \textit{the simplest problem in the calculus of variations} \cite{Burns_Optimal_Control_Book}.

We will not discuss its formal solution here. Instead, we evaluate the action integral $S$ for several candidate trajectories, shown in Fig. \ref{Fig:Projectile_Results}: (i) the well-known quadratic trajectory $y^*(x)=\frac{v_0^2}{2gL^2}\left(Lx-x^2\right)$; (ii) a linear candidate $y_l$; (iii) a quartic candidate $y_q= \frac{4v_0^2}{gL^4}x^2(L-x)^2$; (iv) a sine candidate $y_s=\frac{v_0^2}{4g} \sin\left(\pi\frac{x}{L}\right)$; (v) a cosine candidate $y_c=\frac{v_0^2}{8g} \left[1-\cos\left(2\pi\frac{x}{L}\right)\right]$; and (vi) a rational candidate $y_r=\frac{v_0^2}{4(3-2\sqrt{2})gL} \frac{x(L-x)}{L+x}$. Taking, for simplicity, unit values for $m$, $L$ with $g=10$ and $v_0=\sqrt{2g}$, we obtain the following values of the action integral:
\[ S[y^*]=2.68, \;\; S[y_l]=2.98, \;\; S[y_q]=4.06, \;\; S[y_s]=2.72, \;\; S[y_c]=3.01, \;\; \mbox{and} \;\; S[y_r]=2.70,\]
which clearly demonstrates the statement of the principle of least action: among all kinematically admissible trajectories with the same endpoints, Nature selects the one that renders the action integral stationary (actually minimum in this case). We now make the following important remark.

\textbf{Remark 1.} Given a classical mechanics problem, one can solve it either using Newtonian mechanics ($F = m a$) or using variational mechanics. When the latter approach is adopted, there is no need to invoke $F = m a$; the variational formulation alone is sufficient to determine the solution, independently from Newtonian mechanics. In other words, although the resulting trajectory is guaranteed to coincide with that obtained from $F = m a$, it is calculated without any regard to $F = m a$. It is therefore essential to emphasize that, in variational mechanics, one must consider \textit{\textbf{all}} possible trajectories---both physical and unphysical. Note that we do not know the physical trajectory \textit{a priori} because we have not solved $F = m a$, nor will we do so, since the variational approach is fully independent of Newton's equation. This is why, in the projectile problem, one must admit all conceivable paths connecting the endpoints: not only the physical parabolic trajectory $y^*$, but also other manifestly unphysical candidates; the deliberately unrealistic trajectory $y(x)$ shown in the schematic diagram of Fig. \ref{Fig:Projectile_Schematic} is presented for this reason. The physical trajectory then emerges solely as the one that minimizes the action integral, through a process of pure minimization among all possible trajectories, entirely independent of Newton's equations. However, it is guaranteed to satisfy Newton's equation.

When we refer to ``all possible trajectories," we specifically mean all \textit{kinematically admissible} ones; i.e., those satisfying the prescribed kinematic and geometric constraints. But most of these trajectories are dynamically incorrect. Only the one that minimizes the action integral (equivalently the one satisfying Newton's equation) is dynamically realizable.

This reasoning extends directly to the lift problem, where Newton's equation is represented by the Navier-Stokes equation (or Euler's). If we choose to pursue any variational approach for this problem, we must consider \textit{\textbf{all}} possible flows---physical and unphysical alike. At the outset,
the physical flow is not known, because the governing equations have not yet been solved, nor are they intended to be solved explicitly within a variational framework. Hence, with a variational approach, we must consider all kinematically admissible flows; i.e., those satisfying the continuity constraint and boundary conditions. These are infinitely many. Only one flow occurs in reality: the one selected by the variational principle, or equivalently,  the one satisfying the governing equation (Navier-Stokes or Euler).

This observation clearly refutes the criticism that circulatory flows should be excluded from the family of admissible fields in a variational formulation of lift \cite{Liu_Steady_Lift,Peters_Gauss}. Indeed, a rigorous application of any variational approach to the lift problem must consider all kinematically admissible flows; and this family necessarily includes circulatory flow fields. Excluding them is akin to assuming the conclusion in advance, or to unnecessarily restricting the search space.

\subsection{Gauss's Principle of Least Constraint}
During The Age of Enlightenment, a prevalent belief emerged (particularly after Maupertuis) that the universe was optimally designed and that natural phenomena take place in some optimal sense. This line of thought has led to the description of natural phenomena through principles of optimality, giving rise to several such formulations: Fermat's principle of shortest time, Maupertuis's principle of least action, Gauss's principle of least constraint, Hertz's principle of least curvature, among others. Inspired by this view, Gauss postulated one of the fundamental principles in mechanics in his four-page philosophical paper \cite{Gauss_Least_Constraint}, published in a journal that still exists today.

\begin{wrapfigure}{l}{0.50\textwidth}
\vspace{-0.15in}
 \begin{center}
 \includegraphics[width=7.5cm]{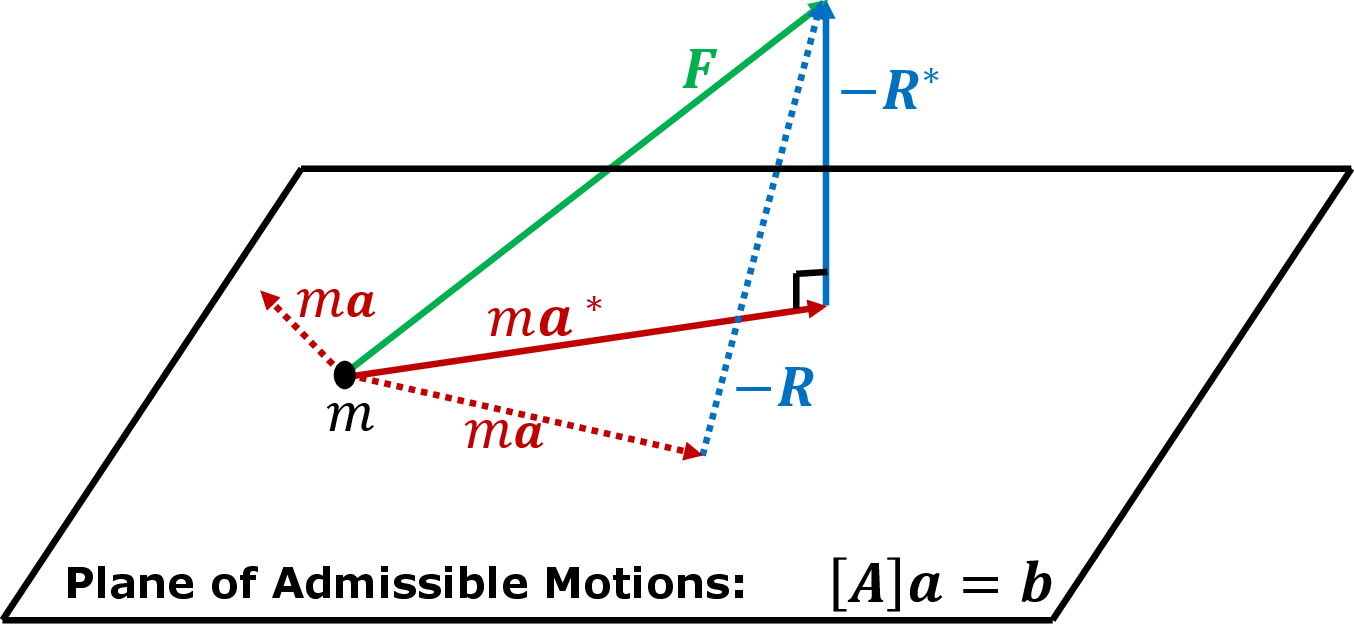}  \vspace{-0.2in}
 \caption{Schematic illustration of the tangent plane to the configuration manifold, defined by (\ref{eq:Constraints_General}), at a particular configuration. Infinitely many instantaneous motions lie this plane (dotted vectors), each satisfying the linear constraint (\ref{eq:Constraints_Linear}) at the expense of a constraint force $\bm{R}$.  Gauss's principle asserts that, among all kinematically admissible motions, Nature selects the one that requires the least constraint force. This minimization is simply equivalent to projecting the impressed force $\bm{F}$ onto the plane of admissible motions.}
 \label{Fig:Gauss_Schematic} \vspace{-0.1in}
 \end{center} \vspace{-0.1in}
\end{wrapfigure} \noindent If a particle of mass $m$ is subject to an \textit{impressed} force $F$, as shown in Fig. \ref{Fig:Gauss_Schematic}, it would simply accelerate in the direction of the force with acceleration $\bm{a}^{\rm{free}} = \frac{\bm{F}}{m}$. Gauss referred to this motion as the \textit{free motion}---the motion that would occur in the absence of constraints. However, if the particle is constrained to accelerate along some instantaneous plane of admissible motion, defined by the constraint $[\bm{A}]\bm{a} = \bm{b}$, the actual motion will necessarily deviate from the free motion. Since this deviation arises solely because of the constraint, Gauss's profound insight postulated that, it must be the \textit{least} deviation that satisfies the constraint. Nature will not overdo it. He wrote:
\begin{center}
``\textit{The motion of a system of $N$ material points takes place in every moment in maximum accordance with the free movement or under least constraint, the measure of constraint, is considered as the sum of products of mass and the square of the deviation to the free motion.}"
\end{center}

Gauss's principle can be described in a different but equally illuminating interpretation. On the instantaneous admissible plane of motion, there exists infinitely many candidates, as illustrated in Fig. \ref{Fig:Gauss_Schematic}. Each candidate requires a specific \textit{constraint} force $\bm{R}$ such that, when combined with the impressed force $\bm{F}$, the resulting motion
\[ m\bm{a}=\bm{F}+\bm{R}\]
satisfies the constraint. Note that the force $\bm{R}$ exists solely to enforce the constraint; its \textit{raison d'être} is the constraint itself---if the constraint is removed, $\bm{R}$ ceases to exist. Gauss then asserted that Nature selects the motion that requires the smallest magnitude of the constraint force necessary to ensure the constraint, hence the name \textit{Least Constraint}. Any other candidate would demand an unnecessarily larger constraint force to ensure the constraint, which is unphysical.

Jacobi \cite{Jacobi} later gave Gauss's principle an explicit mathematical form by introducing the quadratic cost function:
\begin{equation}\label{eq:Gauss}
  Z = \frac{1}{2} \sum_{i=1}^N m_i \left| \bm{a}_i- \frac{\bm{F}_i}{m_i} \right|^2,
\end{equation}
where $N$ is the number of particles, $\bm{a}_i$  is the inertial acceleration of the $i^{\rm{th}}$ particle. According to Gauss' principle, $Z$ must be minimum at every instant, provided that the constraints are satisfied. Assume that the particles evolve in a $d$-dimensional space, so the array $\bm{a}=[\bm{a}_1^T, ..., \bm{a}_N^T]^T\in\mathbb{R}^{dN}$ includes all inertial accelerations of the $N$ particles. Suppose that the system is subject to $c\leq dN$ constraints that may be arbitrarily nonlinear in positions and velocities:
\begin{equation}\label{eq:Constraints_General}
\Psi_\ell(\bm{x},\bm{v}) = 0, \; \ell=1,...,c,
\end{equation}
where $\bm{x}, \; \bm{v}\in\mathbb{R}^{dN}$ denote the arrays of positions and velocities of the $N$ particles, respectively. Differentiating Eq. (\ref{eq:Constraints_General}) with respect to time yields a form that is linear in accelerations:
\begin{equation}\label{eq:Constraints_Linear}
A_{\ell j}(\bm{x},\bm{v}) \; a_j = b_\ell
\end{equation}
for some $\bm{A}\in\mathbb{R}^{c\times dN}$, $\bm{b}\in\mathbb{R}^c$.

To solve this dynamics problem using Newtonian mechanics, $dN+c$ equations must be solved in $dN+c$ unknowns. The equations are $dN$ Newton's equations of motion:
\begin{equation}\label{eq:Newton}
  m_i \bm{a}_i = \bm{F}_i + \bm{R}_i \;\;\; \forall i=1,..,N,
\end{equation}
together with the $c$ constraint equations (\ref{eq:Constraints_Linear}). The unknowns are the $dN$ accelerations $\bm{a}$ and the $c$ constraint forces constituting the $\bm{R}_i$'s.

Gauss's principle transforms this dynamics problem into the following minimization problem:
\begin{equation}\label{eq:Gauss_QP_Problem}
\min_{\bm{a}} \; Z(\bm{a}) \;\;\; \mbox{s.t.} \;\;\; [\bm{A}(\bm{x},\bm{v})] \; \bm{a} = \bm{b},
\end{equation}
In our recent effort \cite{NS_QP_IEEE}, we showed that this minimization problem is a strongly convex quadratic programming problem and therefore admits a unique solution. Moreover, its first-order necessary conditions of optimality are precisely Newton's equations of motion. That is, the unique solution of Gauss's minimization problem (\ref{eq:Gauss_QP_Problem}) naturally satisfies Newton's equation of motion.

\begin{wrapfigure}{l}{0.30\textwidth}
\vspace{-0.15in}
 \begin{center}
 \includegraphics[width=4.5cm]{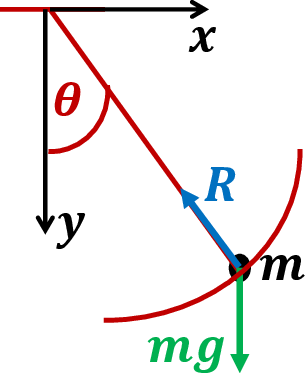}  \vspace{-0.2in}
 \caption{Pendulum Schematic.}
 \label{Fig:Pendulum_Schematic} \vspace{-0.1in}
 \end{center} \vspace{-0.1in}
\end{wrapfigure} \noindent \textbf{Pendulum Example:} Because Gauss's principle is rarely taught in engineering graduate curricula, it may be useful to illustrate it through a simple example, which will also play a role later in our response to Ref. \cite{Peters_Gauss}. Consider a simple pendulum, shown schematically in Fig. \ref{Fig:Pendulum_Schematic}. The mass $m$ is constrained to move along a circular path. This constraint gives rise to a constraint force $R$. The \textit{free motion}---that would occur in the absence of the constraint---is simply:
\[ \bm{a}^{\rm{free}} = (0,g). \]
The Gaussian cost is then written as
\[ Z = \frac{1}{2} m \left| \bm{a}- \bm{a}^{\rm{free}} \right|^2 = \frac{1}{2} m \left[ (\ddot{x}-0)^2 + (\ddot{y}-g)^2 \right]. \]
Since the cost is expressed in terms of the \textit{design variables} $(\ddot{x},\ddot{y})$, the constraint must likewise be written in terms of these variables. The holonomic constraint of the pendulum is
\[ x^2+ y^2 = \ell^2, \]
where $\ell$ is the pendulum length. Differentiating this relation twice with respect to time yields a form linear in the accelerations:
\begin{equation}\label{eq:Pendulum_Constraint}
\psi = x\ddot{x} + y\ddot{y} +\dot{x}^2 + \dot{y}^2 = 0.
\end{equation}

The dynamics problem is thus transformed, via Gauss's principle, to the constrained minimization problem
\[ \min_{(\ddot{x},\ddot{y})} Z = \frac{1}{2} m \left[ \ddot{x}^2 + (\ddot{y}-g)^2 \right] \;\; \mbox{s.t.} \;\; x\ddot{x} + y\ddot{y} +\dot{x}^2 + \dot{y}^2 = 0.\]
This problem can be solved by augmenting the cost $Z$ with the constraint through a Lagrange multiplier $\lambda$:
\[ \mathcal{L} = Z + \lambda \psi. \]
The first-order necessary conditions for optimality are then:
\begin{equation}\label{eq:Pendulum_Gauss}
\begin{array}{lll}
\frac{\partial \mathcal{L}}{\partial \ddot{x}} &=& m \ddot{x} + \lambda x = 0 \\
\frac{\partial \mathcal{ L}}{\partial \ddot{y}} &=& m \ddot{y} -mg + \lambda y = 0,
 \end{array} \end{equation}
which are identical to Newton's equations of motion:
\[ m \ddot{x} = - \frac{x}{\ell} R, \;\; m \ddot{y} = mg - \frac{y}{\ell} R, \]
upon identifying $\lambda \ell = R$---the constraint force is directly associated with the Lagrange multiplier enforcing the constraint.

Gauss's principle, as is standard in analytical mechanics, allows the elimination of holonomic constraints together with their associated constraint forces. By introducing a reduced set of generalized coordinates that automatically satisfy the constraint, the dynamics is projected onto the configuration manifold. Consequently, the constraint force, which is typically orthogonal to that manifold, does not appear; no Lagrange multiplier is needed to enforce the constraint \cite{Lanczos_Variational_Mechanics_Book,Schaub_Junkins}. In this setting, Gauss's principle yields an unconstrained minimization problem for the accelerations on the configuration manifold.

In the pendulum example, this elimination can be demonstrated by formulating the problem in the angular coordinate $\theta$ instead of the Cartesian coordinates $(x,y)$. The relation
\[ (x,y) = \ell(\sin\theta,\cos\theta) \]
automatically satisfies the holonomic constraint. As such, the inertial acceleration is written in terms of $\ddot\theta$ as:
\[ \bm{a} = (\ddot{x},\ddot{y}) = \ell\left(\cos\theta \ddot\theta - \sin\theta \dot\theta^2,-\sin\theta \ddot\theta-\cos\theta \dot\theta^2 \right). \]
Consequently, Gauss's minimization problem becomes
\[ \min_{\ddot\theta} Z = \frac{1}{2} m \left[ \ell^2\left(\cos\theta \ddot\theta - \sin\theta \dot\theta^2\right)^2 + \left(\ell\sin\theta \ddot\theta+\ell\cos\theta \dot\theta^2 + g \right)^2 \right] \]
subject to no constraints. The first-order necessary condition is simply $\frac{dZ}{d\ddot\theta}=0$, which immediately yields the familiar pendulum dynamics
\begin{equation}\label{eq:Pendulum_Dynamics}
\ddot\theta = -\frac{g}{\ell}\sin\theta .
\end{equation}

\textbf{Remark 2.} When the dynamics is projected onto the configuration manifold, any motion generated by the projected equations automatically satisfies the constraint. In the pendulum example, any solution $\theta(t)$ of the equation of motion (\ref{eq:Pendulum_Dynamics}) induces a trajectory
\[ (x(t),y(t)) = \ell(\sin\theta(t),\cos\theta(t)) \]
that identically satisfies the constraint (\ref{eq:Pendulum_Constraint}). However, this fact, together with the elimination of the constraint force from the projected dynamics, should not be misinterpreted to conclude that the motion \textit{naturally} satisfies the constraint without the need for a constraint force. In other words, it does not imply that the constraint force $R$ vanishes. The constraint force can be recovered by projecting the dynamics onto the direction normal to the manifold---a step that may be carried out \textit{a posteriori} after solving for the motion. This remark, while elementary, will play an important role later to address the claim made in Ref. \cite{Peters_Gauss} regarding the role of pressure in incompressible flows.

In summary, Gauss's formulation casts dynamical evolution as an instantaneous minimization problem in which the accelerations serve as the \textit{design variables}. It is the minimization formulation of projecting the impressed forces $\bm{F}$ onto the instantaneous tangent plane of admissible motions, defined by the constraints, as illustrated in Fig. \ref{Fig:Gauss_Schematic}. This projection is equivalent to minimizing the magnitude of the deviation between the impressed (free) motion $\bm{F}$ and the actual motion $m\bm{a}$, which is proportional to the magnitude of the constraint force $\bm{R}$.

Finally, we conclude by noting that, in the absence of impressed forces, the free motion is simply a uniform motion along a straight line; i.e., a \textit{geodesic}. Any deviation from this straight path is therefore measured by curvature. Hence, in this special case, Gauss's principle of least constraint reduces to Hertz's principle of least curvature, and the Guassian cost $Z$ reduces to the \textit{Appellian} 
\begin{equation}\label{eq:Appellian}
  S = \frac{1}{2} \sum_{i=1}^N m_i \left| \bm{a}_i \right|^2.
\end{equation}

\section{The Variational Theory of Lift}\label{Sec:Variational_Theory}
Inspired by Gauss's principle of least constraint, Gonzalez and Taha \cite{Variational_Lift_JFM} proposed a variational theory of lift that relaxes the overly constricted assumption of classical airfoil theory concerning sharp trailing edges. Rooted in Gauss's principle as a general framework of mechanics, the theory also admits the possibility of further generalizations. Its central postulate is that, among all kinematically admissible flows, Nature selects the one that minimizes the Gaussian cost $Z$.

For incompressible flows, the Gaussian cost admits a natural extension to the continuum setting:
\begin{equation}\label{eq:Gauss_Continuum}
  Z = \frac{1}{2} \int_\Omega \rho \left| \bm{a}- \bm{f} \right|^2 d\bm{x},
\end{equation}
where $\Omega$ denotes the spatial domain, $\bm{f}$ is the impressed force per unit mass, and $\bm{a}=\bm{u}_t + \bm{u}\cdot\bm\nabla \bm{u}$ is the total inertial acceleration of a fluid particle.

Incompressible flows are subject to pressure forces, viscous forces, and other body forces (such as gravity or electromagnetism, which are typically ignored in airfoil theory). These flows are constrained to satisfy continuity (\ref{eq:Continuity}) and the no-penetration boundary condition. As discussed above, any constraint gives rise to a constraint force whose sole role is to enforce that constraint. For the continuity constraint (\ref{eq:Continuity}), the Helmholtz decomposition is particularly revealing of the geometry of incompressible flows and the associated constraint force \cite{Chorin_Marsden_Book,Geometric_Control_Fluid_Dynamics,Helmholtz_Decomposition_Review}.

\begin{figure*}
\begin{center}
$\begin{array}{cc}
\subfigure[Schematic of the Helmholtz decomposition.]{\label{Fig:Helmholtz_Schematic}\includegraphics[width=7.5cm]{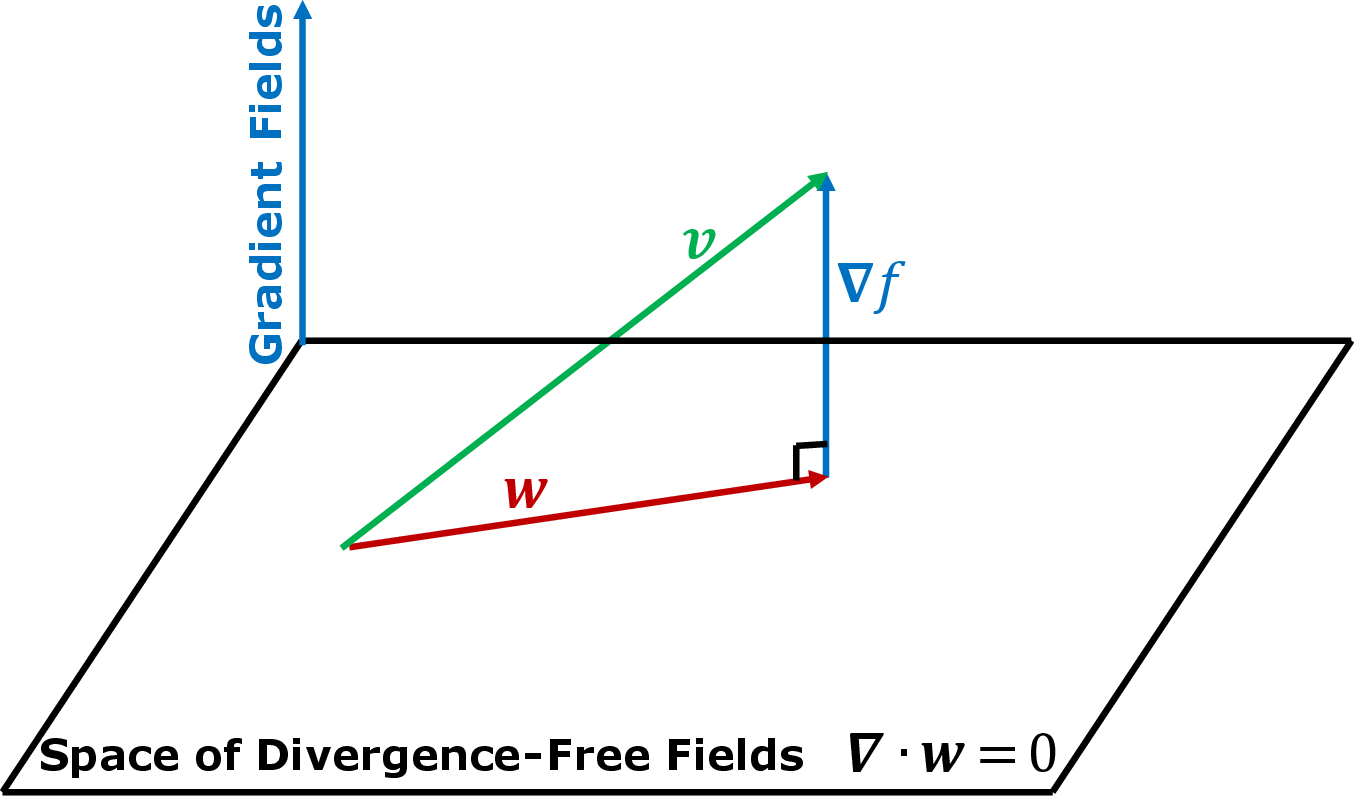}} & \subfigure[Geometry of incompressible flows.]{\label{Fig:Incompressible_Schematic}\includegraphics[width=7.5cm]{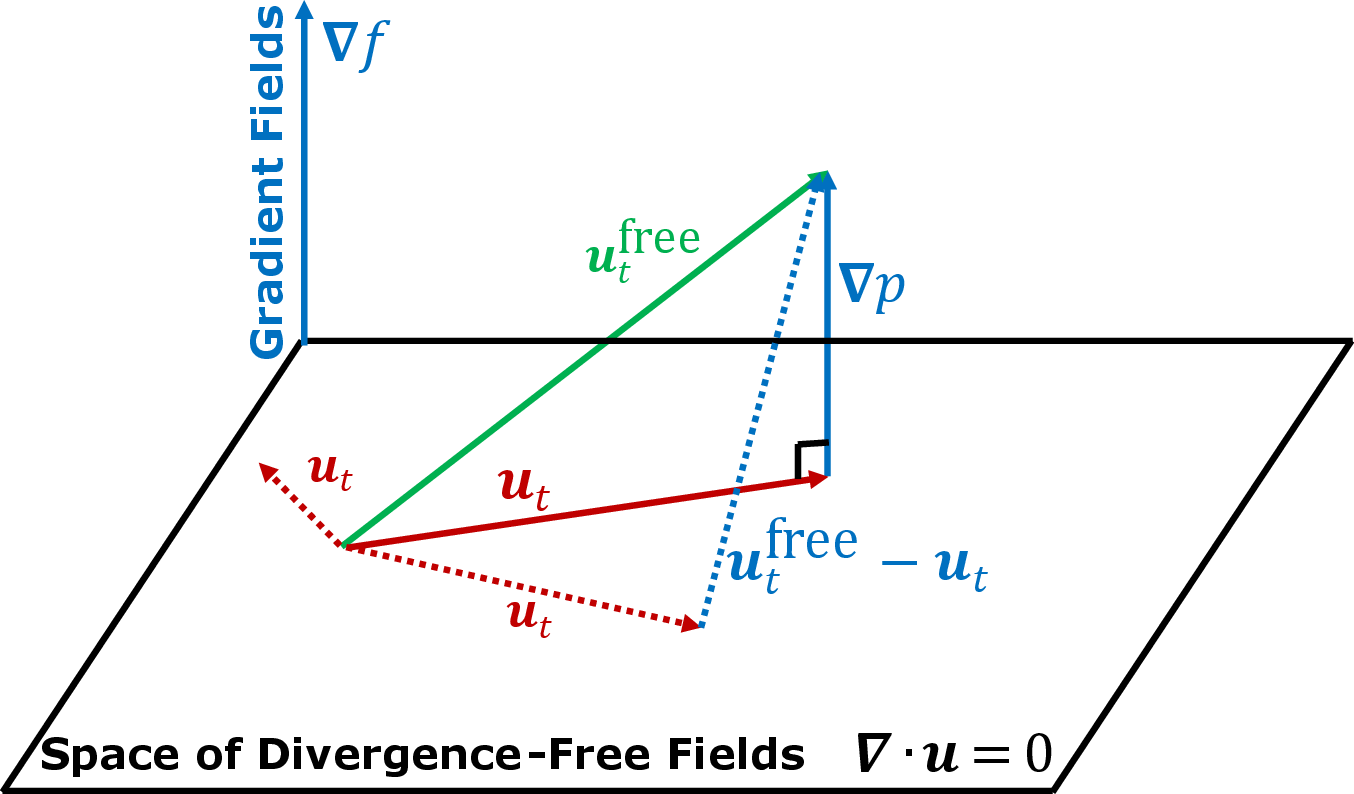}}
\end{array}$
\caption{A Schematic diagram illustrating the Helmholtz-Leray projection and geometry of incompressible flows.}
\label{Fig:Helmholtz}
\end{center}
\end{figure*}

Given a square integrable vector field $\bm{v}(\bm{x})$ over a smooth domain $\Omega\subset\mathbb{R}^n$, it can be \textit{uniquely} decomposed into two orthogonal components: (i) a divergence-free field $\bm{w}$ that satisfies the no-penetration boundary condition $\bm{w}\cdot\bm{n}=0$ on $\partial\Omega$, and (ii) a gradient field $\bm\nabla f$, for some scalar function $f$, as illustrated schematically in Fig. \ref{Fig:Helmholtz_Schematic}. Accordingly,
\[ \bm{v}(\bm{x}) = \bm{w}(\bm{x}) + \bm\nabla f (\bm{x}) \;\; \forall \; \bm{x}\in\Omega. \]
The orthogonality is understood in the $\mathbb{L}^2$ sense:
\[ \int_\Omega \left(\bm\nabla f \cdot \bm{w}\right) d\bm{x} = 0.\]
This orthogonality follows directly from integration by parts:
\begin{equation}\label{eq:Helmholtz_Orthogonality}
\int_\Omega \left(\bm\nabla f \cdot \bm{w}\right) d\bm{x} = -\int_\Omega \left( f \underbrace{\bm\nabla\cdot \bm{w}}_{=0 \; \mbox{in}\; \Omega}\right) d\bm{x} + \oint_{\partial\Omega} \left( f \underbrace{\bm{w}\cdot\bm{n}}_{=0 \; \mbox{on}\; \partial\Omega}\right) d\bm{x} = 0.
\end{equation}

\textbf{Remark 3.} Equation (\ref{eq:Helmholtz_Orthogonality}) clearly shows that Helmholtz orthogonality holds between a divergence-free field $\bm{w}$ and a gradient field $\bm\nabla f$, \textit{only if} the divergence-free field satisfies the no-penetration boundary condition $\bm{w}\cdot\bm{n}=0$ on $\partial \Omega$ (or an equivalent boundary condition ensuring vanishing boundary flux). In other words, being divergence-free in the interior of $\Omega$ is \textit{not} sufficient for orthogonality; the boundary condition plays an essential role.

This perspective is particularly illuminating for the geometry of incompressible flows. From this viewpoint, the Navier-Stokes equation may be rearranged as
\begin{equation}\label{eq:NS_Rearranged}
  \underbrace{\bm{f}-\bm{u}\cdot \bm\nabla \bm{u}}_{\bm{u}_t^{\rm{free}}} = \bm{u}_t + \frac{1}{\rho}\nabla p,
\end{equation}
Equation (\ref{eq:NS_Rearranged}) describes the dynamic evolution of an incompressible flow from one instant to the next as a projection---Helmholtz (or Leray) projection---onto the space of divergence-free fields \cite{Chorin_Marsden_Book,Geometric_Control_Fluid_Dynamics,Helmholtz_Decomposition_Review}. For a given smooth velocity field $\bm{u}$ at some instant, both the convective acceleration $\bm{u}\cdot \bm\nabla \bm{u}$ and the impressed forces (e.g., viscous or gravitational) are known. Together, they define a vector field $\bm{u}_t^{\rm{free}}$ that admits a unique Helmholtz decomposition into two orthogonal components: (i) a divergence-free component satisfying the no-penetration boundary condition, represented by the local acceleration $\bm{u}_t$, and (ii) a gradient component, represented by the pressure gradient $\bm\nabla p$, as illustrated in the schematic diagram of Fig. \ref{Fig:Incompressible_Schematic}.

The schematic shows that, on the instantaneous plane of admissible motion defined by the divergence-free constraint, there exists infinitely many candidates. Each candidate has some deviation from the free motion $\bm{u}_t^{\rm{free}}$ and requires a constraint force to ensure continuity---to project $\bm{u}_t^{\rm{free}}$ on the space of divergence-free fields. According to Gauss's principle, Nature selects the motion with \textit{least} deviation from the free one; equivalently, the motion that requires the smallest magnitude of the constraint force required to ensure continuity.  Geometrically, this selection corresponds to the orthogonal projection of the free motion $\bm{u}_t^{\rm{free}}$ onto the space of divergence-free fields. Clearly, the constraint force with smallest magnitude must therefore be orthogonal to the plane of admissible motions. That is, it lies in the space of gradient fields; i.e., it is the gradient of a scalar function---the pressure. The pressure force is thus the constraint force associated with the contiguity constraint and the no-penetration boundary condition, while the free motion is the motion that would occur in the absence of such a force, which is given by the left hand side of Eq. (\ref{eq:NS_Rearranged}).

This interpretation is well known in the mathematical fluid mechanics community (e.g., \cite{Chorin_Marsden_Book,Temam_Projection,Geometric_Control_Fluid_Dynamics,Morrison2020lagrangian}) and in computational fluid dynamics (e.g., \cite{Pressure_BCs,Chorin_Projection,Projection_Book,Projection_Review,Hirsch_Book2}), though it is perhaps less familiar within the aeronautics literature.

On the basis of the preceding discussion, the Gaussian cost for incompressible flows (ignoring gravitational and electromagnetic forces) can be written as:
\begin{equation}\label{eq:Gauss_Incompressible}
  \mathcal{A} = \frac{1}{2} \int_\Omega \rho \left| \bm{u}_t + \bm{u}\cdot \bm\nabla \bm{u} - \nu \nabla^2 \bm{u} \right|^2 d\bm{x},
\end{equation}
where $\nu$ denotes the kinematic viscosity. This cost was proposed by Taha et al. \cite{PMPG_PoF} as the basis for the \textit{Principle of Minimum Pressure Gradient} (PMPG), which is the continuum-mechanics analogue of Gauss's principle of least constraint. It asserts that an incompressible flow evolves from one instant to the next by minimizing the magnitude of the pressure force required to ensure the continuity constraint. Any alternative evolution would require an unnecessarily larger pressure force to ensure continuity, which is against physical considerations as conceived by Gauss.

Similar to the fact that the first-order necessary condition of optimality of the Gaussian cost $Z$ in (\ref{eq:Gauss}) recovers Newton's equation of motion in particle mechanics, the first-order necessary condition of optimality of the PMPG cost functional $\mathcal{A}$ in (\ref{eq:Gauss_Incompressible}) yields Navier-Stokes' equation. In particular, any incompressible flow whose evolution minimizes the PMPG cost functional $\mathcal{A}$ at every instant is guaranteed to satisfy the Navier-Stokes equation, as established in Theorem 1 of \cite{PMPG_PoF,VPNS_PRF} and Proposition 2 of \cite{NS_QP_IEEE}.

\begin{wrapfigure}{l}{0.50\textwidth}
\vspace{-0.15in}
 \begin{center}
 \includegraphics[width=7cm]{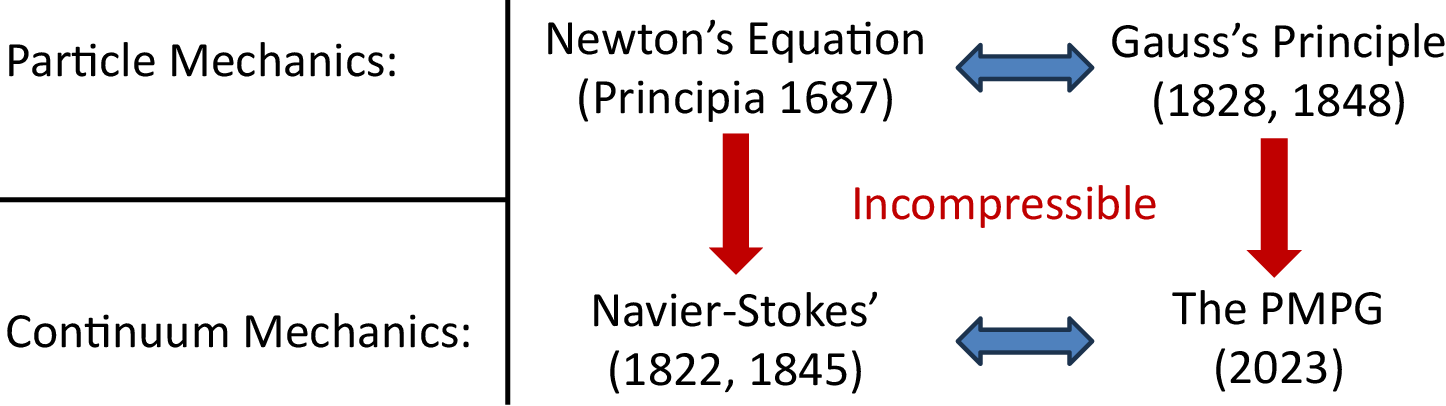}  \vspace{-0.2in}
 \caption{Schematic illustrating the equivalence between Newton's equations and Gauss's principle in particle mechanics, and the corresponding equivalence between the Navier-Stokes equation and the Principle of Minimum Pressure Gradient (PMPG) in incompressible continuum mechanics.}
 \label{Fig:Gauss_Newton_PMPG} \vspace{-0.1in}
 \end{center} \vspace{-0.1in}
\end{wrapfigure} \noindent In other words, the Principle of Minimum Pressure Gradient occupies in incompressible fluid mechanics the same foundational role that Gauss's principle occupies in particle mechanics. Figure \ref{Fig:Gauss_Newton_PMPG} illustrates this correspondence by showing the equivalence between Newton's equations and Gauss's principle in particle mechanics, and the analogous relationship between the Navier-Stokes equation and the PMPG in the continuum mechanics of incompressible flows.

For ideal fluids (without viscosity), the fluid particle is subject to no impressed forces---only constraint forces. Its free motion (in the absence of the continuity and no-penetration constraints) is simply a uniform motion along straight lines (i.e., geodesics). And in the presence of such constraints, its actual motion is the closest to straight lines, in accordance with Hertz's principle of least curvature. In this inviscid case, the PMPG cost functional reduces to the Appellian
\begin{equation}\label{eq:Appellian_Continuum}
  S = \frac{1}{2} \int_\Omega \rho \left| \bm{u}_t + \bm{u}\cdot \bm\nabla \bm{u}\right|^2 d\bm{x}.
\end{equation}
As expected, the first-order necessary condition of optimality of the Appellian $S$ in (\ref{eq:Appellian_Continuum}) recovers Euler's equation, as shown in the appendix of Ref. \cite{Variational_Lift_JFM}.

Inspired by this rich corpus of analytical mechanics, Gonzalez and Taha \cite{Variational_Lift_JFM} adopted the steady counterpart of the Appellian (\ref{eq:Appellian_Continuum}):
\begin{equation}\label{eq:Steady_Appellian}
  S_s[\bm{u}] = \frac{1}{2} \int_\Omega \rho \left|\bm{u}\cdot \bm\nabla \bm{u}\right|^2 d\bm{x}
\end{equation}
and proposed a variational criterion for selecting a unique solution to the airfoil problem: Among all steady solutions of Euler, Nature selects the one that minimizes the steady Appellian $S_s$.

Membership in the set $\Theta$ of steady Euler solutions entails four conditions, namely the conditions (i)-(iv) listed in Sec. \ref{sec:Classical_Theory}. These include: (i) the momentum equation (\ref{eq:Euler}), (ii) the continuity constraint (\ref{eq:Continuity}), (iii) the no-penetration boundary condition on the body surface, and (iv) the far-field boundary condition. In its full generality, the variational theory of lift, therefore, poses the following minimization problem
\begin{equation}\label{eq:Variational_Theory_General}
\min_{\bm{u}\in\Theta}  S_s[\bm{u}],
\end{equation}
The set $\Theta$ is very large; it is an infinite-dimensional set and may include discontinuous flows (understood as \textit{weak} solutions). If, however, one exploits \textit{a priori} knowledge of the expected form of the solution, the resulting optimization problem may be greatly simplified.

From this perspective, classical airfoil theory already provides a reasonable ansatz (\ref{eq:Family}) for attached-flow configurations, where one can confine the analysis to smooth solutions involving a single undetermined parameter---the circulation $\Gamma$. Hence, restricting the optimization problem (\ref{eq:Variational_Theory_General}) to the family
\[ \mathcal{U}=\{ \bm{u}_0(\bm{x})+\Gamma \bm{u}_1(\bm{x}) \}\subset\Theta, \]
the infinite-dimensional problem reduces to a one-dimensional minimization over $\Gamma$. Accordingly, the variational theory of lift, in the restricted formulation presented in \cite{Variational_Lift_JFM} (valid only for attached flows where the family $\mathcal{U}$ is appropriate) selects a unique circulation through the minimization problem:
\begin{equation}\label{eq:Variational_Theory_Special}
\min_{\Gamma}  S_s[\bm{u}_0 +\Gamma \bm{u}_1] = \frac{1}{2} \int_\Omega \rho \left|\left(\bm{u}_0 +\Gamma \bm{u}_1\right)\cdot \bm\nabla \left(\bm{u}_0 +\Gamma \bm{u}_1\right)\right|^2 d\bm{x}.
\end{equation}

Since any member $\bm{u}\in\mathcal{U}$ depends linearly on $\Gamma$, the associated convective acceleration is quadratic in $\Gamma$, and hence, the steady Appellian $S_s$ is a quartic polynomial in $\Gamma$. The resulting one-dimensional optimization problem is remarkably tractable: the first-order necessary condition for optimality, $\frac{d S_s}{d\Gamma}=0$, yields three roots, one real and two complex (discarded). This analysis was performed on a modified Zhukovsky airfoil, which can be constructed from a disc of radius $a$ in the $\zeta$-domain via the conformal map \cite{Karamcheti}
\begin{equation}\label{eq:Modified_Zhukovky}
 z =  \zeta+\mu + \frac{1-D}{1+D}\frac{C^2}{\zeta+\mu},
\end{equation}
where
\[ a=C(1+\epsilon), \; \mu=-\epsilon C ,\; \epsilon=\frac{4\tau}{3\sqrt{3}},\; C=\frac{b(1+D)(1+2\epsilon)}{2(1 + 2\epsilon + \epsilon^2+ D\epsilon(1 + \epsilon))}, \]
Here $b$ denotes the half-chord length, $\tau$ controls the degree of fore-aft asymmetry ($\tau=0$ corresponds to an ellipse), and $D$ governs the smoothness of the trailing edge: $D=0$ recovers the classical Zhukovsky airfoil with a sharp trailing edge, whereas $D=1$ reduces the conformal map to the identity, which corresponds to a circular cylinder.

It should be noted that, for an airfoil with a mathematically sharp trailing edge ($D\equiv0$), the optimization problem becomes trivial: the steady Appellian $S_s$ is infinite for all values of $\Gamma$ except Kutta's value. This observation is fully consistent with the discussion in Sec. \ref{sec:Classical_Theory} regarding the nature of the Kutta condition. Indeed, the essence of the Kutta condition is not viscous in origin; it is merely a singularity removal condition, selecting the only bounded-velocity flow field. However, this observation does not imply that lift can be generated without viscosity (although this may, in fact, be the case). Rather, it simply clarifies that the Kutta condition itself is not a viscous condition, as is often stated in the aeronautics literature. This mathematical fact stands independently of the variational theory of lift.

\begin{figure}
$\begin{array}{cc}
\subfigure[Zhukovsky airfoil with a sharp trailing edge ($D\simeq0$).]{\label{Fig:Sharp_Edge}\includegraphics[width=8.5cm]{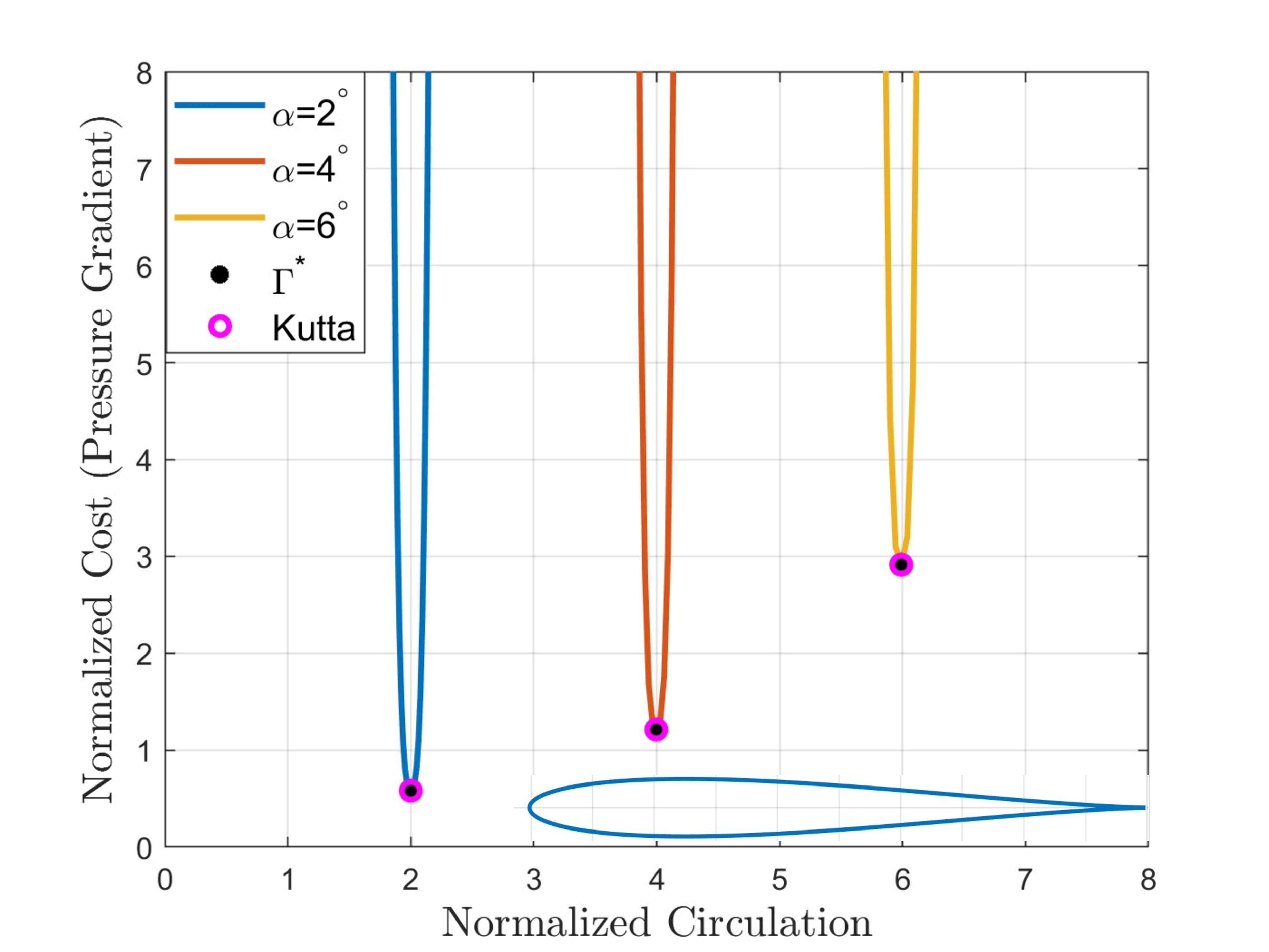}} \hspace{-0.1in} & \hspace{-0.1in}  \subfigure[Zhukovsky airfoil with a smooth trailing edge ($D=0.05$).]{\label{Fig:Smooth_Shape}\includegraphics[width=8.5cm]{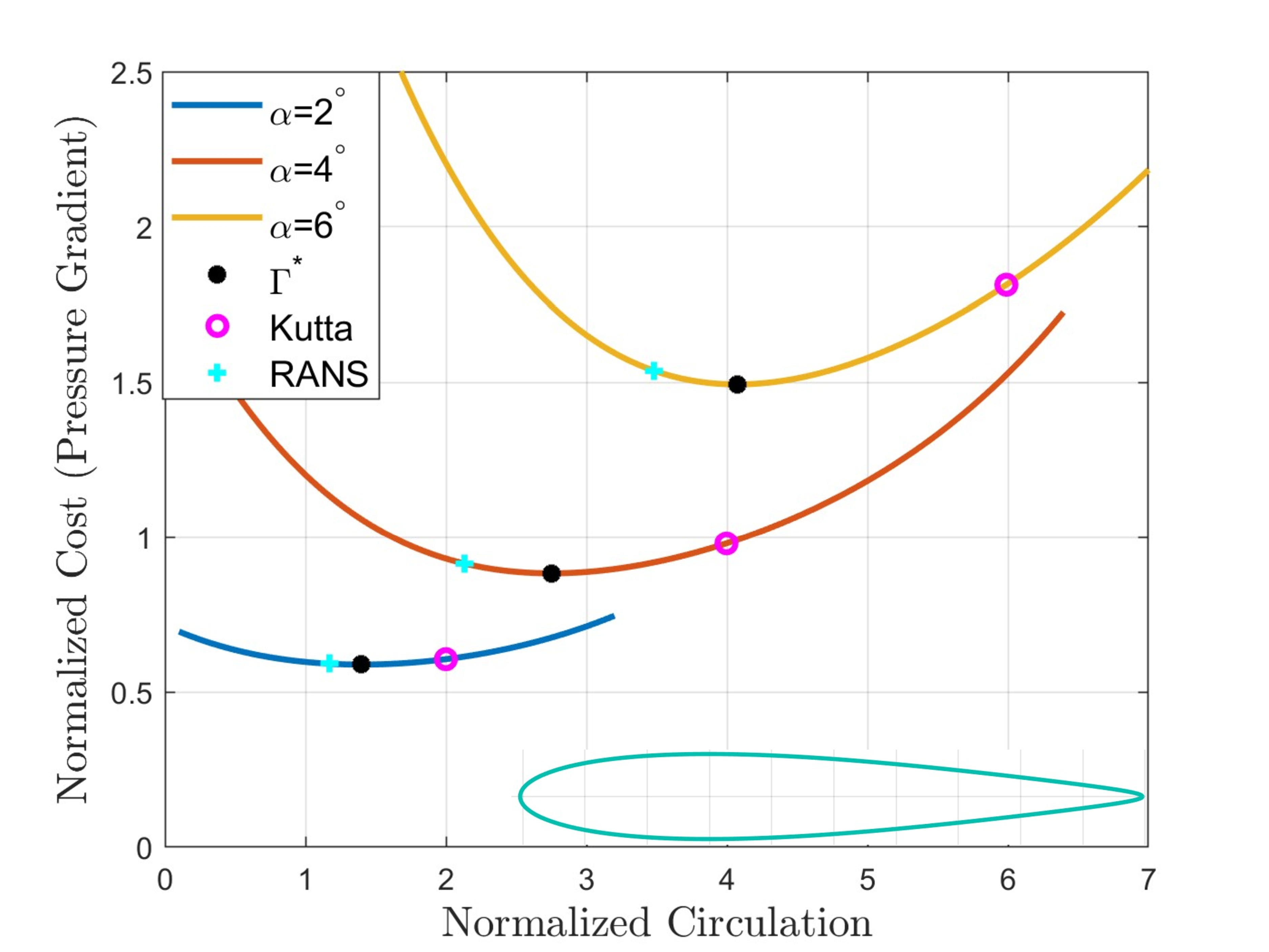}}
\end{array}$ \vspace{-0.1in}
\caption{Variation of the normalized Appellian $\hat{S}=\frac{S}{\rho U_\infty^4}$ with the normalized circulation $\hat\Gamma=\frac{\Gamma}{4\pi U_\infty a}$ (in degrees) for a Zhukovsky airfoil with an almost-sharp trailing edge ($D=0.001$) and a smooth trailing edge ($D=0.05$ corresponds to a radius of 0.1\% chord length) at various angles of attack. In both cases, $\tau=0.1$. This figure is reproduced from \cite{PMPG_PoF}.}
\label{Fig:Results}
\end{figure}

Given the singular nature of a mathematically sharp trailing edge, such geometries are treated within the variational theory as limiting cases $D\simeq0$ (instead of $D\equiv0$). Figure \ref{Fig:Sharp_Edge} (reproduced from \cite{PMPG_PoF}) shows the variation of the normalized steady Appellian ($\hat{S}=\frac{S_s}{\rho U_\infty^4}$) with the normalized circulation $\hat\Gamma=\frac{\Gamma}{4\pi U_\infty a}$ at several angles of attack for a Zhukovsky profile with $\tau=0.1$ and $D=0.001\simeq0$. The figure also shows Kutta's circulation $\Gamma_K=4\pi a U_\infty\sin\alpha$ (equivalently $\hat\Gamma_K=\sin\alpha\simeq\alpha$ for small angles), which enforces stagnation at the most rearward point of the airfoil---the point of maximum curvature. The figure shows that, for each angle of attack, the Appellian possesses a unique minimum at a specific value of the circulation, and this minimizing circulation coincides with Kutta's value for this case of a sharp-edged airfoil. \textit{In this sense, the variational theory of lift recovers the Kutta condition as a special case in the limit of a sharp trailing edge}.

Moreover, the variational theory naturally extends classical theory to smooth geometries without sharp edges; i.e., to cases where the Kutta condition is not applicable because there is no singularity to remove. Figure \ref{Fig:Smooth_Shape} shows the variation of the normalized Appellian with $\hat\Gamma$ at several angles of attack for flow over a modified Zhukovsky airfoil with $\tau=0.1$ and $D=0.05$ (a trailing edge radius of 0.1\% of the chord). The figure also shows the resulting $\hat\Gamma$ from a Reynolds-Averaged Navier-Stokes (RANS) simulations, as detailed in \cite{Kutta_Flat_Plate}. In this case, the Appellian again possesses a unique minimum for each angle of attack; and the minimizing circulation is consistently closer to the RANS predictions than Kutta's values. Perhaps unexpectedly, even a modest smoothing of the trailing edge leads to a significant reduction in circulation at small angles of attack. This observation underscores that a brute-force extension of classical theory---via enforcement of the Kutta condition beyond its domain of applicability---may lead to substantial quantitative errors.

\section{Statement of the Main Argument in Ref. \cite{Peters_Gauss}}
The central argument advanced in Ref. \cite{Peters_Gauss} concerns the nature of the pressure gradient in incompressible flows in the absence of external body forces (e.g., electromagnetic forces). Specifically, the authors claim that the pressure gradient cannot be interpreted solely as a constraint force, but instead contains an impressed component. In this section, we present this argument as it is articulated in Ref. \cite{Peters_Gauss}, adopting their assumptions and point of view, and the following section explains why this reasoning is incorrect.

In their main demonstration, the authors of Ref. \cite{Peters_Gauss} focus themselves with the linearized dynamics about a uniform freestream $(U_\infty,0)$:
\[ \bm{u} = \tilde{\bm{u}} + (U_\infty,0) \;\;\mbox{and} \;\; p=\tilde{p} + p_\infty .\]
Under this linearization, the continuity constraint becomes $\bm\nabla \cdot \tilde{\bm{u}} = 0$, and the steady Euler momentum equation is written as
\begin{equation}\label{eq:Linearized_Steady_Euler}
\rho U_\infty \partial_x \tilde{\bm{u}} = -\bm\nabla \tilde{p}.
\end{equation}
Taking the divergence of the linearized momentum equation and substituting by the continuity constraint yields a Laplace equation in the pressure $\tilde{p}$
\begin{equation}\label{eq:Laplace_Pressure}
\nabla^2 \tilde{p} = 0.
\end{equation}

This equation is typically solved subject to a Neumann boundary condition, as discussed in detail by Gresho and Sani \cite{Pressure_BCs}. In contrast, the authors of Ref. \cite{Peters_Gauss} seek solutions subject to a Dirichlet boundary condition. This choice may not be physically plausible, particularly for an ideal fluid where tangential information on the boundary is not prescribed. In fact, the pressure on the boundary is the ultimate quantity of interest in any lifting analysis. While this point deserves careful consideration, we do not pursue it further here and instead focus on the crux of the authors' argument.

The authors further restrict their analysis to the framework of thin airfoil theory. In this setting, Peters's interesting formulation of potential functions \cite{Peters2} becomes particularly convenient. They, therefore, construct an analytical solution of the Laplace equation (\ref{eq:Laplace_Pressure}), subject to a Dirichlet boundary condition:
\begin{equation}\label{eq:Laplace_Pressure_Solution}
 \tilde{p}(\bm{x}) = \tau_s \Phi_s(\bm{x}) + \tau_a \Phi_a(\bm{x}) + \sum \tau_k \Phi_k(\bm{x}),
\end{equation}
where $\Phi$'s are known harmonic modes and the coefficients $\tau$'s are to be determined from the no-penetration boundary condition. The authors show that the associated velocity component normal to the surface can be written as
\begin{equation}\label{eq:Normal_Velocity_Peters}
w (\zeta) = \left[- \tau_a + \sum \tau_k \cos(k\zeta)\right]/(\rho U_\infty),
\end{equation}
where $\zeta$ is related to the chordwise coordinate $x$ through $x=b\cos\zeta$.

Equation (\ref{eq:Normal_Velocity_Peters}) may be viewed as a Fourier cosine representation of the function $w(\zeta)$. Hence, the coefficients $\tau_a$, $\tau_n$ are determined by matching the prescribed variation of $w$ along the chord; i.e., they guarantee full satisfaction of the no-penetration boundary condition for an arbitrary camber distribution or surface deformation.

This line of analysis leads to the central claim advanced in Ref. \cite{Peters_Gauss} concerning the nature of the pressure gradient force. The solution (\ref{eq:Laplace_Pressure_Solution}), together with its associated velocity field $\tilde{\bm{u}}$ (obtained, for example, from Bernoulli's equation), satisfies the linearized momentum equation (\ref{eq:Linearized_Steady_Euler}) for arbitrary values of the coefficients $\tau$'s. Moreover, the velocity field $\tilde{\bm{u}}$ satisfies the continuity constraint, again for arbitrary values of $\tau$'s; both properties hold by construction. The no-penetration boundary condition then determines the coefficients $\tau_a$, $\tau_k$, while leaving $\tau_s$ undetermined.

The authors' argument then proceeds as follows. Since both the governing momentum equation and all kinematic constraints are satisfied for any value of $\tau_s$, the corresponding pressure component $\tau_s \Phi_s$ cannot be a constraint force; it must be an impressed one. In other words, a constraint force, by definition, must be determined by satisfying the associated constraint. The fact that the strength $\tau_s$ of the pressure mode $\Phi_s$ remains arbitrary after all constraints and governing equations have been enforced precludes its candidature as a constraint force.

We fully agree with the foregoing analysis, yet disagree fundamentally with the conclusion drawn from it. The presented reasoning indeed appears as logical, internally consistent, and at first sight perhaps irrefutable. In this regard, it is difficult not to recall the words of the German physicist and philosopher Georg Christoph Lichtenberg (1742–1799), who stated that ``\textit{the most dangerous of all falsehoods is a slightly distorted truth}." In the next section, we clarify why the authors' conclusion does not follow. For the remainder of this section, however, we continue to adopt their line of reasoning: under the conclusion stated above, minimizing the Appellian cannot provide closure for potential flow.

Treating the pressure component $P_F=\tau_s \Phi_s$ as an impressed force, the Gaussian cost (\ref{eq:Gauss_Continuum}) can then be written as
\[ Z = \frac{1}{2} \int_\Omega \rho \left| \bm{u}_t + \bm{u}\cdot\nabla \bm{u} + \bm\nabla P_F \right|^2 d\bm{x}. \]
Accordingly, its steady, linearized form becomes
\begin{equation}\label{eq:Steady_Linearized_Gaussian}
Z = \frac{1}{2} \int_\Omega \rho \left| U_\infty \partial_x \tilde{\bm{u}}+ \bm\nabla P_F \right|^2 d\bm{x}.
\end{equation}
Also, for any member $\bm{u}$ of the family (\ref{eq:Family}), the perturbation velocity $\tilde{\bm{u}}$ is written as
\[ \tilde{\bm{u}} = \tilde{\bm{u}}_0 + \Gamma \bm{u}_1, \]
and the corresponding pressure gradient follows form the linearized momentum equation (or equivalently, Bernoulli's equation) as
\[ \nabla \tilde{p} = - U\infty \partial_x\left(\tilde{\bm{u}}_0 + \Gamma \bm{u}_1\right).\]
The term proportional to $\Gamma$ is identified in Ref. \cite{Peters_Gauss} as the impressed component $P_F=\tau_s \Phi_s$; i.e., we have
\[ \nabla P_F = - U\infty  \Gamma \partial_x \bm{u}_1. \]
Substituting by $\tilde{\bm{u}}$ and $\bm\nabla P_F$ into (\ref{eq:Steady_Linearized_Gaussian}) yields
\[ Z = \frac{1}{2} \int_\Omega \rho \left| U\infty \partial_x\left(\tilde{\bm{u}}_0 + \Gamma \bm{u}_1\right) + \left(- U\infty  \Gamma \partial_x \bm{u}_1\right) \right|^2 d\bm{x}, \]
which simplifies to
\begin{equation}\label{eq:Steady_Linearized_Gaussian_Final}
Z = \frac{1}{2} \int_\Omega \rho \left| U\infty \partial_x \tilde{\bm{u}}_0 \right|^2 d\bm{x},
\end{equation}
Notably, the resulting Gaussian cost is independent of $\Gamma$.

The component $P_F$, when incorporated into the integrand of the Gaussian cost as an impressed force, precisely cancels the dependence of $Z$ on the circulation. Consequently, minimization of $Z$ with respect to $\Gamma$ becomes trivial and can no longer provide closure for potential flow.

\section{Response to the Main Argument in Ref. \cite{Peters_Gauss}}\label{Sec:Responnse_Main}
\subsection{Constraint Forces Versus Impressed Ones}
In analytical mechanics---whether one aims to apply Gauss's principle, Lagrangian mechanics and the principle of least action, or any equivalent formulation---it is essential to decompose forces into two distinct categories: impressed forces and constraint forces. This decomposition is unique. It is not a matter of interpretation or convenience and leaves no room for ambiguity\footnote{An exception arises in the presence of working constraint forces, such as Coulomb friction. This situation is the opposite of the case considered in Ref. \cite{Peters_Gauss}, where a workless force is claimed to be impressed.}. This classification is standard in analytical mechanics and is discussed in detail in classical textbooks (see, e.g., \cite{Lanczos_Variational_Mechanics_Book,Papastavridis}), as well as in graduate-level mechanics courses in mechanical engineering.

Constraint forces exist solely to enforce kinematical or geometrical constraints. If a constraint is removed, the corresponding constraint force necessarily ceases to exist. Moreover, a constraint force is uniquely determined by the governing dynamics together with the constraint equations. In contrast, impressed forces are entirely agnostic to constraints. In the language of Gauss, they drive the \textit{free motion}---the motion that would occur in the absence of any constraints. Their existence, and any possible functional dependence on the motion through constitutive laws, is completely independent of the constraints, in sharp contrast to constraint forces whose very \textit{raison d'etre} is the constraints themselves.

There is no ambiguity in such a decomposition. Given a force, its classification is straightforward: does it depend on the constraint in any way, or is it entirely independent of it? Consider, for example, the simple pendulum shown in Fig. \ref{Fig:Pendulum_Schematic}. The force $R$ in the rod is a constraint force, because if the rod is removed, the force will no longer exist. Moreover, for a given instantaneous motion $(\theta,\dot\theta)$, the constraint force is uniquely determined from the dynamics and constraint equation as
\begin{equation}\label{eq:Pendulum_Radial_EOM}
R = mg\cos\theta + mL \dot\theta^2 .
\end{equation}
The force $R$ exists solely to ensure the circular trajectory of the mass, and to do so, it must satisfy the balance equation (\ref{eq:Pendulum_Radial_EOM}). In other words, the dynamics and constraint jointly impose a precise restriction (\ref{eq:Pendulum_Radial_EOM}) on the constraint force $R$. Any constraint force in mechanics is expected to be restricted in a way similar to (\ref{eq:Pendulum_Radial_EOM}). In contrast, the gravity force is entirely independent of the constraint. If the rod is broken, it will remain as is: $mg$ acting downward. No restriction can ever be imposed on gravity by the constraint, precisely because gravity is not a constraint force.

\begin{wrapfigure}{l}{0.30\textwidth}
\vspace{-0.15in}
 \begin{center}
 \includegraphics[width=5cm]{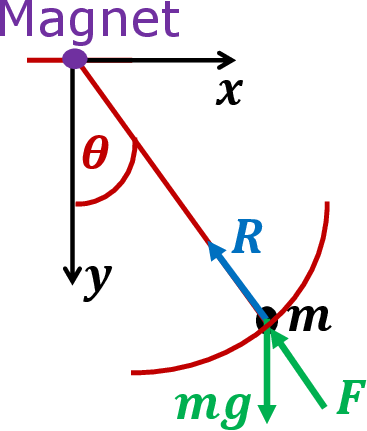}  \vspace{-0.2in}
 \caption{Pendulum schematic with a magnet at the hinge.}
 \label{Fig:Pendulum_Schematic_Magnet} \vspace{-0.1in}
 \end{center} \vspace{-0.1in}
\end{wrapfigure} \noindent To clarify this distinction and address more nuanced situations, assume that the pendulum mass is ferromagnetic, and that there is a magnet placed at the hinge, which is attracting it by an electromagnetic force $F$, as shown in Fig. \ref{Fig:Pendulum_Schematic_Magnet}. Although the magnetic force $F$ acts in precisely the same direction as the constraint force $R$, and may even assist in maintaining the constraint (by reducing the magnitude of the constraint force required to ensure the circular trajectory constraint), it is nonetheless an impressed force. If the rod is broken, the force will remain unchanged, acting radially with the same constitutive relation from electromagnetism; its existence and characteristics are entirely independent of the constraint. There is no restriction whatsoever imposed by the constraint on its characteristics.

Indeed, \textbf{\textit{impressed forces are arbitrary}} by definition. One cannot use the governing dynamics or the constraint equation to impose any restriction on the form, direction, or magnitude of an impressed force. \textbf{\textit{The pendulum dynamics and constraint cannot, in any way, impose restrictions on the characteristics of the gravitational and magnetic forces}} in Fig. \ref{Fig:Pendulum_Schematic_Magnet}. These forces define the free motion that would occur in the absence of the constraint; their characteristics are completely specified independent of it.

Let us now reflect this discussion on incompressible flows. The space of admissible motions ($\simeq$ configuration manifold) consists of divergence-free velocity fields satisfying the no-penetration boundary condition; it is depicted as the horizontal plane in Fig. \ref{Fig:Incompressible_Schematic}. The associated constraint force is orthogonal to this space and therefore lies in the space of gradient vector fields---it is the pressure gradient $\nabla p$. As in any mechanics problem (e.g., the pendulum example), one may use the governing dynamics (i.e., the Navier-Stokes equation) together with the continuity constraint to derive an equation governing the constraint force. In incompressible fluid mechanics, this step is well known: taking the divergence of the Navier–Stokes equation (\ref{eq:NS_Rearranged}) and substituting by the continuity constraint, one obtains the familiar Poisson equation for the pressure:
\begin{equation}\label{eq:Poisson_Pressure}
\nabla^2  p = \rho \nabla\cdot \bm{u}_t^{\rm{free}} =  \rho \nabla\cdot \left(\bm{f}-\bm{u}\cdot \bm\nabla \bm{u}\right).
\end{equation}
This equation represents the projection of the Navier-Stokes dynamics onto the direction normal to the admissible plane of incompressible motions; i.e., onto the space of gradient vector fields. It is directly analogous to projecting the pendulum dynamics onto the radial direction. In both cases, the resulting equation governs the constraint force.

In the linearized, potential-flow formulation of Ref. \cite{Peters_Gauss}, the Poisson equation for the pressure reduces to the Laplace equation (\ref{eq:Laplace_Pressure}). Table \ref{Tab:Taxonomy} juxtaposes the structure of the full Navier–Stokes dynamics with its linearized potential-flow counterpart and with the dynamics of the simple pendulum.

\begin{table*}
\begin{centering}
\begin{tabular}{l|c|c|c}
& Pendulum & Navier-Stokes & Linearized Potential Flow \\ \hline

Full Dynamics &
$\begin{array}{l} m\ddot{x} = -R\frac{x}{L} \\ m\ddot{x} = mg -R\frac{y}{L} \end{array}$ &
$\rho \left(\bm{u}_t +\bm{u}\cdot\nabla \bm{u}\right)= \rho\bm{f}-\bm\nabla p$ &
$\rho \left(\tilde{\bm{u}}_t +U_\infty\partial_x \tilde{\bm{u}}\right)= -\bm\nabla \tilde{p}$\\

Constraint & $x^2+y^2=L^2$ & $\nabla\cdot \bm{u}=0$ \& no-penetration & $\nabla\cdot \tilde{\bm{u}}=0$ \& no-penetration \\

Tangential Dynamics & $\ddot\theta = -\frac{g}{L}\sin\theta$ & No Explicit Formula & $\nabla^2\phi=0$\\

Normal Dynamics & $R = mg\cos\theta + mL \dot\theta^2$ & $\nabla^2  p = \rho \nabla\cdot \bm{u}_t^{\rm{free}}$ & $\nabla^2 \tilde{p}=0$\\

\end{tabular}
\caption{Juxtaposing Navier-Stokes' dynamics with the linearized potential flow and the simple pendulum dynamics. Here $\phi$ denotes the velocity potenital.}
  \label{Tab:Taxonomy}
\end{centering}
\end{table*}

With this technical background, the flaw in the reasoning of Ref. \cite{Peters_Gauss} regarding the role of pressure becomes evident. The authors decompose the pressure into two components:
\[ \tilde{p} = P_R + P_F, \]
where $P_R$ is identified as a constraint component and $P_F$ as an ``impressed" one. Since $\tilde{p}$ must satisfy the Laplace equation (\ref{eq:Laplace_Pressure}), it follows that
\[ \nabla^2 \left(P_R + P_F\right)=0. \]
The authors, then, proceeded to impose the Laplace equation separately on each component:
\begin{equation}\label{eq:Laplace_Habal}
\nabla^2 P_R = 0 \;\;\mbox{and} \;\; \nabla^2 P_F=0,
\end{equation}
which corresponds to equation (17) in Ref. \cite{Peters_Gauss}.

This restriction on $P_F$ is in clear contradiction with the very definition of impressed forces in analytical mechanics, which are independent of the constraints. The condition imposed on $P_F$ arises solely from combining the governing dynamics with the incompressibility constraint. In analytical mechanics, such a restriction is the defining hallmark of a constraint force. When a force is required to satisfy an equation obtained by enforcing the constraints, it cannot---by definition---be regarded as impressed. The very act of subjecting $P_F$ to (\ref{eq:Laplace_Habal}) therefore establishes it as a constraint force, not otherwise.

A potential source of confusion arises in the nuanced situation where an impressed force (e.g., electromagnetic) takes the form of a gradient field $-\bm\nabla \varphi$. In incompressible fluid mechanics, it is common to \textit{absorb} such a force into the pressure force; i.e., to define a modified pressure
\[ \hat{p} = p + \varphi,\]
and to treat $\bm\nabla \varphi$ as if it did not exist in the momentum equation, since it is already included in $\hat{p}$. After solving the Poisson equation for $\hat{p}$, the physical pressure is then recovered via $p = \hat{p} - \varphi.$ This bookkeeping practice may create the impression that the \textit{total} pressure consists of two components: an impressed contribution $\varphi$ and a constraint one $p$.

First, the situation described above is fundamentally different form the case considered in Ref. \cite{Peters_Gauss}, which is concerned with the airfoil problem in the absence of  external body forces (e.g., electromagnetic). Second, although the above absorption procedure is mathematically legitimate, its interpretation can be misleading. There is no ``total" pressure in a physical sense; there is only a single pressure field $p$, whose gradient $\bm\nabla p$ acts as the constraint force required to ensure continuity and the no-penetration boundary condition. The user-defined quantity $\hat{p} = p + \varphi$ is a purely mathematical artifact, introduced for convenience, and should not be interpreted as a physical decomposition of pressure into impressed and constraint components.

The situation can be further elucidated with the aid of the pendulum example with a magnet, shown in Fig. \ref{Fig:Pendulum_Schematic_Magnet}. In this case, the electromagnetic force $F$ happens to act in the same direction as the constraint force $R$. It may therefore be tempting to define a combined force
\[\hat{R} = R+F, \]
and solve the problem while ignoring the impressed force $F$. After solving for $\hat{R}$ from
\[ \hat{R} = mg\cos\theta + mL \dot\theta^2, \]
we can easily recover the actual constraint force in the rod $R$ from $R = \hat{R} - F$. Clearly, this bookkeeping procedure does not imply that force in the rod has two physical components: one constraint $R$ and one impressed $F$.

Indeed, although the electromagnetic force $\bm\nabla \varphi$ may resemble a pressure force, it is fundamentally different in nature. The pressure force, as a constraint force, must satisfy the Poisson equation in order to enforce the continuity constraint. In contrast, $\bm\nabla \varphi$ is arbitrary; it does not have to satisfy the Poisson equation or the Neumann boundary condition. So, even if one were to accept the mathematically permissible (but physically artificial) idea of decomposing the pressure into two components, their identification remains unambiguous. Any pressure component that is restricted by the continuity constraint is, by definition, a constraint force.

The role of pressure in incompressible fluid mechanics is well established in the mathematical fluid mechanics community (e.g., \cite{Chorin_Marsden_Book,Temam_Projection,Geometric_Control_Fluid_Dynamics,Morrison2020lagrangian,DeVoria_Hamiltonian_JFM}) as well as in computational fluid dynamics (e.g., \cite{Pressure_BCs,Chorin_Projection,Projection_Book,Projection_Review,Hirsch_Book2}). For example, Gresho and Sani \cite{Pressure_BCs} explicitly described the pressure in incompressible flows as the ``\textit{Lagrange multiplier that constrains the velocity field to remain divergence-free.}" Sanders et al. \cite{DeVoria_Hamiltonian_JFM} referred to this fact as the ``\textit{well-known result that the pressure usually serves as a Lagrange multiplier for the incompressibility constraint (see p. 361 of Lanczos \cite{Lanczos_Variational_Mechanics_Book} and pp. 137 and 141 of \cite{badin2018variational})}." Later in the same work, they stated even more explicitly: ``\textit{In this case, the pressure is determined last of all, and is whatever it needs to be to enforce the incompressibility constraint ... (again consistent with the role of pressure as Lagrange multiplier \cite{Lanczos_Variational_Mechanics_Book,badin2018variational}).}"

Interestingly, this interpretation appears to be as old as analytical mechanics itself. Morrison \cite{Morrison2020lagrangian} noted that ``\textit{It is worth noting that Lagrange knew the Lagrange multiplier turns out to be the pressure, but he had trouble solving for it.}" However, this fact may be less familiar within the aeronautics literature and to a practicing mechanical engineer who is accustomed to viewing the pressure force as a \textit{driving} force (such as in pipe flows) rather than as a \textit{reaction} enforcing a constraint. Two important points must therefore be emphasized. First, a reaction force may indeed drive the motion. For example, a double pendulum in a horizontal plane is driven entirely by the constraint forces in the two rods. Yet, it can exhibit remarkably rich dynamics even in the absence of gravity or any impressed forces. Constraint forces alone can thus generate complex motion, depending on the nature of the constraint and curvature of the configuration manifold. Clearly, the divergence-free constraint in incompressible flows is, in this regard, vastly richer than the constraint imposed by a double pendulum.

Second, the engineering intuition drawn from pipe flow is, in fact, precisely correct. In that setting, the pressure difference across the pipe acts as an \textit{impressed} force. To clarify this nuanced distinction, we recall Remark 3, which emphasizes the essential role of the no-penetration boundary condition in Helmholtz orthogonality between gradient vector fields and divergence-free vector fields. It should be noted that such orthogonality holds if $\bm{u}\cdot\bm{n}$ is prescribed on the boundary; it need not necessarily vanish. In a pipe flow, this requirement is satisfied on the solid walls. It may also be satisfied at the inlet, since the normal velocity $\bm{u}\cdot\bm{n}$ (i.e., the inlet velocity profile, often taken as uniform) is typically specified. However, at the outlet, the velocity profile is not prescribed \textit{a priori}; it emerges as part of the solution. Consequently, a gradient field $\bm\nabla \varphi$ is not orthogonal to the space of admissible incompressible motions in a pipe flow and therefore can generate accelerations along that space.

Again, this situation is fundamentally different from an external flow over a body, where the velocity normal to the surface is prescribed everywhere on the boundary. In that case, any gradient field is orthogonal to the space of admissible incompressible motions and therefore cannot, in any way, change the resulting motion. It can only modify the pressure force required to ensure the constraint.

If the pressure gradient in incompressible flow over a body is unequivocally a constraint force, how should one interpret the logical reasoning and analysis provided in Ref. \cite{Peters_Gauss}, and summarized in the preceding section, which appears to identify a pressure component that remains undetermined after enforcing the governing dynamics and all constraints?

\subsection{Reconciliation: How should the analysis of \cite{Peters_Gauss} be interpreted?}
To address this question, two facts must be recalled. First, it is widely known the conditions (i)-(iv) listed in Sec. \ref{sec:Classical_Theory} are not sufficient to yield a unique solution to the airfoil problem. These conditions comprise: (i) momentum conservation (\ref{eq:Euler}), (ii) continuity (\ref{eq:Continuity}), (iii) the no-penetration boundary condition, and (iv) the far field boundary condition. Together, they constitute the governing dynamics and all kinematical constraints. Yet, even after enforcing all of them, the classical formulation admits a one-parameter family (\ref{eq:Family}) of solutions:
\[ \bm{u}(\bm{x})=\bm{u}_0(\bm{x})+\Gamma \bm{u}_1(\bm{x}), \]
in which the circulation $\Gamma$ remains undetermined.

Second, within the linearized formulation adopted in \cite{Peters_Gauss}, the perturbation pressure $\tilde{p}$ is linearly related to the perturbation velocity $\tilde{\bm{u}}=\tilde{\bm{u}}_0+\Gamma \bm{u}_1$ through the linearized steady Euler equation (\ref{eq:Linearized_Steady_Euler}), or Bernoulli's equation
\begin{equation}\label{eq:Pressure_Velocity_Dependence}
\tilde{p} = -\rho \bm{U}_\infty \cdot \tilde{\bm{u}},
\end{equation}
where $\bm{U}_\infty = (U_\infty,0)$. Accordingly, the perturbation pressure can be written as
\[ \tilde{p}=-\rho\bm{U}_\infty \cdot \left(\tilde{\bm{u}}_0+\Gamma \bm{u}_1\right) \equiv \tilde{p}_0 +\Gamma p_1, \]
where the term proportional to $\Gamma$ is precisely the undetermined component $\tau_s \Phi_s$ identified in Ref. \cite{Peters_Gauss} and subsequently interpreted as an impressed component:
\[ -\rho\Gamma \bm{U}_\infty \cdot \bm{u}_1  = \Gamma p_1  = \tau_s \Phi_s, \]
where $\tau_s\equiv - \frac{\rho U_\infty \Gamma}{2\pi b}$.

In this setting, the correct interpretation of the analysis in Ref. \cite{Peters_Gauss} is straightforward. When the airfoil problem is formulated in the standard way in terms of velocity $\bm{u}$, there is one component of the velocity field that remains undetermined even after enforcing the dynamics and all constraints. If one instead chooses to formulate the problem in terms of pressure, as in Ref.  \cite{Peters_Gauss}, there is no reason for this situation to change. The non-uniqueness is inherent to the problem itself. In mathematical terms, the problem is not well posed. Consequently, it is certainly expected that one component of the pressure field remains undetermined after enforcing the dynamics and all constraints. This undetermined pressure component is simply the one associated with the undetermined velocity component. Its existence has nothing to do with the nature of the pressure force, nor does it imply that this component is impressed rather than a constraint force.

To clarify this point further, consider first the interpretation of the component $\tilde{p}_0=-\rho\bm{U}_\infty \cdot \tilde{\bm{u}}_0$. This component is identified in Ref. \cite{Peters_Gauss} as a constraint contribution. Indeed, it represents the constraint force \textit{exerted} so that the non-circulatory velocity component $\tilde{\bm{u}}_0$ satisfies the continuity and no-penetration constraints. By the same reasoning, the component
\begin{equation}\label{eq:Pressure_Velocity_Dependence_Undetermined}
\tau_s \Phi_s=\Gamma p_1=-\rho\Gamma\bm{U}_\infty \cdot \bm{u}_1
\end{equation}
must likewise be interpreted as the constraint force exerted so that the circulatory velocity component $\Gamma \bm{u}_1$ satisfies those same constraints. Its magnitude $\tau_s$ (or $\Gamma$) depends directly on the magnitude $\Gamma$ of the corresponding velocity component, which itself remains undetermined.

The authors of Ref. \cite{Peters_Gauss} argued, however, that the circulatory motion $\bm{u}_1$ \textit{naturally} satisfies the continuity and no-penetration constraints, and therefore concluded that this conformity occurs automatically, without the intervention of a Lagrange multiplier or a constraint force. On this basis, they asserted that the component $\tau_s \Phi_s$ (equivalently $\Gamma p_1$) is not required to enforce any constraints. This reasoning is circular, and its flaw becomes transparent once Remark 2 is recalled. The fact that a solution \textit{emerges} in a form that satisfies the constraints does not imply that this conformity occurred without the action of a constraint force. As emphasized in Remark 2, the constraint force \textit{exerted} to achieve such conformity is revealed by the equation governing the motion in the normal direction: the radial equation in the pendulum example, the Poisson equation in the full Navier-Stokes system, or the Laplace equation in the linearized formulation, which precisely yields Eqs. (\ref{eq:Pressure_Velocity_Dependence}, \ref{eq:Pressure_Velocity_Dependence_Undetermined}).

Indeed, the component $P_F$ in Ref. \cite{Peters_Gauss} is the constraint force exerted to impose the continuity and no-penetration constraints on the circulatory motion. Its magnitude remains undetermined solely because the magnitude of the circulatory velocity field itself is undetermined---a direct consequence of the well-known non-uniqueness (i.e., ill-posedness) of the classical airfoil problem.

This point becomes even clearer when the non-linear theory is considered, in contrast to the linearized formulation emphasized in Ref. \cite{Peters_Gauss}. In the nonlinear setting, the steady Euler's equation yields
\[ \bm\nabla p = -\rho\bm{u}\cdot \bm\nabla \bm{u} = -\rho\left[\bm{u}_0\cdot \bm\nabla \bm{u}_0 + \Gamma \left(\bm{u}_0\cdot \bm\nabla \bm{u}_1 + \bm{u}_1\cdot \bm\nabla \bm{u}_0 \right) + \Gamma^2 \bm{u}_1\cdot \bm\nabla \bm{u}_1\right].\]
In Appendix B of Ref. \cite{Peters_Gauss}, the authors identified the terms involving $\Gamma$ as constituting an impressed component:
\[ \bm\nabla P_F = - \rho \left[\Gamma \left(\bm{u}_0\cdot \bm\nabla \bm{u}_1 + \bm{u}_1\cdot \bm\nabla \bm{u}_0 \right) + \Gamma^2 \bm{u}_1\cdot \bm\nabla \bm{u}_1\right], \]
which leads to
\begin{equation}\label{eq:Poisson_PF}
\nabla^2 P_F = - \rho \bm\nabla \cdot \underbrace{\left[ \Gamma \left(\bm{u}_0\cdot \bm\nabla \bm{u}_1 + \bm{u}_1\cdot \bm\nabla \bm{u}_0 \right) - \Gamma^2 \bm{u}_1\cdot \bm\nabla \bm{u}_1 \right]}_{-\bm{u}_t^{\rm{free}}}.
\end{equation}
In contrast to the linearized case, where the Laplace equation $\nabla^2 P_F=0$ obscures the role of $P_F$ (since the right hand side is zero), the nonlinear formulation makes that role explicit. Equation (\ref{eq:Poisson_PF}) unambiguously shows that $P_F$ is governed by the Poisson equation whose source term is the divergence of a specific component of the convective acceleration. That component is precisely the part of the \textit{free motion} whose conformity with the incompressibility constraint is enforced by the pressure $P_F$, see Eq. (\ref{eq:Poisson_Pressure}). In this sense, $\bm\nabla P_F$ is not an impressed force at all; it is the constraint force associated with projecting this part of the nonlinear free acceleration.

Finally, it is important to emphasize a point that was overlooked in Ref. \cite{Peters_Gauss}: for a given velocity field $\bm{u}$, the pressure field required to enforce incompressibility and no-penetration is \textbf{\textit{uniquely}} determined. This is, in fact, the defining property of a constraint force in analytical mechanics. A constraint force is uniquely specified once a kinematically admissible motion is prescribed (whether or not that motion is the physical one). This principle holds universally. In the pendulum example, once $(\theta,\dot\theta)$ is given, the radial reaction force is uniquely fixed. Likewise, for incompressible flows, once a velocity field satisfying continuity and boundary conditions is specified, the corresponding pressure field is uniquely determined. The non-uniqueness encountered in classical airfoil theory therefore resides entirely in the dynamics---namely, in the fact that the governing equations do not uniquely determine the motion. It does not reflect any ambiguity in the nature of pressure itself. Once the motion is fixed, the pressure is fixed, just as any other constraint force in mechanics.


\section{On Misconceptions Concerning Basic Concepts in Analytical Mechanics and the Calculus of Variations}
Although we have several concerns regarding statements made in Ref. \cite{Peters_Gauss}, we deliberately avoid nitpicking. Instead, we focus on core issues that pertain to basic concepts in analytical mechanics and the calculus of variations. In particular, this section addresses two such misconceptions, rather than attempting an exhaustive critique of all conceptual errors in Ref. \cite{Peters_Gauss}.

\subsection{On Misconceptions Concerning Boundary Conditions in the \textit{Simplest Problem in the Calculus of Variations}}
In Ref. \cite{Variational_Lift_JFM}, the authors established a theorem asserting that if a flow field evolves from a given initial condition by following the local acceleration $\bm{u}_t$ that minimizes the Appellian (\ref{eq:Appellian_Continuum}) at every instant, subject to the continuity and no-penetration constraints, then the resulting evolution necessarily satisfies Euler's dynamics. This result was subsequently extended to the Navier-Stokes equation through the principle of minimum pressure gradient, see Theorem 1 of \cite{PMPG_PoF,VPNS_PRF} and Proposition 2 of \cite{NS_QP_IEEE}. This result is, in essence, the variational analogue of the Helmholtz-Leray projection. It represents the continuum-mechanics counterpart of the relation beween Gauss's principle and Newton's equations in particle mechanics, as illustrated schematically in Fig. \ref{Fig:Gauss_Newton_PMPG}. In all cases, Newton's equations of motion---Euler or Navier-Stokes---arise as the first-order necessary condition of optimality for minimizing the Gaussian cost (the Appellian $S$ or the PMPG cost $\mathcal{A}$) subject to the kinematic constraints.

The authors of Ref. \cite{Peters_Gauss} raised objections to the proof of the theorem presented in Ref. \cite{Variational_Lift_JFM}. The raised objections, however, are inconsistent with the standard treatment of, what Burns referred to as, the \textit{Simplest Problem in the Calculus of Variations} \cite{Burns_Optimal_Control_Book}. To show this inconsistency, we first recall the elementary treatment of constrained optimization.

\subsubsection{Elementary Treatment of Constrained Optimization}
Consider the simple optimization problem
\[ \min_{x,y} f(x,y) = x^2+y^2 \;\;\mbox{s.t.}\;\; y=1+x^2.\]
This problem can be solved in one of two standard ways. One approach is to use the constraint to eliminate one of the degrees of freedom, thereby reducing the problem to an unconstrained minimization in a single variable:
\[ \min_{x} F(x) = f(x,y(x)) = x^2+(1+x^2)^2,\]
with the first-order necessary condition:
\[ \frac{d F}{dx} = 2x+4x(1+x^2) = 0.\]

Alternatively, one may augment the cost $f$ with the constraint
\[ \psi=y-1-x^2=0 \]
using a Lagrange multiplier $\lambda$:
\[ \mathcal{L} = f + \lambda \psi = x^2+y^2 +\lambda \left(y-1-x^2\right). \]
In this formulation, the first-order necessary conditions for optimality are:
\[ \begin{array}{lll}
\frac{\partial \mathcal{L}}{\partial x} &=& 2x-2x\lambda = 0 \\
\frac{\partial \mathcal{L}}{\partial y} &=& 2y+\lambda = 0 \\
\frac{\partial \mathcal{L}}{\partial \lambda} &=& y-1-x^2 = 0.
 \end{array} \]
This system is equivalent to the reduced formulation obtained by eliminating the constraint directly.

We now turn to the statement of the \textit{Simplest Problem in the Calculus of Variations} \cite{Burns_Optimal_Control_Book}.

\subsubsection{Statement of the \textit{Simplest Problem in the Calculus of Variations}}
A standard problem in the calculus of variations is to minimize a cost \textit{functional}
\[ J[v(t)] = \int_{t_0}^{t_1} f(t,v(t),\dot{v}(t)) dt \]
over a class of admissible functions (say continuously differentiable functions $C^1 [t_0,t_1]$ for simplicity) that satisfy prescribed boundary conditions
\begin{equation}\label{eq:SPCOV_BCs}
v(t_0)=v_0 \;\; \mbox{and} \;\; v(t_1)=v_1.
\end{equation}
This problem is referred to by Burns \cite{Burns_Optimal_Control_Book} as the \textit{Simplest Problem in the Calculus of Variations} (SPCOV), and is formally stated as
\[ \min_{ v(t)\in \mathcal{V}} J[v(t)] = \int_{t_0}^{t_1} f(t,v(t),\dot{v}(t)) dt , \]
where the set $\mathcal{V}$ of admissible functions is defined by
\[ \mathcal{V} = \left\{ v(t) \in C^1 [t_0,t_1] | \; v(t_0)=v_0, v(t_1)=v_1 \right\}. \]
That is, the SPCOV consists of finding a function $\bm{v}^*\in \mathcal{V}$ such that
\[ J[v^*(t)] \leq J[v(t)] \; \forall v(t) \in \mathcal{V}. \]

To solve this class of infinite-dimensional optimization problems, Lagrange invented the concept of a \textit{variation} $\delta v(t)$. According to this concept, a candidate function $v(t)$ is perturbed by some perturbation function---the variation---$\delta v(t)$, such that the perturbed function $(v+ \delta v)(t)$ must remain within the admissible set $\mathcal{V}$. We then compare the value of the functional $J[v(t)]$ against $J[(v+ \delta v)(t)]$. We emphasize that, in order for this comparison to be meaningful and mathematically legitimate, the perturbed function $v+ \delta v$ must be admissible; i.e., $v+ \delta v \in \mathcal{V}$. Otherwise, the comparison is made against an \textit{infeasible design}.

Since the defining characteristics of $\mathcal{V}$ is to satisfy the prescribed boundary conditions, we then have
\[ (v+ \delta v)(t_0)=v_0 \;\; \mbox{and} \;\; (v+ \delta v)(t_1)=v_1, \]
but $v$ itself must satisfy the same boundary conditions (\ref{eq:SPCOV_BCs}). Hence, we obtain the well-known fact that the variations $\delta v$ must vanish at the end points:
\[ \delta v(t_0)=0 \;\; \mbox{and} \;\; \delta v(t_1)=0. \]

We now proceed to derive the first-order necessary condition of optimality for the SPCOV. However, we recognize that the boundary conditions (\ref{eq:SPCOV_BCs}) formally appear as constraints imposed on the optimization problem; and indeed they are. As in elementary constrained optimization, discussed in the previous subsection, these constraints may be handled in one of two equivalent ways: substituting by the constraints or via Lagrange multipliers. In the calculus of variations, the standard---and overwhelmingly adopted---approach (see, e.g., \cite{Bolza_COV,Caratheodory_COV,bliss1925calculus,Lanczos_Variational_Mechanics_Book,COV_Ewing,COV_Dacorogna,Burns_Optimal_Control_Book}) is the former: the boundary conditions (\ref{eq:SPCOV_BCs}) are enforced directly by restricting the admissible variations, rather than by introducing Lagrange multipliers.

Accordingly, the first-order necessary condition, obtained from the first variation of $J$ with respect to $v$ in the direction of any admissible variation $\delta v$, is given by
\[ \delta J = \int_{t_0}^{t_1} \left[ f_v \delta v + f_{\dot{v}} \frac{d}{dt}(\delta v)\right]  dt = 0. \]
Integrating the second term by parts yields
\[ \delta J = \int_{t_0}^{t_1} \left[f_v \delta v - \frac{d}{dt}(f_{\dot{v}}) \delta v\right] dt + \left[f_{\dot{v}} \delta v\right]_{t_0}^{t_1}= 0. \]
It is precisely at this stage that the boundary conditions are enforced directly---by restricting the admissible variations---to eliminate the boundary term, resulting in
\[ \delta J = \int_{t_0}^{t_1} \left[f_v - \frac{d}{dt}(f_{\dot{v}}) \right]\delta v dt = 0. \]
Since this must hold for all admissible $\delta v$, the familiar Euler-Lagrange differential equation is obtained as the first-order necessary condition:
\begin{equation}\label{eq:SPCOV_FONC}
\frac{d}{dt}\left(f_{\dot{v}}(t,v^*(t),\dot{v}^*(t))\right) - f_v(t,v^*(t),\dot{v}^*(t)) = 0.
\end{equation}
The boundary conditions are also imposed \textit{a posteriori} on the Euler-Lagrange equation (\ref{eq:SPCOV_BCs}); as a differential equation, it requires boundary conditions to solve for the optimal function $v^*(t)$.

The above procedure is standard in the calculus of variations, as discussed extensively in the classical literature (e.g., \cite{Bolza_COV,Caratheodory_COV,bliss1925calculus,Lanczos_Variational_Mechanics_Book,COV_Ewing,COV_Dacorogna,Burns_Optimal_Control_Book}). When the function $v(t)$ is vector-valued, $\bm{v}(t)$, and is required to satisfy prescribed end conditions analogous to (\ref{eq:SPCOV_BCs}), then the admissible variations $\delta\bm{v}$ must vanish componentwise at the end points. If, however, only one component of $\bm{v}$ (e.g., the normal component $\bm{v}\cdot\bm{n}$) is prescribed  on the boundary, then the corresponding component (e.g., $\delta \bm{v}\cdot\bm{n}$) of the variation must likewise vanish on the boundary.

It is understandable, however, that some authors might choose to follow a different, though non-standard, approach for enforcing the boundary conditions (\ref{eq:SPCOV_BCs}) as side constraints using Lagrange multipliers. In particular, in variational formulations of solid mechanics, these boundary conditions are geometric constraints on the structure (e.g., fixed or simply supported boundaries). In that setting, the associated Lagrange multipliers acquire a clear physical interpretation; they represent reaction forces at the supports. It is, therefore, not surprising that researchers with background in solid mechanics, such as the authors of Ref. \cite{Peters_Gauss}, may find this approach more natural. Nevertheless, although legitimate, this treatment is neither the only valid approach, nor the one conventionally adopted in the calculus of variations.


We are now ready to present the theorem stated in Ref. \cite{Variational_Lift_JFM}, together with its proof.

\subsubsection{On the Equivalence Theorem in Ref. \cite{Variational_Lift_JFM}}
The theorem asserts that if the local acceleration $\bm{u}_t(\bm{x};t)$ is continuously differentiable in $\Omega\subset\mathbb{R}^3$ and, at each instant $t\in\mathbb{R}$, minimizes the Appellian functional
\[ S[\bm{u}_t] = \frac{1}{2}\int_\Omega \rho \left(\bm{u}_t(\bm{x})+\bm{u}(\bm{x})\cdot\bm\nabla \bm{u}(\bm{x})\right)^2 d\bm{x}\]
subject to the incompressibility constraint
\[ \bm\nabla \cdot\bm{u}=0 \;\; \forall \; \bm{x}\in\Omega, \; t\in\mathbb{R},\]
and the no-penetration boundary condition
\[ \bm{u}\cdot \bm{n} = 0, \;\; \forall \bm{x}\in\partial\Omega, \; t\in\mathbb{R} \]
then the resulting acceleration $\bm{u}_t(\bm{x})$ necessarily satisfies Euler's dynamics:
\[\rho\left(\bm{u}_t+\bm{u}\cdot\bm\nabla \bm{u}\right)=-\bm\nabla \lambda \;\; \forall \; \bm{x}\in\Omega, \; t\in\mathbb{R} \]
for some scalar field $\lambda(\bm{x},t)$.

The proof follows directly from standard arguments in the calculus of variations. Since the constraint $\bm\nabla \cdot\bm{u}=0$ must hold for all times $t$, it follows immediately (by differentiation with respect to time) that the local acceleration field must satisfy $\bm\nabla \cdot\bm{u}_t=0$. Consequently, minimizing the Appellian $S$ subject to the constraint $\bm\nabla \cdot\bm{u}_t=0$ for all $\bm{x}\in\Omega$ constitutes a constrained variational problem posed over space (with time acting only as a parameter). To tackle this problem, we introduce a Lagrange multiplier field $\lambda(\bm{x})$ at every instant, and construct the augmented functional
\[ \mathcal{L} = S - \int_\Omega \lambda(\bm{x}) \left(\bm\nabla \cdot\bm{u}_t(\bm{x})\right) d\bm{x}. \]

A necessary condition for optimality of the constrained  minimization problem is that the first variation of the augmented functional $\mathcal{L}$ vanish with respect to all admissible $\delta\bm{u}_t(\bm{x})$ that \textit{satisfy the no-penetration boundary condition}. The first variation of $\mathcal{L}$ with respect to $\bm{u}_t$ is written as
\[ \delta\mathcal{L} = \int_\Omega \left[\rho \left(\bm{u}_t+\bm{u}\cdot\bm\nabla \bm{u}\right)\cdot\delta \bm{u}_t - \lambda \bm\nabla \cdot\delta\bm{u}_t \right] d\bm{x}=0. \]
The last term $\lambda \bm\nabla \cdot\delta\bm{u}_t$ can be integrated by parts to give
\begin{equation}\label{eq:First_Variation}
\delta\mathcal{L} = \int_\Omega \left[\rho \left(\bm{u}_t+\bm{u}\cdot\bm\nabla \bm{u}\right)\cdot\delta \bm{u}_t + \bm\nabla \lambda \cdot\delta\bm{u}_t \right] d\bm{x} - \oint_{\partial\Omega} \left(\lambda \delta \bm{u}_t \cdot \bm{n} \right) d\bm{x}=0.
\end{equation}

Starting from an admissible initial condition, in order to preserve the no-penetration boundary condition at all subsequent times, the local acceleration must satisfy $\bm{u}_t\cdot \bm{n}=0$ on the boundary $\partial\Omega$ of the \textit{stationary} domain $\Omega$. Consequently, the corresponding component $\delta\bm{u}_t\cdot \bm{n}=0$ of the variations must vanish on the boundary, which yields
\[ \delta\mathcal{L} = \int_\Omega \left[\rho \left(\bm{u}_t+\bm{u}\cdot\bm\nabla \bm{u}\right) + \bm\nabla \lambda \right]\cdot \delta \bm{u}_t d\bm{x}=0 \]
Since this must hold for all admissible variations $\delta \bm{u}_t$, we obtain Euler's equation as a necessary condition for optimality:
\[\rho\left(\bm{u}_t+\bm{u}\cdot\bm\nabla \bm{u}\right)=-\bm\nabla \lambda \;\; \forall \; \bm{x}\in\Omega, \; t\in\mathbb{R} \; \blacksquare \]

It should be noted that the above theorem does not necessarily require $\bm{u}\cdot \bm{n}$ to vanish on the boundary. Rather, it requires that this quantity be prescribed. Equivalently, the normal component $\bm{u}_t\cdot \bm{n}$ of the local acceleration must be known a priori.This is the essential requirement. Accordingly, although the above theorem has been presented here for a fixed domain $\Omega$, it extends naturally to problems with time-varying boundaries, as demonstrated in the recent work \cite{VPNS_PRF}.

As shown in the proof, the Lagrange multiplier field $\lambda$ enforcing the incompressibility constraint is naturally identified with the pressure. This result is fully consistent with the classical projection-based interpretation of incompressible flow and is well documented in the literature( e.g., \cite{Chorin_Marsden_Book,Temam_Projection,Pressure_BCs,Chorin_Projection,Projection_Book,Hirsch_Book2,Geometric_Control_Fluid_Dynamics,Projection_Review,Morrison2020lagrangian,DeVoria_Hamiltonian_JFM} ). Moreover, as the preceding development makes clear, both the theorem and its proof follow directly from standard practice in the calculus of variations. The authors of Ref. \cite{Peters_Gauss}, however, reached a different conclusion, as discussed in the next subsection.

\subsubsection{The Claim in Ref. \cite{Peters_Gauss}}
The authors of  \cite{Peters_Gauss} raised objections to the proof of the equivalence theorem, presented above. In particular, they insisted that the only way to account for the no-penetration boundary condition $\bm{u}_t \cdot\bm{n}=0$ on $\partial\Omega$ is by enforcing it as a side constraint using an additional Lagrange multiplier. Specifically, they stated: ``\textit{Similarly, it is not allowed to set $\delta \bm{u}_t\cdot \bm{n}=0$ in Eq. (28) [equivalently Eq. (\ref{eq:First_Variation}) here] because that would be tantamount to optimizing in the absence of nonpenetration. In the variational method, those constraints must be included with their Lagrange multipliers in $\delta\mathcal{L}$.}"

As demonstrated in the preceding subsection, this claim clearly contradicts standard practice in the calculus of variations, where boundary conditions are routinely enforced by restricting admissible variations rather than by introducing additional multipliers. We see no need for further clarification beyond the analysis already provided.

In their formulation, the authors of  \cite{Peters_Gauss} introduced the cost functional
\[ S =  \frac{1}{2}\int_\Omega \rho \left(\bm{u}_t+\bm{u}\cdot\bm\nabla \bm{u}+\bm\nabla P_F\right)^2 d\bm{x},\]
which contains an additional term $\bm\nabla P_F$, interpreted by the authors as an impressed force. To minimize this functional subject to the constraints $\bm\nabla \cdot \bm{u}_t=0$ in $\Omega$, and $\bm{u}_t \cdot\bm{n}=0$ on $\partial\Omega$, they constructed the augmented functional
\[ \mathcal{L} = S - \int_\Omega P_R(\bm{x}) \left(\bm\nabla \cdot\bm{u}_t(\bm{x})\right) d\bm{x} + \int_{\partial\Omega} \lambda_2(s) \left(\bm{u}_t \cdot\bm{n}\right) ds, \]
where $P_R$ is what they referred to as the constraint component, which enforces the incompressibility constraint; and $\lambda_2$ is the additional Lagrange multiplier that enforces the no-penetration boundary condition.

An instructive outcome of the authors' analysis is equation (12) in Ref. \cite{Peters_Gauss}:
\[ P_R (s) = \lambda_2(s) \]
on the body surface. This relation implies that the Lagrange multiplier $\lambda_2$, introduced by the authors to enforce the no-penetration boundary condition on the surface, is precisely the restriction to the surface of the Lagrange multiplier $P_R$ enforcing incompressibility in the domain. It therefore follows that the additional multiplier $\lambda_2$ is redundant: the field $P_R$ alone is sufficient to enforce both incompressibility in $\Omega$ and the no-penetration condition on $\partial\Omega$.

The above outcome is, in fact, natural from the perspective of Helmholtz orthogonality. In classical mechanics, a constraint force is typically orthogonal to the space of admissible motions. For example, in the simple pendulum, the constraint force acts in the radial direction, which is orthogonal to the admissible tangential motion. In incompressible flows, the constraint force ensuring incompressibility in the domain must be orthogonal to the space of divergence-free fields. However, as emphasized in Remark 3, this orthogonality property does not hold unless the no-penetration boundary condition (or an equivalent one) is satisfied. In particular, the pressure force $\bm\nabla p$ is not orthogonal to arbitrary divergence-free fields; it is orthogonal only to divergence-free fields that also satisfy the no-penetration condition on the boundary, as shown in Eq. (\ref{eq:Helmholtz_Orthogonality}).

From this perspective, the no-penetration boundary condition on $\partial\Omega$ is fully compatible with the incompressibility constraint in $\Omega$. In particular, the pressure force $\bm\nabla p$, which arises as the projection of the free motion $\rho \bm{u}_t^{\rm{free}}$ onto the space of gradient fields, indeed projects $\rho \bm{u}_t^{\rm{free}}$ onto the space of divergence-free fields that satisfy the no-penetration boundary condition. Hence, it enforces both constraints simultaneously, precisely because they are mutually compatible. This geometric-mechanics interpretation provides a clear explanation for why the additional multiplier $\lambda_2$ is unnecessary.

In summary, from the calculus of variations perspective, the no-penetration boundary condition is required to eliminate the boundary term $\delta \bm{u}_t\cdot \bm{n}=0$ in the variational form (\ref{eq:First_Variation}). From the complementary perspective of geometric-mechanics, the same boundary condition is required to ensure orthogonality between the constraint force $\bm\nabla p$ and kinematically admissible accelerations $\bm{u}_t$. The situation is fundamentally different for other boundary conditions, such as the no-slip condition, which are not compatible with the incompressibility constraint in the same sense.

\subsection{On Misconceptions Concerning Constraint and Workless Forces in Classical Mechanics}
\subsubsection{Virtual Work Versus Actual Work}
Mechanics is the science concerned with relating the \textit{cause} of motion to the \textit{quantity} of motion. In Newtonian mechanics, the cause of motion is identified with a force $\bm{F}$, while the quantity of motion is measured by the linear momentum $m\bm{v}$. In this setting, the Newtonian-mechanics relation between cause and effect is simply $\bm{F} =  \frac{d}{dt}(m\bm{v})$.

In analytical mechanics, the corresponding quantities are the kinetic energy (historically refererred to as Leibniz's \textit{vis viva} \cite{Dugas}), and the \textit{work function} \cite{Lanczos_Variational_Mechanics_Book}. On pp. 27-28, Lanczos \cite{Lanczos_Variational_Mechanics_Book} defines the work function for a system of $N$ particles, where the $i^{\rm{th}}$ particle has position vector $\bm{r}_i=(x_i, y_i, z_i)$ and is subject to an impressed force $\bm{G}_i=(X_i,Y_i,Z_i)$. He writes: ``\textit{Let us change the coordinates of each one of the particles by arbitrary infinitesimal amounts $dx_i$, $dy_i$, $dz_i$. The total work of all impressed forces is:"}
\begin{equation}\label{eq:Virtual_Work}
\overline{dW} = \sum_{i=1}^N \left(X_i dx_i + Y_i dy_i + Z_i dz_i\right).
\end{equation}

Moreover, when the Cartesian coordinates $(x_i, y_i, z_i)$ are expressed in terms of $n$ generalized coordinates $q_1$, ..., $q_n$, Lanczos shows that ``\textit{the infinitesimal work $\overline{dW}$ comes out as a linear differential form of the variables $q_i$:"}
\[ \overline{dW} = F_1 dq_1 + F_2 dq_2 + ... + F_n dq_n,\]
where $F_j$ is the generalized force associated with $q_j$, defined by
\[ F_j = \sum_{i=1}^N \bm{G}_i \cdot \frac{\partial \bm{r}_i}{\partial q_j}. \]

The expressions above for $\overline{dW}$ correspond to the \textit{virtual work} done by the impressed forces $\bm{G}_i$, rather than to the actual work. In analytical mechanics, it is virtual work (not actual work) that plays the central role. Equilibrium conditions, d'Alembert's principle, and Lagrange's equations of motion are all formulated in terms of virtual work. Consequently, any force that does no virtual work has no influence on the reduced dynamics on the configuration manifold. Such forces may be omitted altogether without affecting the resulting equations of motion; the same trajectory is obtained. For example, the tangential dynamics of the pendulum shown in Fig. \ref{Fig:Pendulum_Schematic_Magnet} is governed by the exact same equation of motion
\[ \ddot\theta = -\frac{g}{L}\sin\theta \]
regardless of the presence of the electromagnetic force $F$, precisely because this force does no virtual work; it is orthogonal to the space of admissible motions.

The situation is different for \textit{actual work}, which does not play a comparable foundational role in analytical mechanics. Virtual work is the work done through virtual displacements, which are arbitrary admissible displacements, so they span the space of admissible motion. In contrast, actual work is defined through the realized motion and therefore corresponds to a single trajectory within that space. Hence, a force can be orthogonal to that single trajectory (the actual trajectory) without necessarily being orthogonal to other kinematically-admissible trajectories. Therefore, a force can have zero actual work, but with nonzero virtual work. And it is this latter quantity---virtual work---that enters the variational formulation and determines the equations of motion and the resulting dynamics.

It is precisely this central role of virtual work that gave analytical mechanics its decisive advantage over Newtonian mechanics in handling constrained systems, through the postulate that constraint forces perform no virtual work. Lanczos denoted this assumption as Postulate A and described it as ``\textit{the \textup{only} postulate of analytical mechanics.}" It is this postulate that permits constraint forces to be systematically ignored in the derivation of the equations of motion  in analytical mechanics: because they perform no virtual work, they do not influence the reduced dynamics on the configuration manifold.

Because of this central role of virtual work and the comparatively minor role of actual work, many authors use the term \textit{work}, within the framework of analytical mechanics, to mean \textit{virtual work}---the relevant quantity. As noted above, Lanczos himself defines the ``\textit{total work}" as virtual work. This practice is standard in analytical mechanics, particularly in discussions of ``\textit{workless constraints}", where the term refers precisely to the fundamental postulate: constraint forces perform no virtual work.

This distinction has been explicitly noted by several authors. For example, on p. 238 of \cite{Greenwood_Dynamics} Greenwood  writes (emphasis in the original) ``\textit{A workless constraint is any scleronomic constraint such that the virtual work of the constraint forces acting on the system is zero for any reversible virtual displacement which is consistent with the constraint.}" He further emphasizes that ``\textit{It should be realized that workless constraint forces may actually do work on certain particles within the system.}" Greenwood then consistently adopts the terminology workless constraints throughout the remainder of the monograph.

A similar clarification is emphasized by Papastavridis \cite{Papastavridis} who writes ``\textit{In view of these facts, the frequently occurring expression `workless, or nonworking,
constraints' must be replaced by the more precise one, \textup{virtually workless constraints}}." He further adds ``\textit{In sum: in general, the constraint reactions \textup{are} working; even when virtually nonworking. \textup{Actually, that is why the whole concept of virtualness was invented in
analytical mechanics.}}"

\subsubsection{Constraint Forces Are Workless}
As explained in the preceding subsection, within the framework of analytical mechanics, the familiar statement that \textit{constraint forces are workless} is understood to mean that constraint forces perform no virtual work. Equivalently, they are orthogonal to the space of admissible motions. In the simple pendulum, the force in the rod is orthogonal to the tangential direction of motion. In incompressible flows, the pressure force is orthogonal to the space of divergence-free fields that satisfy the no-penetration boundary condition, as discussed in detail at several points throughout this article (see Sec. \ref{Sec:Variational_Theory}, \ref{Sec:Responnse_Main}).

Likewise, an impressed force whose \textit{virtual work} vanishes does not contribute to the reduced dynamics on the configuration manifold and therefore has no effect on the resulting motion. Whether such a force is present or absent, the same trajectory is obtained. As discussed above for the pendulum shown in Fig. \ref{Fig:Pendulum_Schematic_Magnet}, the electromagnetic impressed force $F$ acts in the same direction as the constraint force $R$ and therefore does no virtual work. Consequently, the tangential dynamics of the pendulum is governed by the exact same equation of motion
\[ \ddot\theta = -\frac{g}{L}\sin\theta \]
irrespective of the presence of this force. The force $F$ merely alters the magnitude of the constraint force $R$ required to ensure the constraint along the resulting motion.

Similarly, for an incompressible flow over a body subject to the no-penetration boundary condition, any square-integrable impressed force in the form of a gradient field (i.e., $\bm\nabla \varphi$) does not alter the resulting motion. Indeed, the exact same velocity field $\bm{u}$, in both space and time, is obtained irrespective of the presence of such a force. The sole effect of $\bm\nabla \varphi$ is to modify the pressure field required to ensure the incompressibility and the no-penetration constraints. Consequently, any quantity that depends only on the velocity field (such as circulation) remains unchanged under the action of such an impressed force.

From this perspective, even if one were to accept the claim in Ref. \cite{Peters_Gauss} regarding the presence of impressed pressure forces in the airfoil problem, in the form $\bm\nabla P_F$, such forces would have no effect on the resulting velocity dynamics. This conclusion follows directly from Helmholtz orthogonality for \textit{smooth} incompressible flows, both ideal and viscous. \textit{Accordingly, such a force may be omitted altogether without altering the unsteady evolution of the velocity field.}

On the other hand, a constraint force may perform \textit{actual} work. Consider a pendulum with an extensible rod of length $\ell(t)$, driven by a motor. In this case, the constraint force $R$ in the pendulum rod performs nonzero actual work $\int R(t) \dot{\ell}(t) dt\neq0$. Nevertheless, its \textit{virtual} work remains zero, because the virtual displacements, which must be admissible, are purely tangential
\[ \delta \bm{r} = \ell\delta\theta\bm{e}_\theta,\]
while the radial displacement is fixed by the motor.

Similarly, for an incompressible flow over a deforming body where $\bm{u}_t\cdot\bm{n}\neq0$, the pressure force may perform nonzero \textit{actual} work, while its \textit{virtual} work remains zero. Although the normal component of the acceleration is nonzero ($\bm{u}_t\cdot\bm{n}\neq0$), it is prescribed by the boundary motion. Consequently, admissible variations of the acceleration must satisfy $\delta\bm{u}_t\cdot\bm{n}=0$. Thus, the pressure force remains orthogonal to all admissible virtual variations, even in the presence of boundary deformation.

In conclusion, \textit{virtual} work is a central quantity in analytical mechanics. The familiar statement that \textit{constraint forces are wrokless} is understood to mean that their virtual work vanishes---a principle that forms the bedrock of analytical mechanics. Likewise, if the virtual work of an impressed force vanishes, it has no effect on the motion and may be omitted altogether without altering the motion dynamics. From this perspective, the actual work plays no fundamental role. Constraint forces, which do not affect the reduced dynamics on the configuration manifold, may perform nonzero actual work. Conversely, an impressed forces may perform zero actual work while having non-vanishing virtual work, and therefore, impact the motion dynamics.

\subsubsection{Claims in Ref. \cite{Peters_Gauss}}
Upon close examination of Ref. \cite{Peters_Gauss}, it becomes apparent that several of its statements conflate the concepts of \textit{virtual} work and \textit{actual} work. In particular, the authors interpret the statement made in Ref. \cite{Variational_Lift_JFM}---that ``\textit{pressure forces are workless through divergence-free velocity fields}"---as asserting that the \textit{actual} work of the pressure force vanishes. This interpretation is inconsistent with the common usage of worklessness in analytical mechanics, where the relevant quantity is virtual work.

In this subsection, we examine the specific claims made in Ref.~\cite{Peters_Gauss} and show that they are directly refutable when assessed in light of the basic definitions of constraint forces and virtual work reviewed above.

The authors state that ``\textit{It is of note that constraint forces do no net work on the system. This is due to the fact that, from Newton's third law of motion, the internal constraint forces must exist in equal and opposite pairs.}" This statement is incorrect. As discussed above, constraint forces may perform nonzero \textit{actual} work on a system, such as in the case of a pendulum with a controlled extensible rod. It does not appear that the authors are referring to \textit{virtual} work in this context, since throughout Ref.~\cite{Peters_Gauss} statements concerning work in Ref.~\cite{Variational_Lift_JFM} are consistently interpreted as referring to actual work. The quoted passage therefore reflects a conflation of actual and virtual work, rather than a statement about worklessness in the variational sense.

Moreover, attributing the vanishing of work by constraint forces to Newton's third law is methodologically incorrect. As clarified by Lanczos \cite{Lanczos_Variational_Mechanics_Book} and emphasized by Papastavridis \cite{Papastavridis}: ``\textit{Certainly the third law of motion, 'action equals reaction,' is not wide enough to replace Postulate A.}". The authors' statement therefore rests on an inappropriate application of Newton's third law and a misunderstanding of the role of virtual work in analytical mechanics.

The authors of Ref.~\cite{Peters_Gauss} reproduce the following passage from Ref. \cite{Variational_Lift_JFM} and raise several objections to its validity:
``\textit{[T]he only acting force on a fluid parcel is the pressure force $\bm\nabla p$. In order to apply Gauss' principle, we must determine whether this force is an impressed force or a constraint force. Interestingly, for incompressible flows, it is the latter. The sole role of the pressure force in incompressible flows is to maintain the continuity constraint: $(\bm\nabla\cdot \bm{u}=0$). It is straightforward to show that if $\bm{u}$ satisfies [continuity and non-penetration], then the integral of ($\bm\nabla p\cdot\bm{u}$) over the volume is zero for any scalar $p$, which indicates that pressure forces are workless through divergence-free velocity fields .... That is, if a velocity field is subject to continuity and the no-penetration boundary condition, the pressure force does not contribute to the dynamics of such a field ... it is clear that the pressure force is a constraint force and the dynamics of ideal fluid parcels are subject to no impressed forces .... Hence, Gauss' principle of least constraint reduces to Hertz' principle of least curvature in this case.}"

In light of the technical background presented in this paper, the quoted passage stands as written. For incompressible flows subject to the no-penetration boundary condition, the pressure forces is a constraint force. Both its actual work and its virtual work vanish when $\bm{u}\cdot\bm{n}=0$ on $\partial\Omega$; when the normal velocity $\bm{u}\cdot\bm{n}$ is prescribed in an arbitrarily time-varying manner, only its virtual work vanishes. In either case, the pressure force, as a constraint force, does not contribute to the reduced dynamics on the configuration manifold, as discussed in detail throughout this paper (e.g., Sec. \ref{Sec:Variational_Theory}, \ref{Sec:Responnse_Main}, and the preceding subsection).

The authors of Ref.~\cite{Peters_Gauss} interpret the statement "\textit{pressure forces are workless through divergence-free velocity fields}" in two ways: first, as asserting that the \textit{actual} work of the pressure force vanishes; and second as serving as evidence that the pressure force is therefore a constraint force in incompressible flows. On this basis, they object that ``\textit{one should not infer from this that every force system that does no net work on the system is consequently a constraint force."} Both interpretations rest on a misreading of the original statement.

%
The authors further assert that ``\textit{it is not the case that if a velocity field satisfies the continuity and nonpenetration boundary conditions, then the pressure force is workless.}" This assertion is incompatible with Helmholtz-Leray orthogonality: for velocity fields that are divergence-free and satisfy the no-penetration boundary condition, the pressure force is orthogonal to the admissible velocity space and therefore performs no virtual work. This result is reviewed in detail in Sec.~\ref{Sec:Variational_Theory}.

The authors also state ``\textit{It is not the case that if a pressure gradient $\bm\nabla p$ that does no net work on the system [i.e., the integral of ($\bm\nabla p\cdot\bm{u}$)=0], then it is necessarily a constraint force."} We agree with this statement. However, two points must be clarified. First, nowhere in Ref. \cite{Variational_Lift_JFM} is the worklessness of the pressure force used to \textit{infer} that it is a constraint force. Rather, this property is invoked to demonstrate consistency with a fundamental quality of constraint forces; namely, Postulate A. Second, the point is not merely regarding the vanishing of work; the statement that \textit{pressure forces are workless through divergence-free velocity fields} asserts that the \textit{virtual} work of the pressure force vanishes; that is, the pressure force is orthogonal to the entire space of admissible velocity fields. This distinction is essential and is not addressed in Ref. \cite{Peters_Gauss}.

The authors' fourth objection to the excerpt quoted above is stated as follows: ``\textit{Fourth, the statement that if a pressure force does no work, 'the pressure force does not contribute to the dynamics of such a field' is incorrect."} Here again, objection rests on a misinterpretation of the statement \textit{pressure forces are workless through divergence-free velocity fields}. As discussed in the preceding subsections, the common usage of \textit{workless forces} in analytical mechanics refers to forces whose \textit{virtual} work vanishes. Forces of this type (whether they are constraint or impressed) do not contribute to the motion dynamics on the configuration manifold.

The authors then state ``\textit{Similarly for the airfoil problem, the forces from both the impressed and constraint pressure gradients influence the system dynamics.}" This claim is inconsistent with basic principles of analytical mechanics: Constraint forces perform no virtual work and therefore do not contribute to the reduced dynamics on the configuration manifold. It is precisely for this reason that constraint forces may be ignored when deriving the equations of motion in analytical mechanics, since their sole role is to enforce the constraint, they do not influence the motion along the constraint surface.

In conclusion, the majority of the claims in Ref. \cite{Peters_Gauss} arise from one or more of the following issues: (1) conflating of virtual work with actual work; (2) failure to recognize Helmholtz-Leray orthogonality; (3) misinterpretation of the statement \textit{constraint forces are workless} as asserting that their \textit{actual} work vanishes, rather than their \textit{virtual} work; and (4) misreading of statements made in Ref. \cite{Variational_Lift_JFM}. In light of these points, the conclusions reached in Ref.~\cite{Peters_Gauss} follow naturally from the underlying misconceptions identified above.

\section{On Predictions of the Variational Theory}
The authors of Ref. \cite{Peters_Gauss} acknowledge that the predictions of the variational theory of lift \cite{Variational_Lift_JFM} are ``\textit{entirely reasonable}" for conventional airfoils with sharp trailing edges ($D\simeq0$) and smooth rear portions ($D$ away from 0) in forward flight. They further note that the theory ``\textit{is particularly appealing when one considers that the Kutta condition is not generally applicable for shapes without sharp trailing edges}". However, the authors regard some of the theory's predictions (such as those for flow over a flat plate or a reversed airfoil) as perplexing and unacceptable. The purpose of this section is to address these concerns.

\subsection{Predictions Outside Its Scope: Reversed Airfoils}
As Thomas Kuhn noted, ``\textit{To be accepted ... a theory must seem better than its competitors, but it need not, and in fact never does, explain all the facts with which it can be confronted.}" (\cite{Kuhn_Structure}, p. 18). He further emphasized that ``\textit{to be admirably successful is never, for a scientific theory, to be completely successful.}" (p. 68).

The crux of the matter is that no theory is complete, nor can any theory account for all conceivable cases. It is therefore essential to delineate clearly the region of validity and domain of application of any theoretical framework. In Ref.~\cite{Variational_Lift_JFM}, the emphasis was placed on relaxing the sharp trailing-edge requirement, while a detailed discussion of the theory's domain of applicability was not the primary focus. Nevertheless, nowhere in Ref.~\cite{Variational_Lift_JFM} is it claimed that the variational theory of lift applies to reversed airfoils.

This limitation is not unique to the variational theory; it applies equally to the classical theory of lift. In fact, it applies to any theory founded on the assumptions of (i) steady flow and (ii) irrotational motion. The reason lies in a reversibility argument, as explained in Birkhoff's \textit{Hydrodynamics: A Study In Logic, Fact, and Similitude} \cite{Birkhoff_Ideal_Lift_Logic}. He asserted that ``\textit{Such a theory will predict by [Bernoulli's equation] that a steady flow and its reverse will give the same pressure thrust on an obstacle, whereas it is a matter of common experience that a flow and its reverse ordinarily give pressure thrusts in approximately opposite directions.}" (\cite{Birkhoff_Ideal_Lift_Logic}, p. 14).

\begin{wrapfigure}{l}{0.50\textwidth}
\vspace{-0.15in}
 \begin{center}
 \includegraphics[width=8cm]{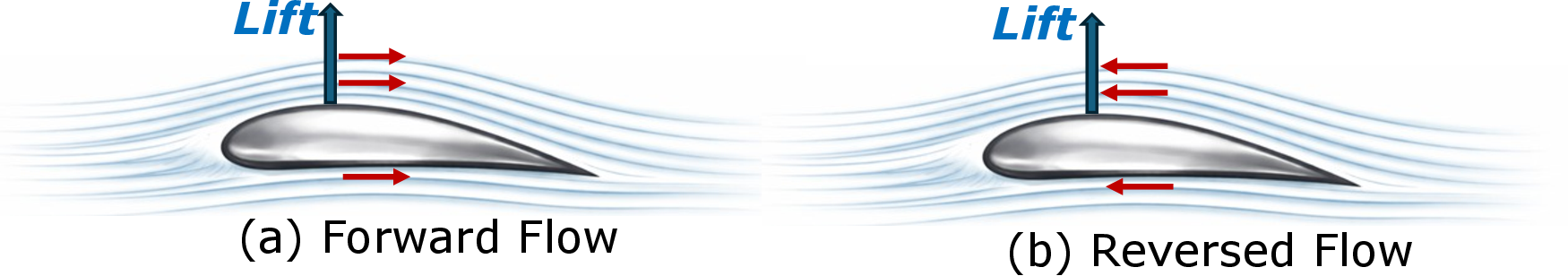}  \vspace{-0.2in}
 \caption{Forward and reversed flows over an airfoil assuming steady, irrotational motion.}
 \label{Fig:Airfoil_Forward_Reversed} \vspace{-0.1in}
 \end{center} \vspace{-0.1in}
\end{wrapfigure} \noindent The reversibility argument noted by Birkhoff asserts that if a steady flow past a body from left to right yields a solution $\bm{u}$, then the flow field $\bm{u}_r$ obtained by reversing the velocity $\bm{u}$ everywhere (i.e., $\bm{u}_r=-\bm{u}$) is also a solution. This situation is illustrated schematically in Fig. \ref{Fig:Airfoil_Forward_Reversed}. In particular, the reversed field $\bm{u}_r$ satisfies the same four conditions of the classical airfoil problem: (i) the momentum equation (\ref{eq:Euler}), (ii) the continuity constraint (\ref{eq:Continuity}), (iii) the no-penetration boundary condition on the body surface, and (iv) the far-field boundary condition. Since the pressure depends quadratically on the speed---from either Bernoulli's equation or Euler's steady dynamics (\ref{eq:Euler})---the resulting pressure distribution, and hence the lift, is identical for $\bm{u}$ and $\bm{u}_r$.

Accordingly, the attempt in Ref. \cite{Peters_Gauss} to apply the variational theory of lift, in the restricted formulation presented in \cite{Variational_Lift_JFM} (which is valid only for attached flows where the admissible family $\mathcal{U}$ is appropriate), to reversed airfoils is an unwarranted stretch beyond its domain of applicability.

This limitation can be overcome in a viscous formulation, which breaks reversibility through dissipation and vitiates the quadratic dependence of pressure on velocity. However, it should be emphasized that this well-known resolution is not the only one. A \textit{proper} unsteady formulation of Euler's dynamics may also break reversibility. This fact, which may be less familiar within the aeronautical community, deserves more elaboration.

It is well known from the classical theory of unsteady aerodynamics \cite{Wagner,Theodorsen,VonKarman_Sears} that, a proper unsteady formulation of Euler's dynamics for the airfoil problem must allow vortex sheets to emanate from the body. These are sheets of discontinuity that may be viewed as the inviscid-limits of vortical shear layers \cite{Sears_Two_Models}. Consequently, the resulting flow field is no longer smooth or differentiable; it must instead be interpreted as a \textit{weak} solution of the Euler equation.

As noted by Birkhoff, ``\textit{It is well-known that, if one is willing to admit discontinuous functions ... the differential equations} [meaning Euler's]\textit{  ... permit many steady flows around an obstacle. These flows are of various topological types, and most of them avoid the d'Alembert paradox.}" (\cite{Birkhoff_Ideal_Lift_Logic}, p. 15).

In fact, none of the familiar conservation principles of Euler's dynamics (e.g., energy, circulation, or helicity) are guaranteed to remain valid for rough (non-smooth) weak solutions of the Euler equation \cite{Eyink_PhysicaD}. Onsager's conjecture \cite{Onsager_Hydrodynamics} concerning the dissipative nature of weak solutions was recently proved \cite{Onsager_Conjecture_Proof,Onsager_Conjecture_Proof2} (see also \cite{Onsager_Conjecture1,Onsager_Conjecture2,Eyink_Onsager_Conjecture}). That is, if a weak solution of the Euler equation lacks sufficient regularity (i.e., is not smooth enough), it may dissipate energy even in the absence of viscosity. The critical regularity (i.e., smoothness) threshold below which energy conservation of Euler's dynamics is no longer guaranteed is Holder continuity with exponent of 1/3. Therefore, the discontinuous solutions that arise in the classical theory of unsteady aerodynamics \cite{Wagner,Theodorsen,VonKarman_Sears} are well below this regularity threshold and are therefore prone to energy dissipation without viscosity.

It then remains to determine both the shedding location at the rear portion of the reversed airfoil (i.e., near the rounded leading edge) and the strength of shed vorticity. One may, in principle, utilize the principle of minimum pressure gradient (PMPG), or an extended unsteady formulation of the variational theory of lift, to determine the \textit{optimal} shedding location and vorticity strength at every instant in time by minimizing the instantaneous total curvature of the flow  (i.e., the Appellian). Interestingly, such an \textit{instantaneous unsteady} application is more faithful to the PMPG and Gauss's principle than the steady formulation adopted in Ref. \cite{Variational_Lift_JFM}. At present, however, this approach should be regarded as a proposal that requires careful testing and validation.

In summary, both the classical theory and the variational theory of lift, as well as any framework based on steady irrotational flow, are inapplicable to reversed airfoils. However, in contrast to the rigidity of the classical theory, the variational theory of lift, rooted in first principles of analytical mechanics, points to a viable path forward: a promising extension that may overcome this limitation.

\subsection{Unexpected Predictions on Symmetric Shapes}
After acknowledging the value of the variational theory in handling conventional airfoils with rounded trailing edges in forward flight, the authors of Ref. \cite{Peters_Gauss} contend that ``\textit{the situation is dramatically different for the case of the flat-plate airfoil ... where the variational theory of lift predicts that the circulation and lift will be zero.}" We agree with the authors that this result is surprising to any aeronautical engineer, whose first encounter with a lift formula is often Kutta's $2\pi\alpha$ for a flat plate. The prediction was, in fact, equally unexpected to the authors of the variational theory of lift themselves.

\begin{wrapfigure}{l}{0.60\textwidth}
\vspace{-0.25in}
 \begin{center}
 \includegraphics[width=8.5cm]{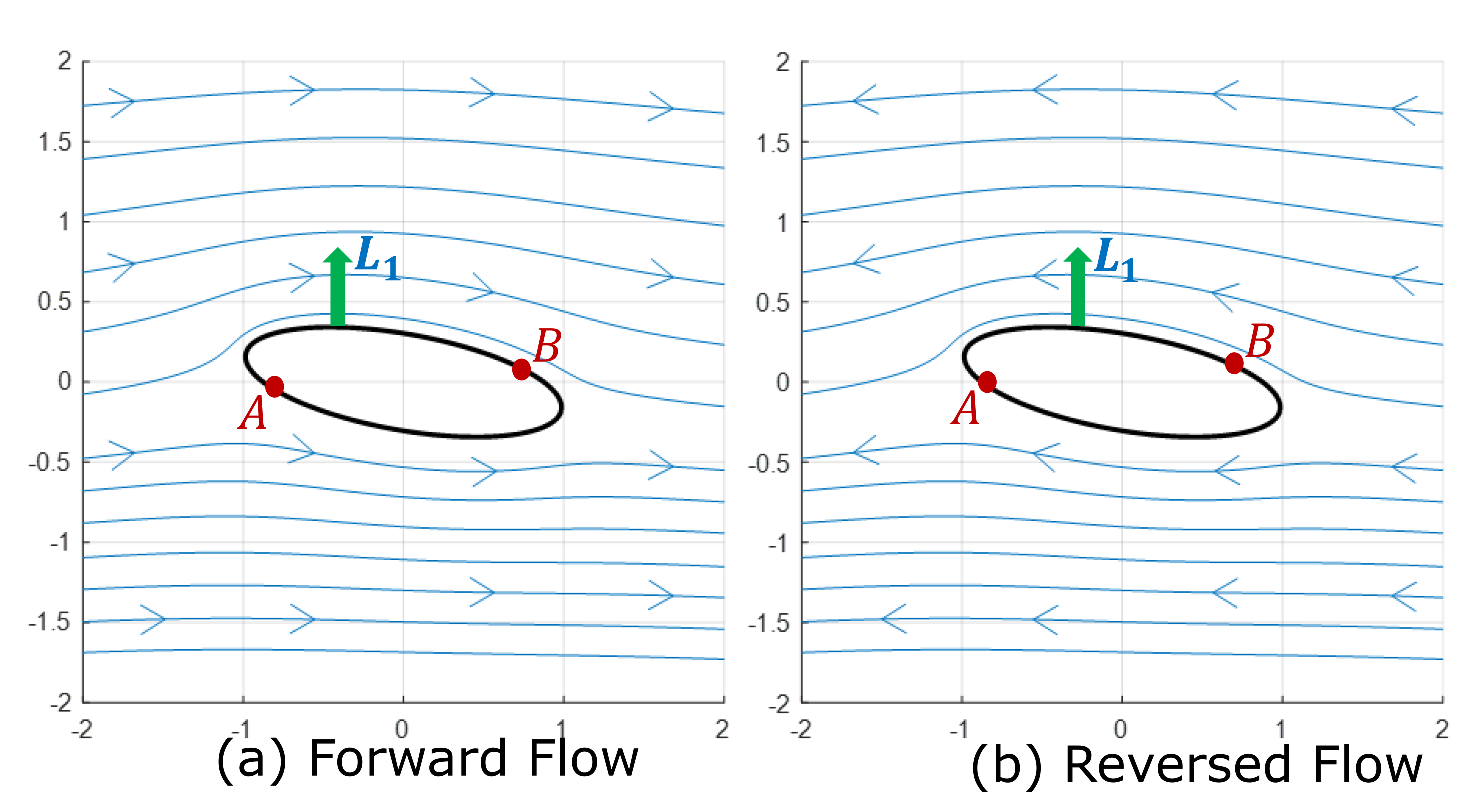}  \vspace{-0.2in}
 \caption{Forward and reversed flows over an ellipse assuming steady, irrotational motion.}
 \label{Fig:Ellipse_Forward_Reversed} \vspace{-0.1in}
 \end{center} \vspace{-0.1in}
\end{wrapfigure} \noindent However, it turns out that this outcome is natural and even required for the internal consistency of the theory, and more generally of any theory based on the assumptions of (i) steady flow and (ii) irrotational motion. For this reason, a dedicated study was undertaken to examine this outcome in detail; its main findings were reported in \cite{Kutta_Flat_Plate} and are briefly summarized below.

Let us first recall that this zero-lift prediction of the variational theory is not confined to the flat-plate case, but applies more generally to any fore-aft symmetric shape, such as flat plates, ellipses, and circular cylinders. The main reason lies in the reversibility argument of Birkhoff \cite{Birkhoff_Ideal_Lift_Logic}, presented in the preceding subsection.

\begin{wrapfigure}{l}{0.40\textwidth}
\vspace{-0.15in}
 \begin{center}
 \includegraphics[width=6cm]{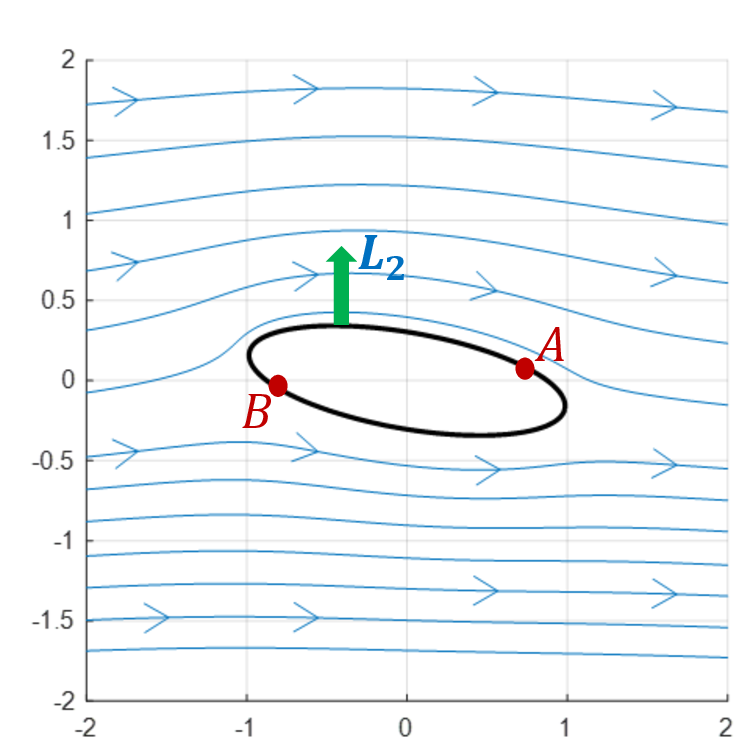}  \vspace{-0.2in}
 \caption{The reversed flow, facing the marker $B$, in Fig. \ref{Fig:Ellipse_Forward_Reversed} (b) with a mirrored orientation.}
 \label{Fig:Ellipse_Foverse} \vspace{-0.1in}
 \end{center} 
\end{wrapfigure} \noindent Consider the flow over any fore-aft symmetric shape (e.g., an ellipse) according to any theory involving steady irrotational motion. Birkhoff's reversibility argument is then illustrated in Fig. \ref{Fig:Ellipse_Forward_Reversed}. Both the forward and reversed flows produce the same lift force $L_1$. The points $A$, $B$ on the ellipse are markers needed for subsequent clarification.

The reversed configuration, shown in Fig. \ref{Fig:Ellipse_Forward_Reversed} (b), may equivalently be viewed from the right, facing the marker $B$. Under this change of viewpoint, the flow appears as that shown in Fig. \ref{Fig:Ellipse_Foverse}, with a mirrored orientation. Consequently, the lift force $L_2$ in Fig. \ref{Fig:Ellipse_Foverse} must be the opposite of the lift $L_1$ in Fig. \ref{Fig:Ellipse_Forward_Reversed} (b); that is,
\[ L_2=-L_1.\]
On the other hand, owing to fore-aft symmetry, the view in Fig. \ref{Fig:Ellipse_Foverse} facing $B$ is indistinguishable from the view facing $A$ in Fig. \ref{Fig:Ellipse_Forward_Reversed} (a). These two configurations must therefore generate the same lift force, implying
\[ L_2=L_1. \]
Taken together, these two conclusions require that $L_1=L_2=0$. \textit{A fore-aft symmetric body cannot be lifting within any internally consistent theory that assumes steady, irrotational motion.}

The foregoing argument rests solely on two ingredients: (1) reversibility, and (2) fore-aft symmetry. The former follows directly from the assumptions of (i) steady flow and (ii) irrotational motion. Therefore, for any theory of lift (classical or variational) that adopts these two assumptions to be internally consistent, it must therefore predict zero lift over fore-aft symmetric bodies.

This conclusion is fully consistent with the predictions of the variational theory of lift. But what about the classical theory? Is it consistent in the same sense? If so (meaning that it too must predict zero lift for ellipses and flat plates), how should the famous result $2\pi\alpha$ be interpreted? And how does this reconcile with the observed lift on a flat plate in a real, high–Reynolds-number flow? These questions, together with their illuminating historical context, were discussed in Ref. \cite{Kutta_Flat_Plate} and are briefly summarized below.

The classical theory of lift is not consistent in the case of a flat plate. The resulting flow field is unbounded; it has a singularity at the leading edge. This issue was so perplexing to Kutta himself that delayed the publication of his work for nearly eight years.

In his first paper on the subject \cite{Kutta_Joukosky_Theorem11}, Kutta considered a circular-arc camber at zero angle of attack. In this case, the circulation indeterminacy problem is readily resolved by requiring that the velocity remain bounded everywhere in the flow field. For this configuration, all members of the admissible family $\mathcal{U}$ are singular at both the leading and trailing edges except for a single solution. Owing to symmetry, the corresponding unique value of circulation removes both singularities simultaneously. This value must therefore be the physically admissible circulation.

However, when Kutta considered the same shell (or a flat plate) at a nonzero angle of attack, no value of the circulation could eliminate both singularities at the leading and trailing edges. He was therefore forced to remove the singularity at only one edge, and he astutely chose the trailing edge. As Kutta himself noted, ``\textit{In contrast, an infinitely fast flow around an edge can be avoided in problems of an angled flow against a plate or shell only at one edge---we call it the rear edge}" \cite{Kutta_Joukosky_Theorem12}. The singularity at the leading edge, however, remained unresolved---an issue that troubled Kutta to such an extent that he did not publish the flat-plate results in his 1902 paper \cite{Kutta_Joukosky_Theorem11}. Instead, he confined his presentation to the flow over a circular arc at zero angle of attack, which does not suffer from this difficulty.

A few years later, Zhukovsky published his paper \cite{Kutta_Joukosky_Theorem21} where he presented the celebrated result $L=\rho U \Gamma$. In fact, this relation already appeared in Kutta's unpublished work and is now universally known as the Kutta–Zhukovsky lift theorem \cite{Schlichting}. At the urging of his adviser, Professor Sebastian Finsterwalder, Kutta was encouraged to publish the remaining results of his habilitation dissertation. Reflecting on this period, Kutta later wrote, ``\textit{The analysis was not published in detail, only a part of the main results was given briefly in the Aeronautical Communications of 1902, p. 133. Following renewed encouragement from Professor Finsterwalder I have taken up the investigation again in the last few months and as a result I have also found flow patterns for angled flows over plates and shells}" (\cite{Kutta_Joukosky_Theorem12}; see also \cite{Kutta_Zhukovsky_History_Book}).

Kutta laid out what is essentially the entire two-dimensional aerodynamic theory still taught in engineering curricula today. Spanning more than 40 pages, the work presents a comprehensive framework that includes conformal mapping between a circular cylinder and a plate or shell, the condition now bearing Kutta's name, lift contributions due to both angle of attack and camber, the location of the center of pressure and its variation with angle of attack, and the notion of a suction force together with its connection to the removal of the leading-edge singularity for rounded noses.

\begin{wrapfigure}{l}{0.50\textwidth}
 \begin{center}
 \includegraphics[width=8cm]{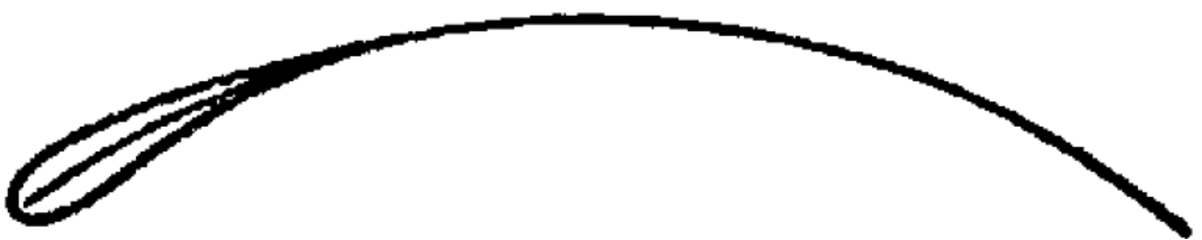}
 \caption{Actual shape considered by Kutta. Taken from Kutta \cite{Kutta_Joukosky_Theorem11}, also see \cite{Kutta_Zhukovsky_History_Book}.}
 \label{Fig:Rounded_LE}
 \end{center}
\end{wrapfigure}
\noindent Of this 40-page paper, Kutta devoted more than 10 pages (Section 5 titled ``\textit{Rounding of the Leading Edge}") to addressing the unresolved leading-edge singularity in his solution. He was fully aware that the flow field associated with this singularity was not physically plausible, describing it as ``\textit{mathematically exact, but admittedly not physically exact.}" Kutta emphasized that rounding the leading edge, even by a small amount, eliminates the singularity and gives rise to a suction force, which both reduces drag and increases lift.

Kutta himself deliberately avoided the singular case of a flat plate or a shell at nonzero angle of attack. Instead, he focused on the \textit{practical} configuration with a slightly rounded leading edge, as shown in Fig. \ref{Fig:Rounded_LE}, reproduced from Kutta's work \cite{Kutta_Joukosky_Theorem11}. In this sense, Kutta's solution for the flow over a thin shell does not correspond to an idealized, infinitely-thin shell but rather to a shell (or plate) with a rounded nose. The celebrated result $2\pi \alpha$ should therefore be interpreted as the lift coefficient of a flat plate with a slightly rounded leading edge, rather than as the lift of an ideal flat plate with a sharp leading edge, as is commonly assumed.

This rounding, however small, destroys the fore-aft symmetry of the flat plate, thereby permitting nonzero lift within a steady, ideal-flow framework (i.e., within the admissible family $\mathcal{U}$). In addition, the remaining singularity at the trailing edge must be eliminated, whether by the Kutta condition or the variational formulation. As a result, the circulation that minimizes the Appellian in this setting is necessarily close to Kutta's value, recovering the familiar $2\pi \alpha$ dependence, as demonstrated in Ref.  \cite{Kutta_Flat_Plate}.

It is noteworthy that the variational theory of lift, by virtue of its sensitivity to body geometry, points directly to this distinction, identifying fore-aft \textit{asymmetry} as a necessary condition for nonzero lift within any steady, ideal-flow framework. This conclusion, though implicit in the structure of the classical theory, was never articulated explicitly from it.

A practical aeronautical engineer might then ask: does this argument imply that the flat plate configuration (much like the reversed-airfoil case) lies outside the scope of both the classical and variational theories? From a \textit{practical} standpoint, the answer is no. No realistic geometry possesses a mathematically sharp leading edge. If such a geometry also features an almost sharp trailing edge, it falls squarely within the domain of applicability of the classical theory. If this condition is not met, the variational theory may offer a reasonable extension, yielding physically plausible predictions of lift.

\subsection{Variational Theory Versus Classical Theory}
The authors of Ref. \cite{Peters_Gauss} assert that ``\textit{Unlike classical airfoil theory, the variational theory of lift offers little practical utility to the airfoil designer as a surrogate for airfoil characteristics in real fluid flow.}" In light of the analysis presented here, this assessment is puzzling. As shown above, the classical theory emerges as a \textit{special case} of the variational theory when the trailing edge is mathematically sharp. Moreover, the reversed-airfoil scenario raised in Ref. \cite{Peters_Gauss} applies equally to both theories; such a configuration lies outside the domain of applicability of any theory that assumes steady, irrotational flow. Finally, the flat-plate issue has been clarified in detail in the preceding subsection, and more fully in Ref.  \cite{Kutta_Flat_Plate}, where it was shown that, for \textit{practical} configurations, both theories yield physically reasonable predictions.

The authors of Ref. \cite{Peters_Gauss} present lift measurements for a NACA 0012 in both forward and reversed configurations, together with Kutta’s classical prediction for the forward case. On this basis, they conclude that ``\textit{the potential flow classical airfoil theory and the real fluid experimental measurements ... are comparable for all of the airfoils discussed."}

It should be emphasized that the classical airfoil theory does not, in fact, provide a meaningful prediction for the reversed configuration. Only a single prediction exists---namely, for the forward flow over a sharp-edged airfoil---and this prediction agrees reasonably well with the corresponding experimental result. The observation that the measured lift in the reversed configuration is not dramatically smaller than that in the forward case may therefore create an impression similar to that implied by the authors' statement, even though no theoretical prediction exists for the reversed case within the classical framework.

Statements of this kind may lead one to claim that a brute-force application of the classical theory to a reversed airfoil yields a reasonable estimate of the lift. First, this application, in fact, contradicts Kutta's classical theory of lift itself, which is explicitly concerned with eliminating singularities at sharp edges in order to obtain a bounded velocity field. In contrast, such an application leaves the actual singularity at the physical trailing edge unresolved, while attempting instead to eliminate a putative singularity at a fictitious trailing edge located on the rear portion of the reversed configuration---—where the true trailing edge, in fact, lies upstream.

Second, such a brute-force application ignores the very nature of the reversed flow configuration and treats the problem exactly as if it were identical to the forward-flow case. Approaches of this kind should be avoided in serious aerodynamic analysis, even when they appear to yield reasonable agreement in isolated instances, as any such agreement is more plausibly attributable to coincidence than to an underlying mathematical fact or physical principle.

Another perplexing statement in Ref. \cite{Peters_Gauss} is the claim that ``\textit{the lift characteristics predicted by the variational theory of lift differ greatly and even discontinuously as a function of airfoil shape."}In fact, one of the principal strengths of the variational theory lies precisely in the \textit{continuity} of the predicted circulation and lift with respect to airfoil geometry. This continuity is guaranteed by the mathematical structure of the minimization problem itself. Each admissible velocity field $\bm{u}$ in the family $\mathcal{U}$ depends smoothly (i.e., in a continuously differentiable manner) on the airfoil shape parameters. In addition, the Appellian depends smoothly on both $\bm{u}$ and $\Gamma$. It follows that the minimizing value of $\Gamma$ must vary continuously with the shape parameters.

\begin{wrapfigure}{l}{0.50\textwidth}
 \begin{center}
 \includegraphics[width=8cm]{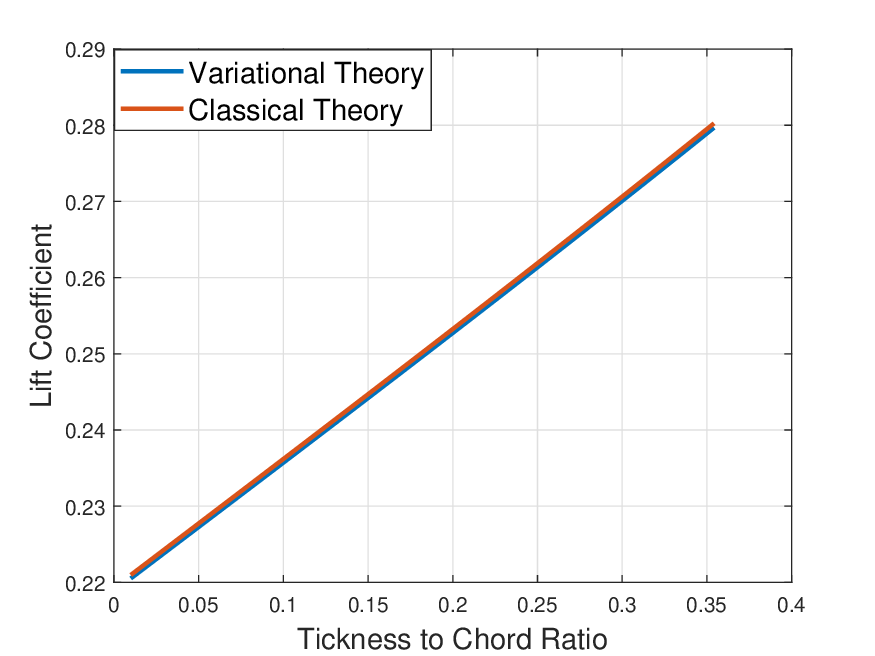}
 \caption{Variation of the lift coefficient with thickness ratio for a Zhukovsky airfoil at $\alpha=2^\circ$, as predicted by the variational and classical theories.}
 \label{Fig:Thickness_Effects}
 \end{center}
\end{wrapfigure}
\noindent The claimed discontinuity appears to stem from a mischaracterization of the predictions of the variational theory. Specifically, the authors state that ``\textit{According to the variational theory, the lift of a conventional airfoil would remain essentially constant as the thickness decreased until it collapses to zero discontinuously at zero thickness}". This conclusion contradicts direct application of the variational theory. In fact, a straightforward evaluation of the minimization formula (\ref{eq:Variational_Theory_Special}) applied to the modified Zhukovsky family (\ref{eq:Modified_Zhukovky}) with $D\simeq0$, $\alpha=2^\circ$ yields a smooth variation of lift with thickness that is indistinguishable from the classical prediction, as shown in Fig. \ref{Fig:Thickness_Effects}.

The authors of Ref.~\cite{Peters_Gauss} further point to another source of purported discontinuity. In figure 3 of Ref. \cite{Peters_Gauss}, they present what they describe as ``\textit{representative
lift characteristics \textbf{expected}} [emphasis added] \textit{from the variational theory of lift for this airfoil family.}" Their conclusions are thus based on stated expectations rather than on an actual implementation of the variational theory. On the basis of these expectations, the authors assert that for an infinitely thin airfoil whose ratio $r_{TE}/r_{LE}$ of trailing-edge to leading-edge radius varies smoothly from below unity to above unity, the circulation predicted by the variational theory would jump discontinuously from a positive value to a negative one as $r_{TE}/r_{LE}$ crosses unity.

First, it should be noted that as $r_{TE}/r_{LE}$ crosses unity, the airfoil transitions from a forward to a reversed configuration. At that point, the predictions of the variational theory are being extended beyond its domain of applicability. Second, the argument above, as well as much of the analysis in Ref. \cite{Peters_Gauss}, focuses on the singular case of an infinitely thin body: $t/c\equiv0$. For any finite thickness, however small, the authors already acknowledge a smooth behavior from the variational theory in their sketches; when $r_{TE}/r_{LE}\equiv 1$ the geometry is fore–aft symmetric, for which the variational theory predicts zero lift. Consequently, the transition from a positive circulation for $r_{TE}/r_{LE}<1$ to a negative circulation for $r_{TE}/r_{LE}> 1$ must pass smoothly through $\Gamma=0$ at $r_{TE}/r_{LE}\equiv 1$---a trend that is, in fact, reflected in the authors’ own sketches. Third, in the singular limit of a zero-thickness shell or flat plate, the notions of leading- and trailing-edge radii $r_{LE}$, $r_{TE}$ are no longer well defined; they lose both their physical meaning and mathematical significance.

Reflecting on the claims of Ref. \cite{Peters_Gauss}, one is naturally led to ask: even if these claims were correct, how would the situation be any different with the classical theory of lift? For example, the argument above discusses effects as $r_{TE}$, $r_{LE}$ change. Yet these very same parameters have little to no influence within the classical framework. The classical theory is primarily applicable to airfoils with sharp trailing edges, for which the removal of the trailing-edge singularity dominates the aerodynamic response, blurring the influence of other geometric details of the body. Indeed, Kutta's circulation
\[ \Gamma_K=4\pi a U_\infty\sin\alpha \]
depends on a single geometric parameter $a$---the radius of the circle in the conformal $\zeta$-domain. All geometrical effects are thus collapsed into one parameter. Clearly, such a restriction is far from capturing many geometric features of the body, including the influence of the leading- and trailing-edge radii $r_{TE}$, $r_{LE}$, emphasized by the authors.

The variational theory, on the other hand, is sensitive to nearly all geometric details of the body, particularly those related to curvature. It is precisely this sensitivity that identifies fore-aft asymmetry as a necessary condition for lift within a steady, irrotational framework. The theory clearly distinguishes between a mathematically sharp leading edge and a slightly rounded one. Likewise, it can handle effects of trailing-edge geometry, which is beyond the capability of the classical theory.

Finally, it may be prudent to recall Kuhn's statement, quoted at the beginning of this section: ``\textit{To be accepted ... a theory must seem better than its competitors, but it need not, and in fact never does, explain all the facts with which it can be confronted.}" (\cite{Kuhn_Structure}, p. 18). In light of this observation, and of the preceding discussion, Table \ref{Tab:Comparison} summarizes and contrasts the respective capabilities of the classical and variational theories of lift.

\begin{table*}
\begin{centering}
\begin{tabular}{l|c|c}
& Classical Theory & Variational Theory \\ \hline

Conventional airfoils with sharp TEs & $\checkmark$ & $\checkmark$ \\

Conventional airfoils with smooth TEs & $\times$ & $\checkmark$ \\

Flat Plate (sharp edges) & unbounded velocity field & zero lift (in the limit). \\

Flat Plate with rounded LE & $2\pi\alpha$ & $2\pi\alpha$. \\

Reversed Airfoils & $\times$ & $\times$ \\

Sensitivity to geometric details & $\times$ & $\checkmark$ \\

Room for extension & $\times$ & $\checkmark$

\end{tabular}
\caption{Comparison between the capabilities of the classical and variational theories of lift. Here, LE and TE denote leading and trailing edges, respectively.}
  \label{Tab:Comparison}
\end{centering}
\end{table*}

The final entry in Table \ref{Tab:Comparison} may require further clarification.

\subsection{Extensions of the Variational Theory}
The principal strength of the variational theory does not necessarily lie in its ability to capture specific geometric details per se, but rather in its flexibility and its capacity for systematic extension, in contrast to the rigidity of the classical theory. A representative example is the rotating-cylinder problem. It has been known since the early experiments of Prandtl \cite{prandtl1931hydro} that sufficiently rapid rotation of a circular cylinder ($\kappa=\frac{\omega a}{U}>>1$) suppresses flow separation and vortex shedding, leading to a simple, fully attached flow. In this regime, the velocity field $\bm{u}$ outside the boundary layer belongs to the admissible family $\mathcal{U}$. Yet, within the classical framework, there is no mechanism by which the resulting circulation and lift can be determined.

\begin{wrapfigure}{l}{0.50\textwidth}
 \begin{center}
 \includegraphics[width=8cm]{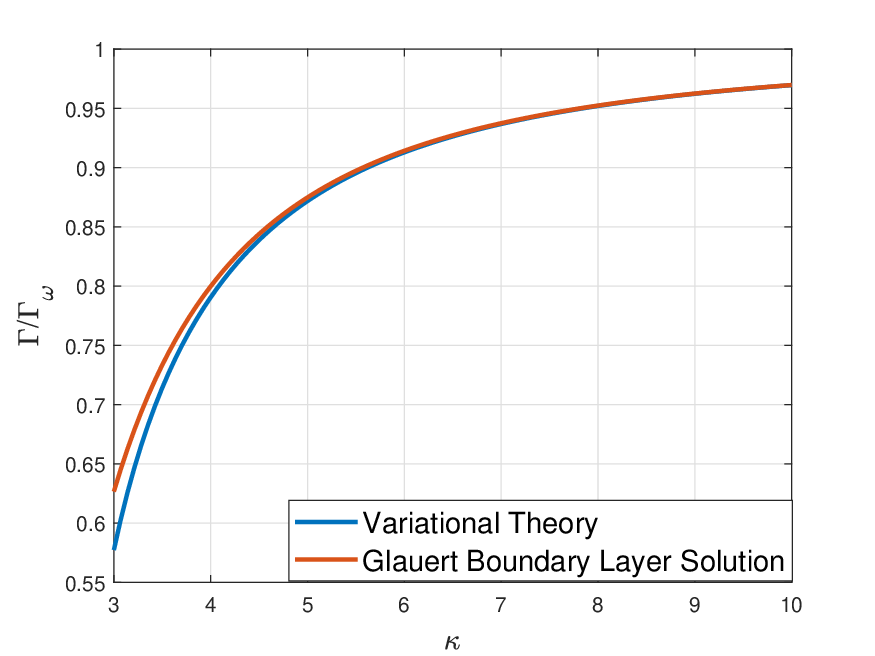}
 \caption{Comparison for the variation of the normalized circulation $\frac{\Gamma}{\Gamma_\omega}$ with the normalized rotational speed $\kappa=\frac{\omega a}{U}$ between Glauert's viscous boundary layer solution \cite{Glauert_Rotating_Cylinder} and the variational solution \cite{Shorbagy_Magnus_AIAA}.}
 \label{Fig:Shorbagy_Magnus}
 \end{center}
\end{wrapfigure}
\noindent Interestingly, this problem was addressed in Ref. \cite{Shorbagy_Magnus_AIAA} using an extended version of the variational theory of lift. In that work, the cost function (\ref{eq:Variational_Theory_Special}) was modified to account explicitly for the rotation of the cylinder. The resulting problem was then reduced to a one-dimensional minimization over the same admissible family $\mathcal{U}$ with respect the circulation $\Gamma$. An analytical expression was obtained for the minimizing circulation
\[\frac{\Gamma^*}{\Gamma_\omega} = \sqrt{1-\frac{6}{\kappa^2}}, \]
where $\Gamma_\omega=2\pi a^2 \omega$. This result is in excellent agreement with Glauert's asymptotic solution for the boundary-layer dynamics \cite{Glauert_Rotating_Cylinder}
\[ \frac{\Gamma}{\Gamma_\omega} = 1-\frac{3}{\kappa^2} -\frac{3.24}{\kappa^4}+O\left(\frac{1}{\kappa^6};\nu\right), \]
over the region of interest ($\kappa>>1$), as shwon in Fig. \ref{Fig:Shorbagy_Magnus}. This ability to recover the correct inviscid limit with such analytical simplicity is not attainable within the classical formulation.

Another, perhaps even more, surprising outcome is the estimation of the separation angle $\beta$ for flow over a circular cylinder in the subcritical regime ($10^4<Re<10^5$), where global characteristics (e.g., mean force coefficients and the separation angle) are known to be largely insensitive to Reynolds number \cite{Zdravkovich_Cylinder}. It is well established that flow separation from a curved surface is fundamentally a viscous phenomenon, governed by the dynamics within the boundary layer and typically characterized by the vanishing of the wall friction coefficient. Consequently, there is no conceivable approach within the classical formulation to address this problem.

\begin{wrapfigure}{l}{0.40\textwidth}
\vspace{-0.35in}
 \begin{center}
 \includegraphics[width=5cm]{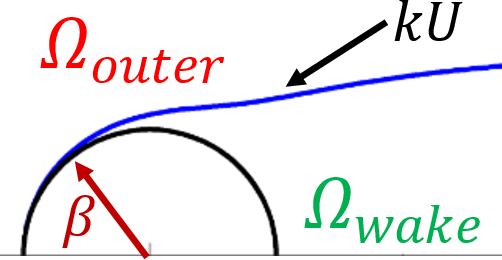}
 \caption{A schematic diagram for the free streamline theory model of the separating flow over a circular cylinder.}
 \label{Fig:FSL_Theory} \vspace{-0.25in}
 \end{center} \vspace{-0.3in}
\end{wrapfigure} \noindent Fortunately, there exists a well-established framework for modeling the outer flow in the subcritical regime---the \textit{free streamline theory}. This theory was developed by early pioneers, including Helmholtz (1868), Kirchhoff (1869), and Rayleigh (1876), initially for flow over a flat plate. It was later extended to the case of a circular cylinder by Brodetsky \cite{Brodetsky}, Roshko \cite{Roshko_Free_Streamline_NACA}, Wu \cite{Wu_FSL_Annual_Review}, among others.

The fundamental concept underlying free streamline theory is that the flow separates from the body in the form of vortex sheets, as illustrated schematically in Fig. \ref{Fig:FSL_Theory}. In the near-wake region, which is of primary interest in the free streamline (FSL) theory, the flow speed along each separating streamline is assumed to be approximately constant. The theory thus provides a potential-flow description of the outer region $\Omega_{outer}$, lying outside the separating streamlines. In addition, the slowly moving flow in the near-wake region $\Omega_{wake}$, bounded by the separating streamlines, is approximated as stagnant. In this sense, the separating streamlines act as sheets of discontinuity that delineate the outer inviscid flow from the wake.

The FSL theory provides a family $\mathcal{U}_s$ of kinematically-admissible velocity fields. Each member $\bm{u}\in\mathcal{U}_s$ can be written as:
\begin{equation}\label{eq:FSL_Family}
\bm{u}(\bm{x};k,\beta) = \left\{\begin{array}{lr}
\bm{u}_p(\bm{x};k,\beta), & \bm{x}\in\Omega_{outer} \\
0, & \bm{x}\in\Omega_{wake} \end{array}\right. ,
\end{equation}
where $\bm{u}_p$ represents the potential-flow model over the outer region $\Omega_{outer}$. This family has two parameters that must be specified through external considerations, such as experimental observations: (i) the constant speed along the free streamlines, represented by the parameter $k$, and (ii) the separation angle $\beta$.

In contrast to the family $\mathcal{U}$ of smooth attached-flow solutions, the family $\mathcal{U}_s$ comprises a collection of \textit{weak} (discontinuous) solutions of the Euler equation. As noted by Birkhoff, such solutions ``\textit{avoid the d'Alembert paradox}" (\cite{Birkhoff_Ideal_Lift_Logic}, p. 15) and therefore admit nonzero drag.

One question that remained open for decades is the following: for a given value of $k$, how should the separation angle $\beta$ be determined? This question is trivial for flat plates or a wedges, where a sharp edge typically dictates the separation location. For curved surface, however, such as a circular cylinder, the issue proved far more elusive and puzzled many researchers for decades \cite{Roshko_Free_Streamline_NACA,Parkinson_Jandali_FSL_JFM,Kiya_Arie_FSL_JFM}.

Brodetsky \cite{Brodetsky} and Roshko \cite{Roshko_Free_Streamline_NACA} proposed a curvature-matching condition to estimate the separation angle $\beta$. According to this criterion, the curvature of the separating streamline at the point of separation must match the local curvature of the cylinder surface. However, this condition predicts a separation angle that is significantly smaller than that observed experimentally. This discrepancy was revisited more recently in Ref.~\cite{Shorbagy_Separation_ArXiv}, where an extended form of the variational theory was employed to predict the separation angle $\beta$.

Shorbagy and Taha \cite{Shorbagy_Separation_ArXiv} adopted the FSL model developed in Ref. \cite{Miller2024}, which yields a family of kinematically admissible flows of the form (\ref{eq:FSL_Family}). For a fixed value of $k=\hat{k}$, the family $\mathcal{U}_s$ reduces to a one-parameter family of weak solutions of the Euler equation, parameterized by the separation angle $\beta$. The authors then posed the following variational problem: Over the family $\mathcal{U}_s$ of kinematically-admissible flows parameterized by $\beta$, minimize the Appellian
\[ S(\beta) = \frac{1}{2}\rho \int_\Omega \left[\bm{u}(\bm{x};\hat{k},\beta)\cdot\bm\nabla \bm{u}(\bm{x};\hat{k},\beta)\right]^2 d\bm{x} \]
with respect to $\beta$, where $\Omega=\Omega_{outer}\cup\Omega_{wake}$.

\begin{wrapfigure}{l}{0.50\textwidth}
\vspace{-0.35in}
 \begin{center}
 \includegraphics[width=7cm]{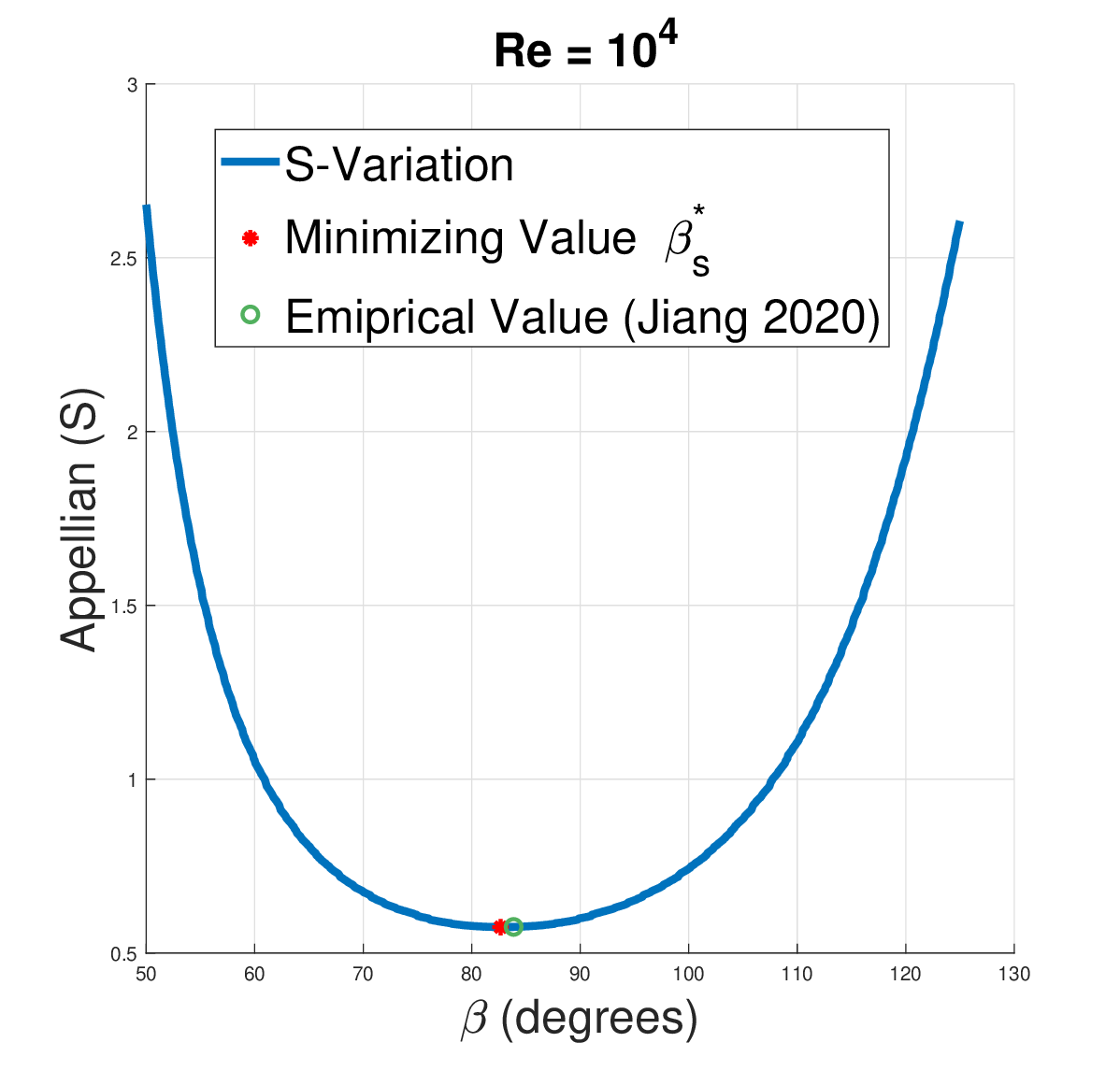}
 \caption{Variation of the Appellian with the separation angle $\beta$ at $Re=10^4$ showing a unique value $\beta^* = 83.7^\circ$ that minimizes the Appellian $S$, which is very close to the empirical prediction from \cite{jiang2020separation} ($\beta_{\text{Empirical}} = 83.85^\circ$).}
 \label{Fig:beta_star} \vspace{-0.1in}
 \end{center} \vspace{-0.1in}
\end{wrapfigure}

Figure \ref{Fig:beta_star} shows the variation of the Appellian with the separation angle $\beta$ at $Re=10^4$. Notably, the Appellian attains a unique minimum at $\beta^* = 83.7^\circ$ This minimizing value is remarkably close to the empirical prediction in \cite{jiang2020separation}: $\beta_{\text{Empirical}} = 83.85^\circ$. This level of agreement is not restricted to $Re=10^4$. The deviation between $\beta^*$ and $\beta_{\rm{Empirical}}$ remains below 5\% over the subcritical regime ($10^4-10^5$).

 \noindent This result warrants several comments. First, it is worth emphasizing the remarkable simplicity with which the variational formulation yields a physically accurate prediction of the separation angle without any explicit modeling of boundary-layer dynamics---a task long regarded as inaccessible within inviscid frameworks. Second, this example underscores that the variational theory is far more general than a theory of lift alone. Here, it is not invoked to determine circulation at all; instead, it is used to fix the separation angle---a fundamentally different quantity arising in a distinct physical setting.

Third, which is one of the key points in generalizing the variational theory, the framework is not confined to the restricted formulation presented in Ref. \cite{Variational_Lift_JFM}, where the admissible family $\mathcal{U}$ is parameterized solely by the circulation $\Gamma$. Rather, the variational formulation is inherently flexible and admits systematic extensions to other kinematically admissible families, such as the separating family $\mathcal{U}_s$, which are more representative of the underlying physics of the problem.

From this perspective, the variational formulation paves the way for the revival of theoretical and reduced-order modeling in aerodynamics---an area that has arguably reached a point of saturation under the rigidity of the classical theory. Consider, for example, the flow over a circular cylinder. If one is concerned solely with circulation and lift, minimization over the family $\mathcal{U}(\Gamma)$ may be sufficient. If, instead, the objective is to address flow separation, then the separating family $\mathcal{U}_s(k,\beta)$ is more physically appropriate. More generally, if additional physical features are of interest, fluid dynamicists may leverage physical insight and modeling intuition to propose richer kinematically admissible families $\mathcal{U}_r (P_1,...,P_n)$ in terms of more parameters $P_1$, ..., $P_n$. The variational formulation then supplies, in a systematic manner, the governing conditions for these parameters through the optimality conditions $\frac{\partial S}{\partial P_i}=0$.

This approach has already been adopted through neural-network parameterization of the flow field in Refs. \cite{PMPG_PINN_Daqaq,PMPG_PINN_PoF1,Hussam_Unsteady_PINNs_CMAME}. In that setting, neural networks are employed to construct a family $\mathcal{U}_{NN}$ of kinematically admissible flows, parameterized by a large set of weights and biases of the network. The variational principle is then applied over this high-dimensional family. This strategy enables the capture of rich flow physics while alleviating the burden of devising a smart parameterization based solely on physical intuition.

Moreover, when viscous effects are essential, the viscous cost functional (\ref{eq:Gauss_Incompressible}) may be employed instead. When the problem is unsteady, the variational formulation becomes even more faithful to Gauss's principle and the principle of minimum pressure gradient. In this unsteady setting, minimizing the instantaneous cost---whether viscous (\ref{eq:Gauss_Incompressible}) or inviscid (\ref{eq:Appellian_Continuum})---with respect to the parameters of a reduced-order model at each instant of time is mathematically equivalent to a Galerkin projection.

In fact, the parameterization need not be restricted to low-dimensional models; it may be taken to be as generic and exhaustive as a brute-force representation in terms of the nodal values of the velocity field on a computational mesh. In this most general setting, the resulting formulation falls squarely within computational fluid dynamics (CFD), rather than reduced-order modeling. The variational framework has been employed in such high-dimensional formulations in Refs.~\cite{NS_QP_IEEE,VPNS_PRF,Hussam_CQP_TAML}. In particular, Anand and Taha \cite{VPNS_PRF} developed a variational projection of the Navier–Stokes equations (VPNS), which enables projection without requiring the solution of a pressure Poisson equation.

Several researchers have adopted the variational formulation of minimum pressure gradient in high-dimensional computational settings across a range of flow problems, including flow over a sphere \cite{PMPG_PINN_PoF1}, steady \cite{PMPG_PINN_Daqaq} and unsteady lid-driven cavity flows \cite{Hussam_Unsteady_PINNs_CMAME,NS_QP_IEEE,VPNS_PRF,Hussam_CQP_TAML}, as well as pipe flow \cite{Hussam_Unsteady_PINNs_CMAME}. These studies make it clear that the formulation is far more general than a theory of lift in potential flow.

In summary, the variational formulation of minimum pressure gradient provides a unified framework encompassing a broad spectrum of models: from one-dimensional formulations in the classical airfoil lift problem to high-dimensional direct numerical simulations of turbulent flows. The aerodynamicist selects the desired level of fidelity through the chosen parameterization. Yet, whether in a one-dimensional reduced-order model or in a billion-cell high-fidelity simulation, the parameters are determined by minimizing the instantaneous magnitude of the pressure-gradient force required to ensure the continuity and no-penetration constraints. The circulation in the airfoil problem \cite{Variational_Lift_JFM}, the separation angle in subcritical separating flows \cite{Shorbagy_Separation_ArXiv}, the neural-network weights and biases in pipe flow \cite{Hussam_Unsteady_PINNs_CMAME}, and the nodal values in numerical simulations of the unsteady lid-driven cavity \cite{NS_QP_IEEE,VPNS_PRF} are all determined by the same underlying principle---the principle of minimum pressure gradient.

\section{On Viscosity}
Given the comprehensive nature of this paper, we feel obliged to address, at least partially, the role of viscosity, which has remained a source of conceptual confusion and debate in the theory of lift since the early controversy between the Göttingen and Cambridge schools \cite{Bloor_Enigma}. This topic is exceptionally rich and cannot be treated adequately within a single section. Interested readers are therefore referred to the author's three-lecture series entitled ``\textit{D'Alembert Paradox: A Tale of Two Schools}" \cite{DAlembert_Paradox_Videos1,DAlembert_Paradox_Videos2,DAlembert_Paradox_Videos3}, where the subject is discussed in depth from historical, mathematical, and physical perspectives. Here, we confine ourselves to three essential points that are directly relevant to the present discussion.

First, even if the variational theory of lift were to suggest that viscosity is \textit{not} necessary for lift generation, and even if compelling evidence existed indicating that viscosity \textit{is} necessary, an unbiased scholarly approach would not discard the former on the basis of the latter alone. Rather, such an approach would treat the following three possibilities on equal footing: (i) the variational theory may be incorrect; (ii) prevailing conclusions regarding the role of viscosity in lift generation may be incomplete or mistaken; or (iii) the two results may be reconcilable within a broader theoretical framework.

Second, even if the variational theory of lift presented in Ref.~\cite{Variational_Lift_JFM} were entirely correct, it cannot, by construction, make definitive statements about the role of viscosity in lift \textit{generation}. Lift generation is inherently an unsteady process, whereas the variational theory addresses the steady-state equilibrium of the flow. As already emphasized in Ref.~\cite{PMPG_PoF} in the context of the variational formulation, ``\textit{It should be noted this result may not imply that the lift mechanism is purely inviscid. It only implies that the \textup{steady state equilibrium} of the lift problem can be obtained without invoking viscosity. The Kutta condition has traditionally been used to provide such a steady state equilibrium. However, the pervasive conventional wisdom suggests that the Kutta condition is a viscous condition. Here it is shown that the Kutta condition is a special case of a minimization principle for momentum conservation: the PMPG.}"

Third, the authors of Ref. \cite{Peters_Gauss} invoke Hele-Shaw flow, which exhibits a non-lifting behavior over an airfoil as ``\textit{analog for inviscid, incompressible, irrotational, steady flow.}" They then interpret the experimental observation that Hele-Shaw flow produces no lift as follows: ``\textit{This 'inviscid-flow-by-analogy' example refutes the variational theory prediction of ideal fluid lift for this airfoil. Hele–Shaw flow thus provides experimental evidence that disproves the variational theory of lift.}"

The first point above already exposes the logical flaw in the authors’ conclusion. More fundamentally, however, it reflects a widespread misconception regarding the status of Hele–Shaw flow as a surrogate for Euler dynamics. It is useful to recall the well-known extremes of the spectrum of Navier-Stokes' solutions: (i) the inviscid limit, represented by Euler's equation, in which the viscous term vanishes; and (ii) the Stokes' limit in which inertial effects vanish and viscous forces dominate. In light of this standard classification, conclusions such as the one drawn above rest on a fundamental misinterpretation of the governing physics.

As acknowledged by the authors of Ref. \cite{Peters_Gauss} themselves, Hele-Shaw flow is obtained ``\textit{by neglecting the unsteady and inertial terms in comparison to the viscous terms.}" It therefore represents the opposite asymptotic limit: a highly viscous, creeping-flow regime in which inertia is negligible. Consequently, if any conclusions are to be drawn from the experimental observation that such flows are non-lifting, they cannot be taken as evidence for an essential role of viscosity in lift generation.

\section{Conclusion}
In this paper, we have examined the claims made in Ref. \cite{Peters_Gauss} concerning the variational theory of lift. We acknowledge the substantial effort devoted in that work to scrutinizing a newly emerging theoretical framework. Such scrutiny is both necessary and welcome, as it plays an essential role in ensuring that flawed ideas do not infiltrate the scientific literature. From this perspective, the efforts in Ref. \cite{Peters_Gauss} must be commended. However, upon examining the specific claims advanced in that paper, we find that several of them rest on misconceptions regarding basic principles of analytical mechanics and the calculus of variations, as well as on misunderstandings of the respective domains of applicability of the classical and variational theories of lift.

This paper has clarified the foundations and scope of the variational theory of lift by revisiting a number of claims advanced in Ref.~\cite{Peters_Gauss} through the lens of classical analytical mechanics and aerodynamics. The present work has sought not merely to respond to specific criticisms, but to place the variational theory of lift within its proper mathematical, physical, and historical context.

At the level of mechanics, we have shown---via the Helmholtz decomposition---that, for incompressible flows subject to the no-penetration boundary condition, the pressure force is orthogonal to the entire space of kinematically admissible velocity fields. Contrary to the interpretation advanced in Ref. \cite{Peters_Gauss}, the \textit{virtual} work of the pressure force therefore vanishes in this setting. As a consequence, the pressure gradient does not influence the dynamics of an incompressible velocity field when projected onto its configuration manifold.

We have further revisited the \textit{Simplest Problem in the Calculus of Variations} and the standard treatment of boundary conditions in such problems through restricting admissible variations, rather than through Lagrange multipliers. This distinction clarifies the misunderstanding underlying the claim in Ref. \cite{Peters_Gauss} that boundary conditions must necessarily be imposed via Lagrange multipliers---a misconception that led to inaccurate conclusions regarding the theorem established in Ref. \cite{Variational_Lift_JFM} on the equivalence between Appellian minimization and Euler’s dynamics.

From an aerodynamic standpoint, we have demonstrated that both the classical theory and the variational theory of lift, as well as any theory based on (i) steady flow and (ii) irrotational motion, must be reversible. Such theories therefore cannot distinguish between forward and reversed flow configurations. As a consequence, none of these theories is applicable to reversed-flow problems. Hence, the application of the variational theory to reversed flows in Ref. \cite{Peters_Gauss} constitutes an extension of the theory beyond its domain of applicability. As a consequence of this misunderstanding, the authors of Ref. \cite{Peters_Gauss} drew inaccurate conclusions regarding discontinuous behavior in the variational theory of lift. On the contrary, we have shown that, within its domain of applicability, the circulation that minimizes the Appellian depends continuously on the geometric parameters of the body. The variational framework makes this continuity explicit, as it is guaranteed by the smooth dependence of admissible velocity fields and the Appellian functional itself on shape parameters.

Finally, we have presented a comparative assessment of the classical and variational theories of lift in terms of their respective domains of applicability, sensitivity to geometric details, and capacity for systematic generalization. We have further shown how the variational formulation naturally extends beyond the prediction of circulation in airfoil problems to the prediction of separation angles over circular cylinders in the subcritical regime, and ultimately to the construction of high-dimensional computational fluid solvers.


\bibliography{Fluid_Dynamics_References,Aeronautical_Engineering_References,Geometric_Control_References,Dynamics_Control_References,Math_References,My_Published_References,Flapping_References,History_Philosophy_References,Viscous_Corrections_Ref}
\bibliographystyle{unsrt}

\end{document}